\documentclass[pra,aps,amssymb,onecolumn,noshowpacs]{revtex4-1}
\usepackage{color}
\usepackage{graphicx}
\usepackage{hyperref}
\usepackage{amsmath}
\usepackage{latexsym}
\usepackage{amssymb}
\usepackage{mathrsfs}
\usepackage{mathtools}
\usepackage{layout}
\usepackage{verbatim}
\usepackage{multirow}
\usepackage{bm}
\usepackage{amsfonts,epsfig}
\usepackage[flushleft]{threeparttable}
    
\begin{document}

\title{Quantum logic and entanglement by neutral Rydberg atoms: methods and fidelity}
\date{\today}
\author{Xiao-Feng Shi}
\affiliation{School of Physics and Optoelectronic Engineering, Xidian University, Xi'an 710071, China}

\begin{abstract}
Quantum gates and entanglement based on dipole-dipole interactions of neutral Rydberg atoms are relevant to both fundamental physics and quantum information science. The precision and robustness of the Rydberg-mediated entanglement protocols are the key factors limiting their applicability in experiments and near-future industry. There are various methods for generating entangling gates by exploring the Rydberg interactions of neutral atoms, each equipped with its own strengths and weaknesses. The basics and tricks in these protocols are reviewed, with specific attention paid to the achievable fidelity and the robustness to the technical issues and detrimental innate factors. 
  
\end{abstract}
\maketitle

\section{Introduction}\label{sec01}
Quantum mechanics~(QM) is both of fundamental interest and beneficial for industries and technologies in, e.g., determination of structures of disease-related biomolecules~\cite{Gremer2017,PhysRevE.87.022712} and high-precision time and frequency standard~\cite{Bloom2014,Ludlow2015}. The study of QM was initialized by Max Planck in 1900 but a complete mathematical formalism was not fully developed until a series of experiments and efforts to explain them took place. From then on, many efforts were directed at the study on QM which led to the invention of lasers that facilitated the study of QM by, e.g., enabling creation of entangled photons via spontaneous parametric down-conversion in laser excited atoms~(for a review, see, e.g.,~\cite{Kok2007,Pan2012}), optical control over quantum states in semiconductors~(for a review, see, e.g.,~\cite{Liu2010,Awschalom2013}), cooling of atoms~\cite{PhysRevLett.55.48,PhysRevLett.61.169,PhysRevLett.61.826}, quantum manipulation of individual atoms~\cite{PhysRevLett.77.4887,RevModPhys.73.565} and ions~\cite{Monroe1995,RevModPhys.75.281,PhysRevLett.76.1796}, and quantum computing~(QC)~\cite{Benioff1980,Benioff1982,Feynman1982}. 

QC aims to solve challenging computational tasks by using quantum mechanical rules to manipulate the information stored in quantum registers~\cite{Benioff1980,Benioff1982,Feynman1982}. The mathematical foundation was laid for QC soon after QC was introduced~\cite{Deutsch1989,DiVincenzo1995,Barenco1995,Sleator1995,Steane1996}, but it was after the introduction of pioneering QC algorithms~\cite{Shor1997,Grover1997,Grover1998} did people recognize its extremely appealing power. Since then, there has been much interest in realizing large-scale QC~\cite{Ladd2010}.

\subsection{Physical implementation of quantum computing}
 One of the biggest challenges in realizing QC is implementing QC on the physical level. Although several physical platforms such as systems of cavity quantum electrodynamics~\cite{Pellizzari1995}, NMR~\cite{Cory1998}, and semiconductor quantum dots~\cite{Loss1998} were often studied in the early days~\cite{DiVincenzo2000,Nielsen2000}, the first quantum entangling gate, an important element in the circuit model of QC, was demonstrated with trapped ions~\cite{Monroe1995}. Trapped ions can be accurately entangled thanks to the strong Coulomb interactions between ions~\cite{Cirac1995,Wineland1998,Molmer1999,Sorensen1999}, and until now two-particle entanglement with the highest fidelity~(about 0.9992) was realized with trapped ions~\cite{Gaebler2016}. Another notable candidate for QC is superconducting systems which include (i) superconducting circuits~(SC) in which a quantum bit~(qubit) can be defined in various ways, from charge qubit, charge-flux qubit, and flux qubit to phase qubit~\cite{You2005}, and (ii) superconducting coplanar waveguides which can be easily coupled with SC and naturally support microwave qubits~\cite{ChengP.Wen1969,PhysRevA.69.062320,Goppl2008,Megrant2012} that enable hybrid quantum systems. SC require a cryogenic environment to operate, but can host a large-scale qubit array and be conveniently manipulated by microwave fields. These motivated many efforts to create entanglement with SC~(for a review, see, e.g.,~\cite{Clarke2008,Chuang2013}) after the first attempt of coherent control over qubits in SC~\cite{Nakamura1999}. Two-qubit entangling operations with SC can reach a fidelity 0.991 in \cite{Sheldon2016} and up to~0.997 in~\cite{Kjaergaard2020}. SC offer a high-level scalability on a chip, but both the relaxation time~($T_1$) and the coherence time~($T_2^{\ast}$) of the qubit are short, and up until now their largest values realized are $(T_1,~T_2^\ast)=(70,~92)~\mu$s~\cite{Rigetti2012}. The fast dissipation means that about $10^3$ cycles of two-qubit entanglement can be implemented if each entanglement operation needs $60$~ns in SC~\cite{Barends2014,Kjaergaard2020}. For trapped ions, about $1.8\times10^6$ entangling cycles can be realized with the setup of, e.g., Ref.~\cite{Ballance2015}, where two-qubit gate times 27.4~$\mu$s were reported with a qubit coherence time 50s . However, the states of trapped ions are difficult to be accurately manipulated when the number of ions in one quantum register is large~\cite{Monroe2013}. Either the scalability or the stability issue seem an unsolvable problem for these two widely-studied QC candidates. Similar issues exist for other QC candidates~\cite{Pan2012,Pla2013,Veldhorst2015,Zhong2015} which motivates the study of hybrid quantum systems in order to bring together to the largest extent only the strengths of each element in the hybrid system~\cite{Xiang2013}. However, building a hybrid system does not eliminate the weaknesses of each system. Though there is interest in the light-matter hybrid~\cite{Forn-Diaz2019,Marpaung2019,Clerk2020,Elshaari2020,Wang2020} in that light is a useful carrier for a flying qubit, it is difficult to use photons for QC because of lack of a protocol to deterministically entangle photons. All these challenges seem to suggest that it is hard to realize realistic and large-scale QC.

\begin{table*}[ht]
     \begin{threeparttable}
  \begin{tabular}{|c|c|c|c|c|}
    \hline
    &  \multirow{2}{*} {Number of qubits} &   \multirow{2}{*} {Coherence time} &  \multicolumn{2}{|c|}{Fidelity and duration of quantum operations } \\
      \cline{4-5}
    &  &  & One-qubit gate\tnote{a} & Two-qubit gate or Bell state\tnote{b} \\\hline
    SC& 53~\cite{Arute2019};~54~\cite{Neill2021}&$70~\mu$s~\cite{Rigetti2012}\tnote{c}   & $0.9992$;~$10$~ns~\cite{Barends2014}\tnote{c}  &$0.997$;~$60$~ns~\cite{Kjaergaard2020}~~(C$_{\text{Z}}$ gates)\\\hline
    Trapped ions &53~\cite{Zhang2017} & 50~s~\cite{Harty2014,Harty2016} &$0.999999$;~$12~\mu$s~\cite{Harty2014}  & $~~0.9992$;~$30~\mu$s~\cite{Gaebler2016}~(Bell states)\\\hline
    Neutral atoms& 209~\cite{Schymik2020};~219~\cite{Semeghini2021};~256~\cite{Ebadi2020} & 7~s~\cite{Wang2016}; 48~s~\cite{Young2020}\tnote{d} & $0.99986$;~$31~\mu$s~\cite{Olmschenk2010} &$~~0.991$;~$59$~ns~\cite{Madjarov2020}\tnote{e}~~(Bell states)\\\hline  
  \end{tabular}
  \caption{ Comparison of the achievable performances between three types of systems regarding QC. Numbers shown above are representative data. For the number $N$ of qubits that can be prepared in one register, other notable results include $N=40$ for trapped ions~\cite{Pagano2020}, and $N\sim50$ in Refs.~\cite{Xia2015,Wang2016,Bernien2017}, $N\sim150$ in Ref.~\cite{Young2020}, $N=184$ in Ref.~\cite{Scholl2020}, and $N=200$ in Ref.~\cite{Bluvstein2021} for neutral atoms; for fidelities $\mathcal{F}_{1(2)}$ of single(two)-qubit gates, other notable results include $\mathcal{F}_2=0.991$~\cite{Sheldon2016} and $0.9944$~\cite{Barends2014} for SC, and $\mathcal{F}_1=0.998$~\cite{Xia2015,Wang2016} and $\mathcal{F}_2=0.974$~\cite{Levine2019} for neutral atoms. Here, results with larger fidelities are shown. Faster gates based on a similar mechanism can have smaller fidelities as studied in Ref.~\cite{Ballance2016}; take trapped ions as example, Ref.~\cite{Gaebler2016} studied single-qubit gates of duration $2~\mu$s and fidelity $0.99996$, and Ref.~\cite{Schafer2018} studied an entangling gate of duration $1.6~\mu$s and fidelity $0.9982$.    \label{table1}  }
  \begin{tablenotes}
        \item[a] The duration for single-qubit gates refers to a Clifford gate such as a $\pi/2$ rotation between the two states of a qubit.
        \item[b] The time here refers to the duration of either implementing a controlled-Z~(C$_{\text{Z}}$) gate or creating a Bell state from a product state.
    \item[c] The coherence time for superconducting qubits refers to the smaller one among the relaxation time~($T_1$) and the decoherence time~($T_2^{\ast}$) of Ref.~\cite{Rigetti2012}; the single-qubit gate data are taken from Table S2 of the supplementary information of Ref.~\cite{Barends2014}.
        \item[d] Unlike that in Ref.~\cite{Wang2016} which studied qubits formed by ground states, the coherence time in Ref.~\cite{Young2020} refers to that of the optical clock state $(5s5p)^3P_0$ of $^{88}$Sr. Ref.~\cite{Young2020} reported an atomic coherence time up to 48s.
          \item[e] A Rabi frequency $\Omega=2\pi\times6-7$~MHz was used in Ref.~\cite{Madjarov2020} so that a $\pi$ pulse for exciting the ground to Rydberg states has a duration $\pi/(\sqrt{2}\Omega)\sim51-59$~ns with $\sqrt{2}$ a many-body enhancement factor.   
    \end{tablenotes}
     \end{threeparttable}
\end{table*}

\subsection{Quantum computing with neutral atoms}
Neutral atoms are promising for QC because they are equipped with the capability to support hundreds of identical qubits in one array~\cite{Ebadi2020,Schymik2020,Semeghini2021,Young2020,Scholl2020,Bluvstein2021}, long coherence times~\cite{Wang2016,Young2020}, high-fidelity single-qubit gates~\cite{Olmschenk2010,Xia2015,Wang2016}, and high-fidelity two-qubit entanglement~\cite{Madjarov2020}. Compared to SC and trapped ions shown in Table~\ref{table1}, neutral atoms have been experimentally demonstrated to be capable to form an array with more qubits, and the ratio between the qubit coherence time and the time to carry out a quantum entangling gate is large. Theoretical analysis estimated that up to a thousand qubits in one register can in principle be prepared with neutral atoms~\cite{Henriet2020}. Most remarkably, a unique advantage of neutral atoms is that it is possible to manipulate trapped atoms in both their internal state space and external space using external fields~\cite{Barredo2016,Endres2016,Barredo2018,Schlosser_2020,Bluvstein2021,Semeghini2021} which is a feature by no means shared by SC or trapped ions. This enables a feasibility to carry out analog quantum computation tasks of specific optimization problems via neutral atoms~\cite{Pichler2018,Serret2020}. It is also possible to simulate condensed matter physics~\cite{Browaeys2020} including quantum phase transitions~\cite{Bernien2017,Keesling2019}, topological spin liquids~\cite{Semeghini2021}, symmetry-protected topological phases~\cite{DeLeseleuc2019}, and quantum many-body scars~\cite{Bluvstein2021} that are difficult to probe in solid-state systems.
\begin{figure*}[ht]
\includegraphics[width=5.0in]
{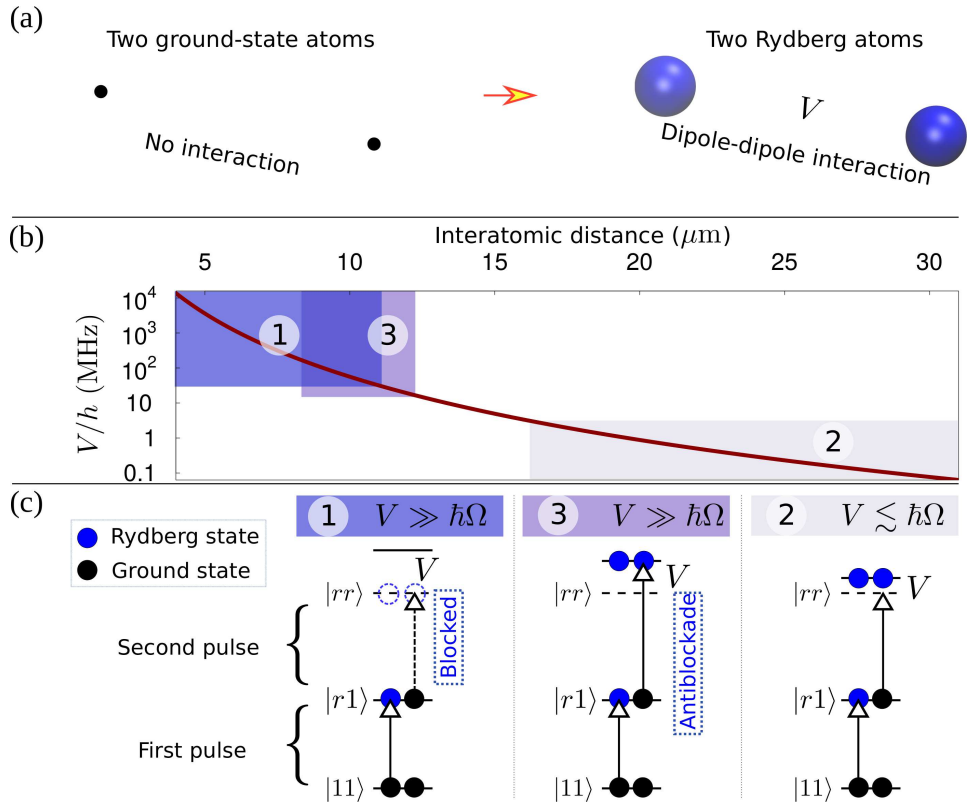}
 \caption{Overview of Rydberg-mediated entanglement in three regimes. (a) shows that when two atoms are both in the Rydberg states, an electric dipole-dipole interaction occurs. (b) shows interaction strength $V$ between two rubidium atoms in an $s$-orbital Rydberg state $|r\rangle$ of a principal quantum number $100$, where $C_6$ is the van der Waals coefficient. $V$ drops off according to $C_6/L^6$ when the interatomic distance $L$ is large compared to a crossover distance $L_{\text{c}}\approx5~\mu$m; when $L\lesssim L_{\text{c}}$, the dipole-dipole interaction couples the state $|rr\rangle$ to many nearby states~[see Fig.~\ref{figure-orientation}(c)], and is not well described by a pure energy shift $C_6/L^6$ so that the interaction shown here is valid for $L$ considerably larger than $L_{\text{c}}$~(for example, at least larger than $1.6L_{\text{c}}$). (c) shows the key steps for three basic classes of entangling gates. After one atom is excited from the ground state $|1\rangle$ to the Rydberg state $|r\rangle$, a nearby atom which is called the target qubit is excited with a laser Rabi frequency $\Omega$ and detuning $\Delta$. There are three regimes: \textcircled{1} when $\hbar\Omega\ll V$ and $\Delta=0$, the excitation of the target qubit is blocked; \textcircled{2} when $\hbar\Omega\gtrsim V$ and $\Delta=0$, the target atom is excited to $|r\rangle$; and \textcircled{3} when $\hbar\Omega\ll V$ and $\hbar\Delta$ exactly compensates the energy shift $V$, the target qubit can be fully excited to $|r\rangle$. For the regimes \textcircled{1} and \textcircled{3}, the presence of $V$ leads to different state dynamics compared to the case when $V=0$, so that entangling gates can be designed based on them. For the regime \textcircled{2}, a dynamic phase can arise when the two-atom state $|rr\rangle$ is left there for a certain time, which can lead to conditional phase gates, and, thus, entanglement for appropriate input states. Rydberg-mediated entanglement in regimes \textcircled{2} and \textcircled{3} requires either isolated dipole-dipole flip-flop processes or a pure energy shift as reviewed in Secs.~\ref{sec04} and~\ref{sec05}, so that the two atoms should be separated large enough~[see, e.g., Fig.~\ref{figure-orientation}(c)]; for the methods in the regime \textcircled{1}, the two atoms can be placed close enough as long as their wavefunctions do not overlap. For this reason, the allowed interval of interatomic distances as denoted by the shaded region in (b) is narrower for regime  \textcircled{3} compared to the regime \textcircled{1}.        \label{figure01} }
\end{figure*}

Neutral atoms that are separated micrometers away have negligible interaction with each other in their ground states, but can exhibit strong dipole-dipole interactions when they are excited to high-lying Rydberg states because of their large electric dipole moments~\cite{Gallagh2005,Saffman2010,Walker2012,Saffman2016,Weiss2017,Saffman2019,Adams2020,Browaeys2020}. Typical alkali-metal isotopes like $^{87}$Rb and $^{133}$Cs have ground-state hyperfine splittings of several gigahertz which make them ideal for state-selective optical excitation to high-lying Rydberg states. Methods for fast entanglement generation based on Rydberg interactions was first proposed about two decades ago~\cite{PhysRevLett.85.2208,Lukin2001}, and experimentally demonstrated ten years later~\cite{Wilk2010,Isenhower2010}. The strength of the method by Rydberg interactions, as compared to those by collisions~\cite{Calarco2000,Hayes2007,Daley2008,Cappellini2014,Scazza2014,Kaufman2015}, lies in the fast speed for turning on and off of interatomic dipole-dipole interactions by external control fields. However, an individual atom always moves, the optical trap has finite lifetimes, and the technique to create and observe entanglement between individual atoms is delicate, making the advance toward high-fidelity entanglement relatively slow~\cite{Wilk2010,Isenhower2010,Zhang2010,Maller2015,Jau2015,Zeng2017,Levine2018,Picken2018,Jo2019,Levine2019,Graham2019,Madjarov2020}. Although the best entanglement fidelity $0.97$~\cite{Levine2019} between ground states of neutral alkali-metal atoms looks inferior to those with trapped ions~\cite{Ballance2016,Gaebler2016,Schafer2018} and superconducting circuits~\cite{Barends2014,Kjaergaard2020}, there are a number of theoretical proposals that are not yet experimentally tested but can lead to more accurate quantum gates in future. Each theoretical protocol has its own requirements, strengths, and weaknesses, which will be reviewed.

\subsection{Methods of quantum entanglement with neutral atoms: an overview}\label{secIC}
 
 The essence of entanglement by the Rydberg interaction $V$ is that there should be certain modifications in the dynamics of the two or multi-atom wavefunction compared to their dynamics without interactions. To achieve this, there are in general three classes of methods with their most basic examples shown in Fig.~\ref{figure01}.

The first method is to use the blockade mechanism~\cite{PhysRevLett.85.2208}. When one atom is already in a Rydberg state, a nearby atom can not be excited to a Rydberg state by a resonant laser with a Rabi frequency $\Omega$ that is small compared to $V/\hbar$, where $\hbar$ is the reduced Planck constant. This method only involves one Rydberg excitation, while a residual multi-atom Rydberg excitation can appear as an error due to the finiteness of the Rydberg interactions. The blockade-based method can tolerate the fluctuation of the qubit spacing. Most Rydberg-mediated entanglement experiments were carried out via the blockade mechanism~\cite{Wilk2010,Isenhower2010,Zhang2010,Maller2015,Jau2015,Zeng2017,Levine2018,Picken2018,Levine2019,Graham2019,Madjarov2020}.

The second method is to excite two or more atoms to Rydberg states, either fully or partially. In other words, this class of methods work with multi-atom Rydberg state. The first gate based on this method was proposed in~\cite{PhysRevLett.85.2208} for a two-qubit phase gate by directly using the dynamical phase accumulation when both atoms are Rydberg excited. The original protocol in~\cite{PhysRevLett.85.2208} requires a pure van der Waals type of interaction $V$ that is small compared to $\hbar\Omega$. This methods works well if $V$ does not exhibit much fluctuation. Later on there are variant extensions that depend on isolated dipole-dipole flip-flop processes as viewed in Sec.~\ref{sec04}. There was one experiment on entanglement of individual atoms with this mechanism~\cite{Jo2019}.  %

The third method to entangle neutral atoms is to explore the antiblockade regime of Rydberg interactions~\cite{Ates2007,Amthor2010}. Because $V$ appears as a blockade interaction when the Rydberg laser field is resonant, one can shift the frequency of the Rydberg lasers to compensate $V$ so that resonance is recovered. This class of methods do not necessarily need to excite more than one atoms simultaneously to Rydberg states, but they do depend on ``frozen'' Rydberg interactions so as to fulfill the antiblockade condition. Entangling gate protocols in this regime mainly focus on the condition of $V\gg\hbar\Omega$ as reviewed in~\cite{Su2020epl}, and up till now entanglement of individual atoms has not been tested in experiments via the antiblockade mechanism. The second and the third classes of methods both involve two or multi-atom Rydberg states, but the latter involves compensating the Rydberg interaction by, e.g., a detuning in the Rydberg Rabi oscillation.

The three classes of methods discussed above involve Rydberg interactions in different regimes, shown in Fig.~\ref{figure01}. As shown later in Secs.~\ref{sec04} and~\ref{sec05}, the second method can be extended to conditions $V> \hbar\Omega$, and there are some extensions of the third method to the condition $V\sim \hbar\Omega$. However, the common feature for those extensions in Sec.~\ref{sec04} is still the excitation of multiqubit Rydberg states, and that in Sec.~\ref{sec05} is still matching the Rydberg interaction~(divided by $\hbar$) with either the frequency of lasers or certain time-dependence patterns of the field amplitudes of lasers.

The above classification is based on the the different roles played by the Rydberg interaction $V$. On the other hand, one can classify the Rydberg gates according to the required types of pulses. First, pulses chopped from continuous lasers by pulse pickers can be used to excite the ground-Rydberg transitions, where there is no specific requirement about the shape or the strength of the field as long as the total pulse area equals a desired value. Almost all entangling gate experiments to date were in this regime~\cite{Wilk2010,Isenhower2010,Zhang2010,Maller2015,Jau2015,Zeng2017,Levine2018,Picken2018,Jo2019,Graham2019,Madjarov2020}. Second, quasi-rectangular pulses are used to excite the ground-Rydberg transitions, but the field strength is expected to be constant and there is some requirement about the ratio between the strength and detuning of the laser fields as in the gate protocols of Refs.~\cite{Shi2018Accuv1,Levine2019}. Third, there are protocols based on shaping the pulses so that the amplitude and frequency of the field change analytically~\cite{PhysRevLett.85.2208,Muller2009,Beterov2013,Muller2014,Keating2015,Beterov2016jpb,Theis2016,Moller2008,Wu2017,Beterov2011,Beterov2013,Beterov2014,Beterov2016jpb,Kang2018,Shen:19,Kang2020,Guo2020,Saffman2020,Beterov2020,Mitra2020,Beterov2016,Petrosyan2017,Beterov2018,Yu2019,Liao2019,Sun2020,Khazali2020,Tian2015,Shao2017ground,WU2020126039,Wu:20,Li2020,Wu2021} or even nonanalytically~\cite{Muller2011,Goerz2014} during the gate sequence. The pulse-shaping method in principle can yield large gate fidelities provided fine pulse shaping techniques are available, and it can suppress errors from certain technical imperfections. For this reason, pulse shaping was frequently used in theoretical proposals for Rydberg quantum gates based on the blockade mechanism~\cite{PhysRevLett.85.2208,Muller2009,Beterov2013,Muller2014,Keating2015,Beterov2016jpb,Theis2016,Moller2008,Wu2017,Beterov2011,Beterov2013,Beterov2014,Beterov2016jpb,Kang2018,Shen:19,Kang2020,Guo2020,Saffman2020,Beterov2020,Mitra2020} or the Rydberg excitation of multi-atoms~\cite{Beterov2016,Petrosyan2017,Beterov2018,Yu2019,Liao2019,Sun2020,Khazali2020}~(see Tables~\ref{table3} and~\ref{table4}); it could also be used for quantum entanglement with the antiblockade mechanism~\cite{Tian2015,Shao2017ground,WU2020126039,Wu:20,Li2020,Wu2021} although there were more entanglement protocols without pulse shaping when using the antiblockade mechanism~(see Table~\ref{table5}).

Another way to sort the different protocols focuses on whether dissipation helps for entanglement generation. Atoms excited to Rydberg state can cascade down to lower states and back to ground states because of the vacuum fluctuation and blackbody radiation. The dissipation involves both population decay and pure dephasing. When a state rotation fast compared to the dissipation is considered, one can ignore the dissipation and design quantum entanglement based on unitary dynamics. But when a very slow state rotation is employed, the dissipation can be fast compared to the speed of ground-Rydberg state rotations. Then, one can also use the dissipation to generate entanglement. The experiments~\cite{Wilk2010,Isenhower2010,Zhang2010,Maller2015,Jau2015,Zeng2017,Levine2018,Picken2018,Jo2019,Levine2019,Graham2019,Madjarov2020} so far did not explore proposals for entanglement via dissipation in neutral atoms. 

Each method of entanglement generation has its strengths concerning the fidelity and robustness against noise. Some can provide large fidelities if the fluctuation of $V$ is suppressed and laser fields have stable strength and phase, and some can yield high accuracy even if $V$ fluctuates substantially, and some are robust against the noise of the field, and there are also some that are robust against the Doppler dephasing of the ground-Rydberg transitions. Most theoretical protocols have not been tested even in proof-of-principle experiments because current efforts still focus on suppressing errors due to atomic motion and noise of the laser fields although remarkable advances were made~\cite{Levine2018,Levine2019,Graham2019}. The cooler the qubits and the shorter the gate duration, the larger the fidelity will be. The highest fidelity $0.995$ of neutral-atom entanglement by Rydberg interactions was achieved in divalent atoms~\cite{Madjarov2020} where qubits were cooled close to the motional ground states in the traps and an very fast state rotations between the Rydberg and the metastable clock states were used. %

Rydberg atoms seem to have become a superstar in quantum science and technology, not only for quantum computing~\cite{Adams2020,Saffman2010,Walker2012,Saffman2016,Weiss2017,Saffman2019} but also for quantum many-body physics as reviewed in Refs.~\cite{Browaeys2020,Morgado2021,Wu2021review}. This review mainly focuses on the theoretical methods for creating quantum gates and entanglement with Rydberg interactions in hope to inspire future study for realizing high-fidelity quantum gates useful in reliable, practical, and large-scale quantum computing with neutral atoms. 

The remainder of this review is organized as follows. Section~\ref{section02} introduces the basics about Rydberg-mediated entanglement, Sec.~\ref{sec03} reviews the Rydberg blockade mechanism for entanglement, Sec.~\ref{sec04} reviews entanglement methods by exciting more than one qubits to Rydberg states, and Sec.~\ref{sec05} reviews entanglement methods by the antiblockade mechanism. Section~\ref{discussions} discusses typical challenges in the study of Rydberg-mediated entanglement. A conclusion is given in Sec.~\ref{concl}.

\section{Basics}\label{section02}
In this section, we review the basic elements in quantum information processing with neutral Rydberg atoms which include the way to store quantum information with neutral atoms, the excitation of Rydberg states, the interactions between Rydberg atoms, and the factors that influence the fidelity of quantum entanglement by Rydberg interactions.

\begin{figure*}[ht]
\includegraphics[width=6.0in]
{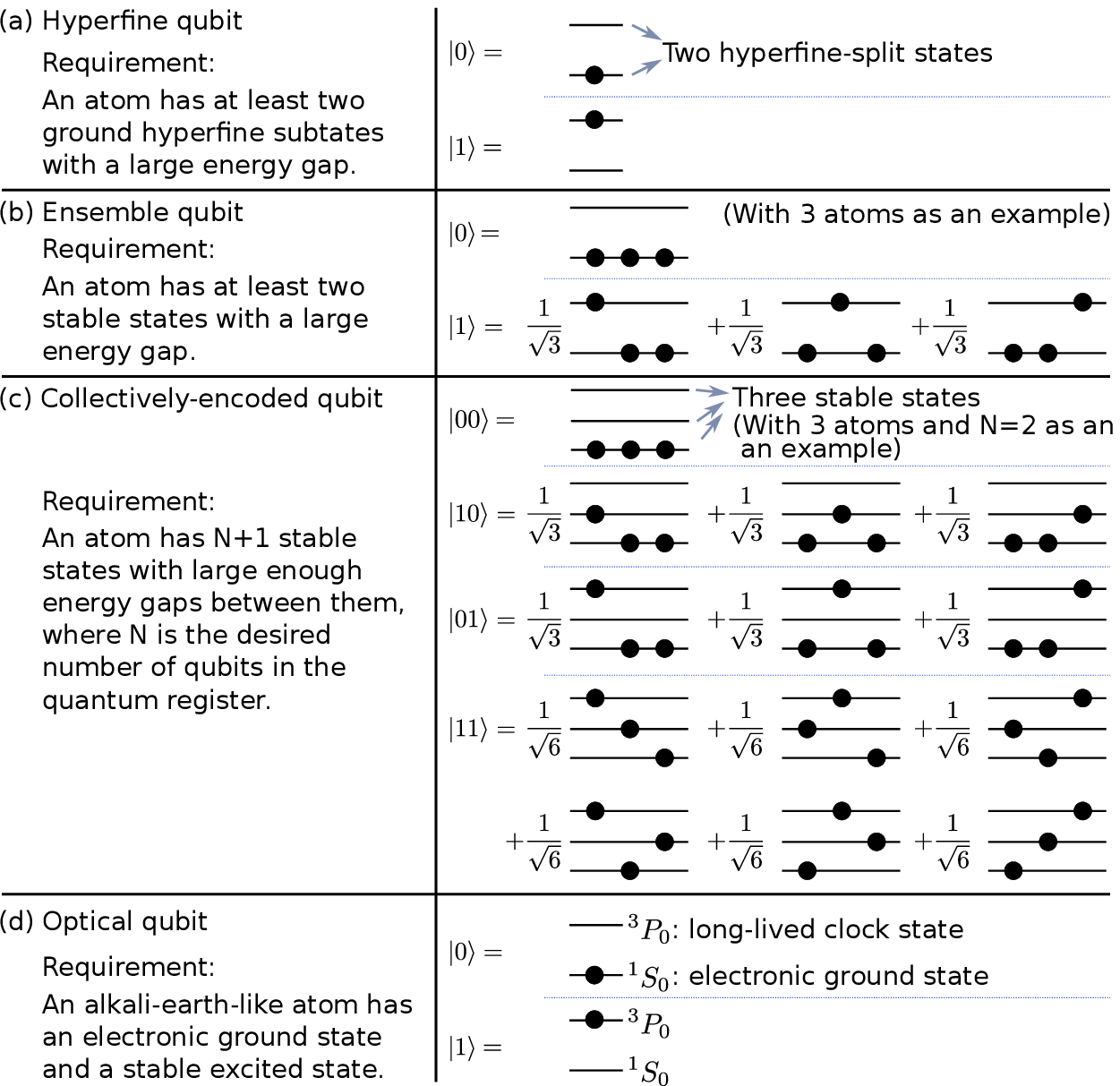}
 \caption{Four typical methods to define qubits with individual neutral atoms. (a) For atoms with a GHz-scale hyperfine splitting in the ground state, two hyperfine-Zeeman substates can define the two states in a qubit. Typical atomic species includes rubidium and cesium. (b) In an ensemble qubit, a collective, symmetrical state defines the qubit state $|1\rangle$ with a three-atom ensemble shown. (c) For an ensemble of $K$ identical atoms, each of which has $N+1$ stable states, an $N$-bit binary can be encoded with one ensemble~\cite{Brion2007prl}. Here, an example of three-atom ensemble with $N=2$ is shown. (d) For an atom of the alkali-earth-metal species that has two electrons in the open shell, the ground state is usually a singlet $^1S_0$ state, above which there is a long-lived $^3P_0$ clock state. These two electronic states can define a qubit with a qubit frequency in the optical domain. Moreover, if the $^1S_0$ ground state has a nonzero nuclear spin, two nuclear spin Zeeman substates can define a qubit in the presence of an external magnetic field.        \label{figure-encoding} }
\end{figure*}

\subsection{Encoding quantum information with neutral atoms}\label{qubitencoding}
Various degrees of freedom can be extracted from the internal Hilbert space of neutral atoms, and some of them are related with states of long lifetimes compared to the time for relevant quantum control. There are at least four general methods to store quantum information in long-lived states of neutral atoms. 

\subsubsection{Hyperfine qubit}\label{qubitencoding01}
Most methods of Rydberg gates use hyperfine-Zeeman ground substates in individual atoms to encode information. Experimental two-qubit entanglement by Rydberg interactions adopted this method with $^{87}$Rb~\cite{Wilk2010,Isenhower2010,Zhang2010,Levine2018,Levine2019}, a combination of $^{87}$Rb and $^{85}$Rb~\cite{Zeng2017}, or $^{113}$Cs~\cite{Maller2015,Picken2018,Graham2019,Jau2015}. In these cases, the two qubit states $|0\rangle$ and $|1\rangle$ are chosen from two hyperfine-Zeeman substates of cesium or rubidium, where for $^{113}$Cs the energy gap between the two qubit states is $h\times 9.2$~GHz, and for $^{87}$Rb it is $h\times 6.8$~GHz, where $h$ is the Planck constant. A schematic is shown in Fig.~\ref{figure-encoding}(a). Besides of heavy alkali-metal atoms, this method is applicable for $^{165}$Ho which has a ground electronic configuration $4f^{11}6s^2$ and 8 hyperfine-split levels with hyperfine numbers in the range $F\in[4,~11]$. These hyperfine levels have distinct nearest splittings $\in h\times [4.3,~8.3]$~GHz~\cite{Saffman2008}, which means that it is possible to choose qubit settings with much flexibility in the case of $^{165}$Ho.

\subsubsection{Ensemble qubit}\label{qubitencoding02}
The hyperfine-split states not only can allow quantum information encoding with single atoms, but also can be used for qubit encoding in ensembles of single atoms~\cite{Lukin2001} as experimentally demonstrated in Refs.~\cite{Ebert2014,Ebert2015}. An example of the ensemble qubit defined by the states of three atoms is shown in Fig.~\ref{figure-encoding}(b). The use of atomic ensemble can also allow a novel method to define a qubit: to load random numbers of atoms in the sites of an atomic array, i.e., as long as there is at least one atom at each site, we will be able to encode quantum information by using each site as a qubit. It is technically achievable but not an easy task to deterministically load a certain number of atoms in one trap, so that loading a random number of atoms in one trap is easier. The difficulty appeared in this method is that because the Rabi frequency for a ground-Rydberg state rotation with $K$ atoms is enhanced by a factor of $\sqrt{K}$~\cite{Urban2009,Gaetan2009}, so that the standard Rydberg excitation with a pulse of a certain pulse area can not work when $K$ fluctuates from site to site. Fortunately, it was shown that by using time-dependent or adiabatic pulses one can achieve high-fidelity Rydberg logic gates with the ensemble encoding when $K$ fluctuates from site to site~\cite{Beterov2013,Beterov2016jpb,Beterov2020}. 

\subsubsection{Collectively-encoded qubit}\label{qubitencoding03}
The third method is to use single-ensemble collective states as qubit states when the atomic ground state of each atom in the ensemble has a multilevel structure~\cite{Brion2007prl}. Consider an ensemble of $K$ identical atoms in which each atom has $N+1$ stable levels where $N\ll K$, where the first state is labeled as $|0\rangle_{\text{P}}$ and the remaining $N$ states are labeled as $\{|1\rangle_{\text{P}},|2\rangle_{\text{P}},\cdots,|N\rangle_{\text{P}} \}$. When all the atoms are in the state $|0\rangle_{\text{P}}$, the N-qubit state is $|000\cdots0\rangle_{\text{L}}$, where P and L distinguish the physical and logical states, respectively. When there is a collective ``excitation'' in the $|1\rangle_{\text{P}}$ states among the $K$ atoms, we have
\begin{eqnarray}
  |100\cdots0\rangle_{\text{L}}&\equiv& \frac{1}{\sqrt{K}}\sum_{j=1}^K |0\cdots 1_j\cdots0\rangle_{\text{P}},\label{ensembleQubit01}
\end{eqnarray}
where the number of digits in the ket on the left~(right) hand side is N~(K). Similarly, when there is a collective ``excitation'' in the $|2\rangle_{\text{P}}$ states among the $K$ atoms, we have
\begin{eqnarray}
  |010\cdots0\rangle_{\text{L}}&\equiv& \frac{1}{\sqrt{K}}\sum_{j=1}^K |0\cdots 2_j\cdots0\rangle_{\text{P}},
\end{eqnarray}
and so on for $|001\cdots0\rangle_{\text{L}}$ and other similar states. This method needs $K\gg N$ so as to employ the Rydberg blockade mechanism with less inhomogeneity, but a schematic with $K=N+1=3$ is shown in Fig.~\ref{figure-encoding}(c) for brevity. As shown in Ref.~\cite{Brion2007prl}, by using the Rydberg blockade mechanism, one can start from Eq.~(\ref{ensembleQubit01}) to prepare the $N$-qubit logical state
\begin{eqnarray}
  |110\cdots0\rangle_{\text{L}}&\equiv& \frac{1}{\sqrt{K(K-1)}}\sum_{m\neq j,m=1}^K\sum_{j=1}^K |0\cdots 1_j\cdots 2_m\cdots0\rangle_{\text{P}}, \nonumber\\
  |101\cdots0\rangle_{\text{L}}&\equiv& \frac{1}{\sqrt{K(K-1)}}\sum_{m\neq j,m=1}^K \sum_{j=1}^K|0\cdots 1_j\cdots 3_m\cdots0\rangle_{\text{P}},  \label{ensembleQubit03}
\end{eqnarray}
and so on. The distinction between this method from those demonstrated in Refs.~\cite{Ebert2014,Ebert2015} lies in that one ensemble is enough to encode an $N$-bit binary here, while the multi-ensemble approach shown in Fig.~\ref{figure-encoding}(b) needs $N$ ensembles.

Another single-ensemble approach was proposed in Ref.~\cite{Saffman2008} by taking $^{165}$Ho as an example. The 8 hyperfine-split levels of $^{165}$Ho have 128 hyperfine-Zeeman substates when an external magnetic field is applied. Excluding eight ``unpaired'' states because the level structure of the 128 states is like an upside-down pyramid, we still have 120 states that can be used for encoding information. Labeling these sates as $i\in\{1,2,\cdots,120\}$, then 60 qubits can be defined, where the two qubit states in the $n$th qubit are defined by the $(2n-1)$th and $(2n)$th hyperfine-Zeeman substates, respectively. The single and two-qubit gates in the single-ensemble encoding method was studied in Ref.~\cite{Brion2007prl}.

\subsubsection{Optical qubit} \label{qubitencoding04}
For most alkali-earth-like~(AEL) atoms, there are both an electronic ground state and a long-lived electronically excited state. The distinction between alkali-metal atoms and AEL atoms is that the outermost shell in an atom of the former species has only one electron, while in the latter case two electrons are in the open shell. As shown in Fig.~\ref{figure-encoding}(d), the ground state and the long-lived $p$-orbital clock state of an AEL atom can define a qubit~\cite{Stock2008}. In the case of $^{171}$Yb, the lifetimes of the $^{3}P_0$ and $^{3}P_2$ states are over $20$ and $10$ seconds, respectively~\cite{Covey2019prappl}, and similar for other AEL atoms. This essentially makes it possible to define the ground state $6s^2~^{1}S_0$ and the metastable clock state $6s6p~^{3}P_{0(2)}$ as the two qubit states for the case of $^{171}$Yb. Other well-studied divalent AEL atoms have similar stable clock states that can be used for defining a qubit.%

\subsubsection{Other approaches}\label{qubitencoding05}
There are other types of neutral-atom quantum registers except those discussed in Secs.~\ref{qubitencoding01},~\ref{qubitencoding02},~\ref{qubitencoding03}, and~\ref{qubitencoding04}.

First, one can define qubits with Rydberg states, either with two Rydberg states~\cite{Cohen2020} or with one Rydberg and one stable states~\cite{Shi2014}. The lifetime of low-angular-momentum Rydberg states is only several hundred $\mu$s~\cite{Beterov2009}, which means that only fast enough quantum operations can make Rydberg-state qubits useful for quantum computing. There have been theoretical entanglement protocols based on this type of qubit via low-angular-momentum Rydberg states~\cite{Shi2014} or circular Rydberg states~\cite{Cohen2020}. Two experiments were put forward for demonstrating two-atom entanglement between Rydberg and ground~(or optical clock) states of individual atoms~\cite{Levine2018,Madjarov2020}.

Second, nuclear spin states in AEL atoms can also define qubits. In most AEL atoms, the ground state has no hyperfine splitting. When the atom has a nonzero nuclear spin, states with different projections of the nuclear spin onto the quantization axis can be used as qubit states. For example, the two nuclear spin Zeeman levels $\pm1/2$ of the $^{171}$Yb ground state can define a qubit. Among the AEL species without ground-state hyperfine splittings, $^{87}$Sr has the largest nuclear spin quantum number $9/2$. In this case, one can choose any two states of nearby spin projections $m_I$ and $m_I+1$~($m_I<9/2$) to define a qubit in the presence of external magnetic fields~\cite{Daley2008}.

Third, one can use both the electronic and nuclear spins as register states~\cite{Gorshkov2009}. Taking $^{171}$Yb as an example, the four states $|6s^2~^{1}S_0,m_I\pm1/2\rangle$ and $|6s6p~^{3}P_{0},m_I\pm1/2\rangle$ are useful for storing quantum information since they are all long-lived states. If three of these states are selected, a qutrit can be defined; if four are chosen, then a qudit can be defined.    %


The categories in Secs.~\ref{qubitencoding01}-\ref{qubitencoding05} highlight the versatility to encode quantum information with neutral atoms. There are several entanglement experiments~\cite{Wilk2010,Isenhower2010,Zhang2010,Levine2019,Zeng2017,Maller2015,Picken2018,Graham2019,Jau2015} by using hyperfine-Zeeman substates for encoding information. It is an open question whether there are experimentally friendly ways to carry out fast, robust, and accurate entanglement suitable for quantum information processing by the other methods of encoding.

\subsection{Excitation of Rydberg states}
To induce quantum entanglement by Rydberg interactions, Rydberg states of atoms should be excited from ground or long-lived register states coherently. To achieve this, coherent laser fields can be sent to atoms which cause electric dipole transitions in the atoms. According to the number of photons absorbed during the transition from the ground state to the desired Rydberg state, we have one, two, three, and four-photon excitations of Rydberg states as schematically shown in Figs.~\ref{figure-excitation}(a),~\ref{figure-excitation}(b),~\ref{figure-excitation}(c), and~\ref{figure-excitation}(d), respectively. We take hyperfine qubits of alkali-metal atoms as an example, and the qubit state that is to be excited is an $s$-orbital ground state unless otherwise specified. 

\subsubsection{One-photon excitation}
\begin{figure*}[ht]
\includegraphics[width=5.0in]
{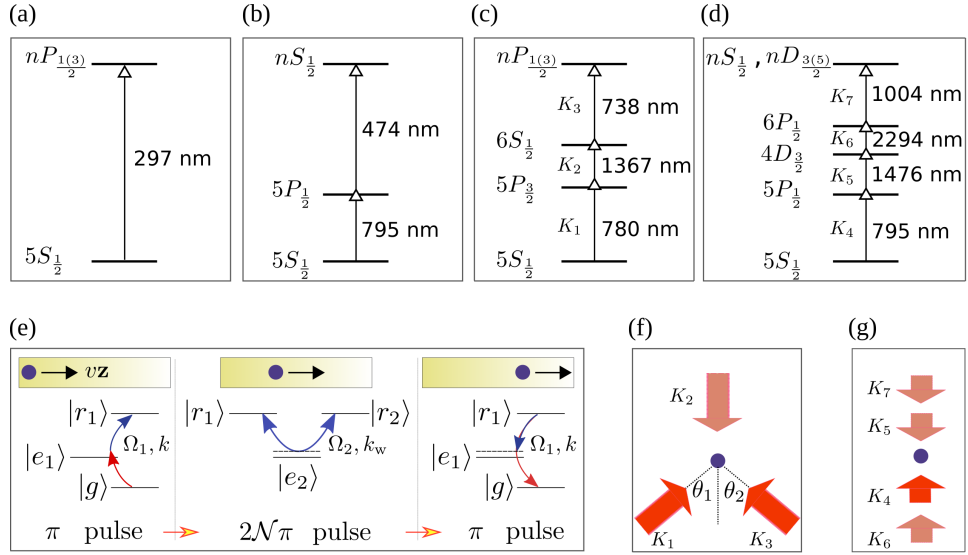}
 \caption{(a), (b), (c), and (d) show examples for the one, two, three, and four-photon Rydberg excitations of a rubidium atom, respectively. (e), (f), and (g) show methods to suppress the Doppler dephasing of the ground-Rydberg transition by the two, three, and four-photon excitations, respectively; these three schemes were proposed in Refs.~\cite{Shi2020},~\cite{Ryabtsev2011}, and~\cite{Bariani2012}, respectively. Numbers beside the arrows denote the wavelengths of the laser fields used for the atomic transitions, and n in the symbols for the Rydberg states denote a principal quantum number around 100. The wavelengths shown are based on the data in Ref.~\cite{Sansonetti2006}.  \label{figure-excitation} }
\end{figure*}

An $s$-orbital ground state of an alkali-metal atom can absorb an ultraviolet~(UV) photon and transit to a $p$-orbital Rydberg state, as demonstrated in~\cite{Hankin2014,Jau2015}. A schematic is shown in Fig.~\ref{figure-excitation}(e). For an AEL atom, the clock state can be excited to an $s$-orbital Rydberg state via the absorption of an UV photon~\cite{Madjarov2020}. To study the one-photon schemes, we use $|g\rangle$ and $|r\rangle$ to denote the qubit and Rydberg states, respectively. By using the dipole approximation in the atom-light coupling and the rotating-wave approximation, the Hamiltonian becomes
\begin{eqnarray}
 \hat{H}_{\text{1-pho}} &=&\hbar \Omega |r\rangle\langle g|/2+\text{H.c.}\label{equation01}
\end{eqnarray}
Here, in principle, there are nonresonant ac Stark shifts from the laser field that can be calculated by the sum over transitions from relevant states~\cite{Maller2015}. But by shifting the laser frequency with an appropriate value, the resonance is recovered and in an appropriate rotating frame we arrive at Eq.~(\ref{equation01}). Starting from the ground state $|g\rangle$, the state evolves to $-i|r\rangle$ at the moment $t=t_1$ when $\int_0^{t_1} \Omega dt=\pi$, i.e., when a $\pi$ pulse is used. Similarly, a $\pi$ pulse can excite the atom from the Rydberg state $|r\rangle$ to the ground state. Usually, we can assume quasi-rectangular pulses and constant Rabi frequency during the excitation. Then, a $\pi$ pulse with duration $t_1=\pi/\Omega$ can achieve a Rydberg excitation. Because the phase change of the wavefunction in the Rydberg excitation $|g\rangle\rightarrow|r\rangle$ or its deexcitation is $\pi/2$, one Rabi cycle $|g\rangle\rightarrow|r\rangle\rightarrow|g\rangle$ results in a $\pi$ phase change which is a crucial factor enabling the Rydberg blockade gate~\cite{PhysRevLett.85.2208}. 

A one-photon Rydberg excitation of $^{133}$Cs atoms was used in~\cite{Jau2015} for entanglement. There was also an entanglement experiment using one-photon excitation of the $^{3}S_1$ Rydberg state from the long-lived $^{3}P_0$ clock state of $^{88}$Sr~\cite{Madjarov2020}. The high fidelity achieved in Ref.~\cite{Madjarov2020} partly benefited from the large Rydberg Rabi frequency which led to fast entanglement generation.

\subsubsection{Two-photon excitation}
A two-photon Rydberg excitation can excite an $s$- or $d$-orbital Rydberg state $|r\rangle$ from the ground state $|g\rangle$. Form the $s$-orbital ground state, a two-photon Rydberg excitation can proceed via an intermediate $p$-orbital state $|p\rangle$ which should be largely detuned to avoid dissipation from it. Most Rydberg-mediated entanglement experiments were based on two-photon excitations~\cite{Wilk2010,Isenhower2010,Zhang2010,Maller2015,Zeng2017,Levine2018,Picken2018,Jo2019,Levine2019,Graham2019}. With similar approximations as used in Eq.~(\ref{equation01}), the  Hamiltonian for a two-photon Rydberg excitation is
\begin{eqnarray}
  \hat{H}_{\text{2-pho;0}} &=&\hbar [\Omega_1 |p\rangle\langle g|/2+ \Omega_2 |r\rangle\langle p|/2+\text{H.c.}]+\hbar \Delta |p\rangle\langle p|+\hbar \Delta_r^{(\text{ac})} |r\rangle\langle r|+ \hbar \Delta_g ^{(\text{ac})}|g\rangle\langle g|,\label{equation02}
\end{eqnarray}
where $\Delta$ is the frequency mismatch between the laser fields and the atomic transition, and $\Delta_r^{(\text{ac})}$ and $\Delta_g^{(\text{ac})}$ are the Stark shifts at the Rydberg and ground states, respectively. According to Ref.~\cite{Maller2015}, they are given by
\begin{eqnarray}
  \Delta_r^{(\text{ac})}&=& -\frac{e^2}{4m_e\hbar}\left( \frac{E_1^2}{\omega_1^2}+ \frac{E_2^2}{\omega_2^2}  \right),~\label{equation04}\\
  \Delta_g^{(\text{ac})}&=& -\frac{1}{4\hbar}\left( \alpha_1 E_1^2+ \alpha_2 E_2^2  \right),~\label{equation05}
\end{eqnarray}
where $e$ is the elementary charge, $m_e$ is the mass of the electron, and $\omega_j$ and $E_j$ are the frequency and electric field amplitude of the laser field, where $j=1$ and $2$ for the lower and upper transitions, respectively, and $\alpha_1$ and $\alpha_2$ are the nonresonant polarizabilities that can be calculated via the sum over transitions to states other than the intermediate state $|p\rangle$.

When $\Delta$ is large compared to the decay rate of $|p\rangle$, the intermediate state can be adiabatically eliminated~\cite{James2007,James2000,Brion2007,Han2013}, leading to 
\begin{eqnarray}
  \hat{H}_{\text{2-pho}} &=&\hbar\{ [\Omega_{\text{eff}} |r\rangle\langle g|/2+ \text{H.c.}]+ \Delta_r |r\rangle\langle r|+ \Delta_g |g\rangle\langle g|\},\label{equation03}
\end{eqnarray}
where $\Omega_{\text{eff}}=-\Omega_1\Omega_2/(2\Delta)$ and
\begin{eqnarray}
  \Delta_r&=& -\frac{\Omega_2^2}{4\Delta} + \Delta_r^{(\text{ac})} ,  \nonumber\\
   \Delta_g&=& -\frac{\Omega_1^2}{4\Delta} + \Delta_g^{(\text{ac})}. ~\label{equation04}
\end{eqnarray}
Here, note that $\Omega_{1(2)}\propto E_{1(2)}$, the effective detunings $\Delta_r$ and $\Delta_g$ are functions of $\Delta$, $E_{1}$, and $E_2$ for any given set of atomic levels. As studied in Ref.~\cite{Maller2015}, resonant conditions with $\Delta_g=\Delta_r$ can be recovered, and it is also possible to recover the conditions so that the phase change to $|g\rangle$ after a $2\pi$ rotation is equal to $\pi$. The establishment of the latter condition requires efforts because the comparable magnitudes of the resonant and the nonresonant ac Stark shifts can lead to any phase shift in the ground state upon the completion of a $2\pi$ rotation; more details can be found in~\cite{Maller2015}.

It is possible to suppress Doppler dephasing in the ground-Rydberg transition in a two-photon Rydberg excitation. Two theories were introduced in Ref.~\cite{Shi2020}, among which one is briefly shown in Fig.~\ref{figure-excitation}(e). The $2\mathcal{N}\pi$ pulse causes a transition between the Rydberg state $|r_1\rangle$~[i.e., the $nS_{\frac{1}{2}}$ state in (b)] and a nearby Rydberg state $|r_2\rangle$ via a low-lying intermediate state $|e_2\rangle$. The choices of the intermediate state $|e_{1}\rangle$ during the Rydberg excitation shall yield a wavevector $k$ that is more than 2 times smaller than the wavevector $k_w$ of the transition $|r_1\rangle\leftrightarrow|r_2\rangle$. By the sequence shown in Fig.~\ref{figure-excitation}(e), the motional dephasing of the ground-Rydberg transition can be suppressed. The wavelengths shown in Fig.~\ref{figure-excitation} are calculated by using the data in Ref.~\cite{Sansonetti2006}. See Ref.~\cite{Shi2020} or a case study in Ref.~\cite{Raina2021} for more details about the suppression of the Doppler dephasing of the ground-Rydberg transition via the $\pi-2\mathcal{N}\pi-\pi$ method; we note, however, that there is another theory to suppress the motional dephasing in Ref.~\cite{Shi2020} which uses a strategy different from the one shown in Fig.~\ref{figure-excitation}(e).

\subsubsection{Three-photon excitation}
Rydberg excitation can be realized with a three-photon excitation via two intermediate states. Like the two-photon method, the intermediate states in the three-photon transition shall be largely detuned to avoid dissipation from them. For the excitation of single atoms, the attraction of a three-photon transition lies in, as proposed in Ref.~\cite{Ryabtsev2011}, that it can offer a complete suppression of the Doppler broadening due to the atomic motion; moreover, it can suppress the recoil effect, i.e., the momentum change of atoms during the Rydberg excitation, as shown in Figs.~\ref{figure-excitation}(c) and~\ref{figure-excitation}(f) where $(\theta_1,~\theta_2)=(1.37,~1.21)$~\cite{Ryabtsev2011}. The configuration in Fig.~\ref{figure-excitation}(f) ensures that when the atomic state is excited by absorption of the three photons, the total momentum of the three photons is zero, leading to no momentum change in the atom. Likewise, there is nearly no {\it position-dependent} phase change to the atomic state since the three wavevectors of the three absorbed photons add up to exactly zero. Until now, there is no experimental demonstration of Rydberg excitation for entanglement generation via the three-photon excitation. But in an experimental study of collective Rydberg excitations in atomic gas, the three-photon method was used for the suppression of Doppler dephasing~\cite{Carr2012}.

\subsubsection{Four-photon excitation}
In Ref.~\cite{Bariani2012}, it was shown that a four-photon excitation chain via three low-lying intermediate states can suppress the Doppler dephasing, where the wavelengths for the four fields $\lambda_j$ with $j=1-4$ satisfy $\sum_j(-1)^j\lambda_j^{-1}\approx0$, as shown in Figs.~\ref{figure-excitation}(d) and~\ref{figure-excitation}(g). This means that by sending the first and third lasers in a direction opposite to that for the second and fourth lasers, Doppler dephasing as well as photon recoil can be avoided. In this four-photon method, one does not need to focus the laser fields on one spot from three directions as in~\cite{Ryabtsev2011}. To suppress control error from the position fluctuation of the qubit, the control fields can be sent along the longitudinal axis of the optical dipole trap (if this type of trap is used). Experiments have not explored the four-photon method for Rydberg-mediated entanglement.    

\subsubsection{Comparison between different schemes of Rydberg excitation}
To simplify, we call an excitation with more than one photons exchanged between the atom and the field a multi-photon excitation. Both the one-photon and multi-photon methods have their strength.

First, high-power laser fields are required in the one-photon Rydberg excitation because the dipole matrix element between the ground and a high-lying Rydberg state is small, but for the multi-photon case a large Rydberg Rabi frequency can be achieved with fields of lower powers. So, a multi-photon excitation is more favorable when fast rotations are desirable if laser powers are limited.

Second, a multi-photon Rydberg excitation suffers from loss via the scattering from the intermediate states, which is absent in a one-photon Rydberg excitation. When ignoring the hyperfine structure of $|p\rangle$ and when $\Omega_1=\Omega_2$ for the two-photon case in Eq.~(\ref{equation02}), the decay probability of the state via $|p\rangle$ is $\pi/(2\tau_p \Delta)$~\cite{Graham2019} in a two-photon Rydberg excitation with a $\pi$ pulse, where $\tau_p$ is the lifetime of the state $|p\rangle$. This essentially means that by using large detuning at $|p\rangle$ the coherence can be preserved. For example, the scattering through the intermediate state in~\cite{Graham2019} was much less detrimental compared to the Doppler dephasing of the ground-Rydberg transition.

Third, the Doppler dephasing in the ground-Rydberg transition is not removable in one-photon Rydberg excitations, but can be suppressed in other schemes. For example, the wavevector in the one-photon Rydberg excitation in Ref~\cite{Hankin2014} is $2.5$ times larger than that of the two-photon transition in typical two-photon excitations~\cite{Graham2019}, one may expect a strong Doppler dephasing in the former case. As shown in Figs.~\ref{figure-excitation}(e),~\ref{figure-excitation}(f), and~\ref{figure-excitation}(g), it is possible to suppress the Doppler dephasing in the two, three, and four-photon methods, respectively. Comparing the three schemes for suppressing the Doppler dephasing, the three-photon method in Fig.~\ref{figure-excitation}(f) requires arranging the three laser fields so that they focus on a specific spot in the three-dimensional space where the atom (or ensemble of atoms) is supposed to reside, while for the two and four-photon methods shown in Figs.~\ref{figure-excitation}(e) and~\ref{figure-excitation}(g) the fields should be nearly collinear.

Fourth, in a colder environment the $p$-orbital Rydberg states excited in the one-photon and three-photon methods can have longer lifetimes than the $s$ or $d$-orbital Rydberg states prepared in the two-photon methods. The Rydberg-state decay is dominated by the blackbody radiation at room temperature, so that the lifetimes of Rydberg states of different angular momenta are similar. For example, the lifetimes of an $s$,~$p$, and~$d$-orbital rubidium Rydberg states of a principal quantum number $100$ are $0.33,~0.38$, and $0.32~$ms, respectively~\cite{Beterov2009}. At lower temperatures, the Rydberg-state decay is dominated by the spontaneous emission which strongly depends on the angular states of the Rydberg atom. For example, the lifetimes are $1.2$,~$2.1$, and $0.9$~ms for an $s$,~$p$, and $d$-orbital rubidium state with a principal quantum number around $100$ at $4.2$~K; at $77$~K these lifetimes are $0.73,~0.99$, and $0.63$~ms, respectively. In this sense, there is an advantage to use the one and three-photon excitations of $p$-orbital Rydberg states so that Rydberg-state decay is less severe when the atomic qubits are hosted in a cryogenic chamber.

Fifth, as in the two-photon case, there can be Stark shifts for the ground and Rydberg states when the intermediate states are adiabatically eliminated. It is a delicate issue to tune to conditions in three and four-photon Rydberg excitations for the induction of $\pi$ phase change in a 2$\pi$ Rydberg pulse that is critical in the Rydberg blockade gate~\cite{PhysRevLett.85.2208}. Because there are multiple terms in the ac Stark shifts in the two, three, and four-photon schemes, it is in principle possible to recover a $\pi$ phase change in a 2$\pi$ Rydberg pulse~\cite{Maller2015}. For one-photon Rydberg excitation the Hamiltonian shown in Eq.~(\ref{equation01}) can easily lead to a $\pi$ phase change to the state with a $2\pi$ pulse.

From the above comparison, one can see that although most experiments on Rydberg atoms used two-photon excitations~\cite{Wilk2010,Isenhower2010,Zhang2010,Maller2015,Zeng2017,Levine2018,Picken2018,Levine2019,Graham2019} and some used one-photon excitations~\cite{Hankin2014,Jau2015}, three-photon excitation methods can suppress the recoil and Doppler dephasing as shown in~\cite{Ryabtsev2011}, and can excite $p$-orbital Rydberg states that possess longer lifetimes in cryogenic chambers. The one-photon method can also excite $p$-orbital Rydberg states, but it is an open question whether there is any method to eliminate the prevailing motional dephasing so as to realize a highly coherent single-photon excitation of a $p$-orbital Rydberg state. 

We have discussed one-photon to four-photon Rydberg excitations. In this sense only a low-orbital-angular-momentum Rydberg state is excited. An atomic Rydberg state with the maximal orbital angular momentum~(equal to $n-1$ with $n$ the principal quantum number) is called a circular Rydberg state. The excitation of circular Rydberg states can be achieved by adiabatic passage~\cite{Anderson2013} or by a fast scheme using microwave and radio-frequency field~\cite{Signoles2017}. Because of the long lifetimes of circular Rydberg states, the latter method for exciting circular Rydberg states can preserve the quantum coherence of the system~\cite{Cortinas2020,Cantat-Moltrecht2020} and may be useful for quantum information processing~\cite{Xia2013}, but it is still based on first exciting a low-orbital-angular-momentum Rydberg state~\cite{Zhelyazkova2016,Morgan2018,Teixeira2020}.

\subsection{Interactions between Rydberg atoms}
\subsubsection{Calculation of Rydberg interactions}\label{interaction01}
Rydberg interactions usually refer to interactions between the electric dipoles of two nearby atoms. There are also three-body F\"{o}rster resonances among three nearby Rydberg atoms~\cite{Faoro2015,Tretyakov2017,Cheinet2020} and even four-body F\"{o}rster resonances can exist~\cite{Liu2020}, where the interaction can be calculated starting from the fundamental dipole-dipole interaction process. Below, we briefly review the basics for calculating the two-body interactions between alkali-metal Rydberg atoms when the distance between their nuclei is larger than the Le Roy radius~\cite{LeRoy1974,Weber2017} so that their Rydberg electrons do not overlap.

\begin{figure*}[ht]
\includegraphics[width=6.0in]
{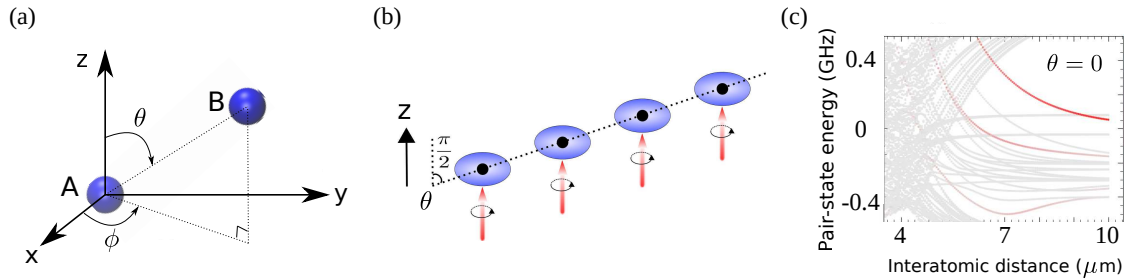}
\caption{(a) Two Rydberg atoms labeled A and B with relative orientation shown by the spherical angles $\theta$ and $\phi$, where $\mathbf{z}$ is the quantization axis. (b) In Rydberg-mediated entanglement experiments in a two-dimensional array~\cite{Maller2015,Graham2019}, the quantization axis is usually perpendicular to the plane of the atomic array, where $\theta=\pi/2$. (c) When the atoms are not well separated, the two-atom Rydberg state is coupled to many other nearby two-atom Rydberg states. Shown here is the energy spectrum in units of $h\times$GHz for the rubidium state $|r;r\rangle=|100S_{1/2}, m_J=1/2;100S_{1/2}, m_J=1/2\rangle$, where the redder the curve is, the more the overlap between the state and $|r;r\rangle$. This spectrum was calculated with the package in Ref.~\cite{Sibalic2017}.    \label{figure-orientation} }
\end{figure*}

The interaction between two Rydberg atoms arises from the electric dipole-dipole interaction while multipole interactions are negligible for most cases~\cite{Walker2008}. The van der Waals interaction denotes the dipole-dipole interaction in the second-order perturbation theory when two Rydberg atoms A and B are well separated. For a highly-excited atom, we use the principal quantum number $n$, the electron angular momentum quantum number $l$, the total angular momentum quantum number $j$, and the magnetic quantum number $m$ to characterize its state. We assume that the quantization axis is along $\mathbf{z}$. The dipole-dipole interaction between two atoms with quantum numbers $(n_{\text{A}}, l_{\text{A}}, j_{\text{A}}, m_{\text{A}})$ and $(n_{\text{B}}, l_{\text{B}}, j_{\text{B}}, m_{\text{B}})$ is an electrostatic interaction given by
\begin{eqnarray}
\hat {V}_{\text{dd}} &=& \mathbf{\hat s}_{\text{A}} \cdot \left( 3\mathbf{\hat r}\mathbf{\hat r}/r^2-\mathbf{\hat I} \right) \cdot  \mathbf{\hat s}_{\text{B}}/(4\pi\epsilon_0L^3),
\end{eqnarray}
where $\mathbf{\hat s}_{{\text{A(B)}}}$ is the dipole moment of atom A(B), $L$ is the distance between the two atoms, $\mathbf{\hat I}$ is the identity operator, and $\epsilon_0$ is the dielectric permittivity in free space. Using the representation of spherical harmonic rank 2 tensor, the matrix element
\begin{eqnarray}
&&V_{ m_{\text{A}} m_{\text{B}};m_am_b} =\langle n_{\text{A}} l_{\text{A}} j_{\text{A}} m_{\text{A}}; n_{\text{B}}  l_{\text{B}} j_{\text{B}} m_{\text{B}} |\hat {V}_{\text{dd}} |n_a l_aj_am_a; n_b l_bj_bm_b\rangle \label{eq101401}
\end{eqnarray}
can be written as
\begin{eqnarray}
&&-\frac{\sqrt6}{4\pi\epsilon_0L^3}\langle n_{\text{A}}l_{\text{A}} j_{\text{A}}||D_{\text{A}}^{[1]}|| n_al_aj_a\rangle \langle n_{\text{B}} l_{\text{B}} j_{\text{B}}||D_{\text{B}}^{[1]}|| n_bl_bj_b\rangle  \sum_{M=-2}^2\Psi_{-M}^{[2]}(\theta,\phi) \sum_{\alpha,\beta=-1}^1 C_{\alpha \beta M}^{112} C_{m_a\alpha  m_{\text{A}}}^{j_a1 j_{\text{A}}}C_{m_b\beta  m_{\text{B}}}^{j_b1 j_{\text{B}}},
\end{eqnarray}
where $D_k^{[1]}$ is the dipole moment of atom $k=A$ or $B$ in terms of rank-1 tensor, $C$ is a Clebsh-Gordan coefficient~\cite{Rose1957}, $\Psi^{[2]}$ is a rank-2 tensor given by the standard spheric harmonics multiplied by a factor of $\sqrt{4\pi/5}$~\cite{PhysRev.62.438}, and $(\theta,\phi)$ gives the angular position of atom B with respect to atom A as schematically shown in Fig.~\ref{figure-orientation}(a). Because all the second-order spheric harmonics, except $\Psi_{0}$, are proportional to $\sin\theta$, the dipole coupling will conserve the total projection of the magnetic angular momentum of the two atoms when $\theta = 0$~(i.e., along the quantization axis). Here, the dipole matrix element $\langle n_{\text{A}}l_{\text{A}} j_{\text{A}}||D_{\text{A}}^{[1]}|| n_al_aj_a\rangle$ can be calculated using the Wigner-Eckart theorem~\cite{Rose1957}.

When the coupling strength between the two electric dipoles is much smaller than the energy difference between the dipole-coupled initial and final two-atom states, the two-atom state can barely be excited away from the degenerate manifold of $(n_{\text{A}}, l_{\text{A}}, j_{\text{A}};~n_{\text{B}}, l_{\text{B}}, j_{\text{B}})$ and $(n_{\text{B}},  l_{\text{B}}, j_{\text{B}};~n_{\text{A}},l_{\text{A}}, j_{\text{A}})$. Even when the principal quantum numbers of the two atoms do not exchange, their magnetic angular momenta $ m_{\text{A}}$ and $ m_{\text{B}}$ can change. In this case, the van der Waals interaction is given by
\begin{eqnarray}
\hat{H}_{\text{vdW}} &=& -\sum_{n_a l_aj_a}\sum_{n_b l_bj_b}\hat{V}_{\text{dd}} \hat{V}_{\text{dd}}^\dag /\delta_{ab}. \label{eq101403}
\end{eqnarray}
Here the matrix elements of $\hat{V}_{\text{dd}}$ are given in Eq.~(\ref{eq101401}), and the energy defect is~\cite{Walker2008}
\begin{eqnarray}
\delta_{ab}&=&    E(n_a l_aj_a) + E(n_b l_bj_b)-[E(n_{\text{A}} l_{\text{A}} j_{\text{A}}) +  E(n_{\text{B}}  l_{\text{B}} j_{\text{B}})] ,
 \end{eqnarray}
where $E(nlj)$ is the atomic energy of a Rydberg atom with quantum numbers $n,l,j$. A remarkable example is that the state labeled by $(n_{\text{A}}, l_{\text{A}}, j_{\text{A}},  m_{\text{A}};~n_{\text{B}}, l_{\text{B}}, j_{\text{B}},  m_{\text{B}})$ can go to the state $(n_{\text{A}}, l_{\text{A}}, j_{\text{A}},  m_{\text{B}};~n_{\text{B}}, l_{\text{B}}, j_{\text{B}},  m_{\text{A}})$ where $m_{\text{A}}$ and $m_{\text{B}}$ exchange. This spin-exchange process is of special interest when the initial and the final two-atom states can be optically coupled to different two-atom ground states~\cite{Shi2014}, giving rise to entanglement of neutral atoms and a quantum gate similar to the SWAP gate. One can also easily find that the state $(n_{\text{A}}, l, j,  m;~n_{\text{B}}, l, j,  m)$ and $(n_{\text{B}}, l, j,  m;~n_{\text{A}}, l, j,  m)$ are degenerate when $n_{\text{A}}\neq n_{\text{B}}$, which means that two atoms in different Rydberg principal levels can exchange their states. This feature can be used to simulate magnetic fields that are useful to realize exotic many-body states~\cite{Shi2018pra_m}.

There are many useful references about the calculation of interactions between two Rydberg atoms, including a detailed analysis on the consequences of Zeeman degeneracies~\cite{Walker2008}, a study about two-atom F\"{o}rster resonance~\cite{Walker2005}, a classification of the interaction according to symmetry~\cite{Singer2005}, and a detailed introduction about the method for the calculation as well as an open-source software~\cite{Weber2017}. Besides, Ref.~\cite{Sibalic2017} provides a comprehensive object-oriented Python library for calculating properties of highly-excited Rydberg states of alkali-metal atoms, including single-atom dipole matrix elements, Rydberg-state lifetimes, Stark maps of atoms in external electric fields, and two-atom interactions accounting for dipole and quadrupole couplings. An update of Ref.~\cite{Sibalic2017} was given in Ref.~\cite{Robertson2021} which treats interactions of several species of AEL Rydberg atoms. Other features in Ref.~\cite{Robertson2021} include the calculation of valence electron wave functions, dynamic polarizabilities, atom-surface interactions, and wavefunctions of atoms in optical traps. 

\subsubsection{Different interactions required in different entanglement protocols}\label{sec02C02}
As briefly mentioned in Sec.~\ref{secIC}, there are mainly three classes of Rydberg gates. These three different methods depend on different characters of the Rydberg interactions. 

The Rydberg gates by the blockade mechanism need Rydberg interaction $V$ to be much larger than the laser Rabi frequency $\Omega$ for the excitation of Rydberg states~\cite{PhysRevLett.85.2208}. The distance between the two atoms we want to entangle can be short so that $V$ is large, but should be larger than the Le Roy radius within which the electronic wavefunctions of the two atoms overlap. This is easily satisfied in the experiments~\cite{Wilk2010,Isenhower2010,Zhang2010,Maller2015,Jau2015,Zeng2017,Levine2018,Picken2018,Jo2019,Graham2019,Madjarov2020} where qubits were separated beyond $3.6~\mu$m for Rydberg states of principal quantum numbers below $100$~\cite{Sibalic2017}. Two-atom Rydberg states are barely populated in this class of Rydberg gates, so that these gates require a minimal condition on the Rydberg interactions: there shall not be accidental zero interactions due to F\"{o}rster resonance or highly anisotropic interactions of, e.g., $d$-orbital states~\cite{Walker2008}; this is in principle not a demanding task since the F\"{o}rster resonance is rare and it requires efforts to find a F\"{o}rster resonant dipole-dipole process~\cite{Tiarks2014}. Moreover, there is in general no specific required relation between the quantization axis (usually set by an external magnetic field) and the two-atom orientation for frequently employed Rydberg states.

For the methods to entangle neutral atoms via the antiblockade or via Rydberg excitation of two or more atoms, the requirement on the Rydberg interactions is either that (i) $V$ shall represent one or several isolated dipole-dipole interactions or (ii) $V$ is a van der Waals interaction in the form a pure energy shift. To achieve (i), usually external fields are required, as studied in Refs.~\cite{PhysRevLett.47.405,PhysRevLett.80.249,PhysRevLett.80.253,Westermann2006,PhysRevLett.108.113001,PhysRevA.93.042505,PhysRevA.93.012703,Tretyakov2014}. Here, we note that a pure energy shift is required in the basic method in regimes \textcircled{2} and \textcircled{3} of Fig.~\ref{figure01}(c), but for the extensions of the basic methods which will be reviewed later on in Secs.~\ref{sec04} and~\ref{sec05}, isolated dipole-dipole interactions were usually used. Below, we detail how to achieve (ii).

The appearance of a pure energy shift depends on two conditions. First, the two-atom separation should be large so that the interaction can be analyzed by a second-order perturbation theory. Second, the choice of principal quantum numbers and angular momenta in $|r\rangle$ shall lead to no state-flip van der Waals interaction. For certain conditions the two-atom separation axis shall be parallel to the quantization axis so that there is no state-flip van der Waals interaction. The first condition can be understood with Sec.~\ref{secIC}. To explain the other condition in detail, we take the state $^{87}$Rb $|nS_{1/2}, m_J=\pm1/2\rangle\equiv|r_\pm\rangle$ as an example. The two-atom state $|r_+;r_+\rangle$ is coupled with
\begin{eqnarray}
 && |n_1P_{1/2}, m_J=\pm1/2; n_2P_{1/2}, m_J=\pm1(3)/2\rangle,~\nonumber\\
&&  |n_3P_{1/2}, m_J=\pm1/2; n_4P_{3/2}, m_J=\pm1(3)/2\rangle,~\nonumber\\
  &&|n_5P_{3/2}, m_J=\pm1(3)/2; n_6P_{1/2}, m_J=\pm1/2\rangle,~\nonumber\\
&&  |n_7P_{3/2}, m_J=\pm1(3)/2; n_8P_{3/2}, m_J=\pm1(3)/2\rangle,\label{ss-pp01}
\end{eqnarray}
where $n_j$ with $j=1-8$ is a principal quantum number near $n$. Here, a semicolon is used between the symbols for the states of the two atoms in the ket so as to indicate that the ket represents a two-atom state. The above equation shows 40 states coupled with $|r_+;r_+\rangle$ for each set of $\{n_1,n_2,\cdots,n_8\}$, and there are many sets of $n_j$. Since the Coulomb interaction is strong, the energy separation between states of different $n$ is large. When the energy gap between $|r_+;r_+\rangle$ and another two-atom state in Eq.~(\ref{ss-pp01}) is too large compared to the dipole-dipole coupling, the coupling barely takes effect and it can be ignored. Then, it is usually sufficient to restrict $|n_j-n|\leq 6$ in the evaluation of Rydberg interactions~\cite{Shi2014,Sibalic2017}, i.e., there will be 13 values for $n_j$. So, there are $40\times13^2$ two-atom Rydberg states coupled with $|r_+;r_+\rangle$. Some states in Eq.~(\ref{ss-pp01}) are mainly coupled with $|r_+;r_+\rangle$, but some states in Eq.~(\ref{ss-pp01}) can couple both $|r_+;r_+\rangle$ and $|r_-;r_-\rangle$. Likewise, when we consider the interaction involving the state $|r_+r_-\rangle$, the two states $|r_+r_-\rangle$ and $|r_-r_+\rangle$ are degenerate and can be coupled. Because of these couplings, it demands efforts to realize a Rydberg interaction in the form of a pure energy shift. We show several examples to clarify this. For $n=100$ and with basis states $\{|r_+;r_+\rangle,|r_+;r_-\rangle,  |r_-;r_+\rangle, |r_-;r_-\rangle\}$, the second-order perturbation leads to the following van der Waals interaction
\begin{eqnarray}
\hat{H}_{\text{vdW}}^{(100s,100s)}&=&h\times  \left(\begin{array}{cccc}56200 &0&0&0\\
0&56980 & 1573 &0\\
0& 1573 &56980 &0\\
0&0&0&56200 \end{array}
\right)\frac{\mu \text{m}^6 \text{GHz}}{L^6},\nonumber
\label{eq101602}
\end{eqnarray}
when $\theta=0$, where $L$ is the two-atom distance. One can see that the two-atom states $|r_+;r_-\rangle$ and $|r_-;r_+\rangle$ are coupled with a strength that is about $3\%$ of the diagonal energy shift. When $\theta=\pi/4$, the interaction becomes,
 \begin{eqnarray}
\hat{H}_{\text{vdW}}^{(100s,100s)} &=& h\times \left(\begin{array}{cccc}
56790 & 	590  &	590  &	-590 \\ 	
590  &	56400 & 	983  &	-590 \\ 	
590  &	983  &	56400 	 &-590 	\\ 
-590  &	-590  &	-590  &	56790 
 \end{array}
\right)\frac{\mu \text{m}^6 \text{GHz}}{L^6}.\nonumber
 \end{eqnarray}
 which shows that the state $|r_+;r_+\rangle$ is coupled to $\{|r_+;r_-\rangle,  |r_-;r_+\rangle, |r_-;r_-\rangle\}$ with a strength that is about $1\%$ of the diagonal energy shift. When $\theta=\pi/2$, the interaction becomes,
 \begin{eqnarray}
\hat{H}_{\text{vdW}}^{(100s,100s)} &=& h\times   \left(\begin{array}{cccc}	
57380 &	0  &	0  &	-1180 \\	
0  &	55800  &	393  &	0 \\	
0  &	393  &	55800  &	0\\	
-1180  &	0  &	0  &	57380	
 \end{array}
\right)\frac{\mu \text{m}^6 \text{GHz}}{L^6}.\nonumber
\end{eqnarray}
 which shows that $|r_+;r_+\rangle$ is coupled to $|r_-;r_-\rangle$ with a strength that is about $2\%$ of the diagonal energy shift. If the Rydberg atoms have different principal quantum numbers, the issue can be more delicate. Consider $^{87}$Rb $|97S_{1/2}, m_J=\pm1/2\rangle\equiv|R_\pm\rangle$, then with the basis 
\begin{eqnarray}
&&\{|R_+ ; r_{+}\rangle, | R_- ;r_{+}\rangle, |R_+ ; r_{-}\rangle, | R_-; r_{-}\rangle , |r_+ ; R_{+}\rangle, | r_-; R_{+}\rangle , |r_+ ; R_{-}\rangle, | r_- ;R_{-}\rangle \} \nonumber
\end{eqnarray}
the two-atom van der Waals interaction of the atoms is given by
\begin{eqnarray}
\hat{H}_{\text{vdW}}^{(97s,100s)} &=& \left(\begin{array}{cc}
\hat V_1 &\hat  V_2\\
\hat V_2 &\hat  V_1\end{array}
 \right),\nonumber
 \end{eqnarray}
 when initially one atom is in the $n=97$ state and the other is in the $n=100$ state, where 
\begin{eqnarray}
\hat V_1&=&h\times   \left(\begin{array}{cccccccc}
- 89180 &0&0&0\\
 0&-59780	  & 	 58800 &0  \\
0& 58800 & -59780 &0 \\
0 &0 &0 &  -89180
\end{array}
 \right)\frac{\mu \text{m}^6 \text{GHz}}{L^6} ,\nonumber\\
\hat V_2&=&h\times  \left(\begin{array}{cccccccc}
-537  &0 &0 &0\\
0 &-375 & 324 &0 \\
0& 324 &-375 &0 \\
0 &0 &0&-537
\end{array}
 \right)\frac{\mu \text{m}^6 \text{GHz}}{L^6}, \label{eq101503}
\end{eqnarray}
when $\theta=0$. Equation~(\ref{eq101503}) shows that the two-atom states $| R_- r_{+}\rangle$ and $|R_+  r_{-}\rangle$ are coupled with each other with a strength that is almost equal to the diagonal energy shift; similarly, the two states $| r_- R_{+}\rangle$ and $|r_+  R_{-}\rangle$ are coupled in a similar way which results in zero energy shift. This is analogous to the well studied F\"{o}rster resonance~\cite{Walker2008}. Because the angular selection rules lead to different couplings for different orbitals, the issue is more complexed if $p$-orbital Rydberg states are employed. We take $| r_{\pm}\rangle = |100P_{\frac{1}{2}}, m_J = \pm 1/2\rangle $ as an example. Then, with the basis $\{|r_+;r_+\rangle,|r_+;r_-\rangle,  |r_-;r_+\rangle, |r_-;r_-\rangle\}$, the van der Waals interaction is given by
\begin{eqnarray}
\hat{H}_{\text{vdW}}^{(100p,100p)} &=&h\times   \left(\begin{array}{cccc}2108 &0&0&0\\
0&-3492 & -11200 &0\\
0& -11200 &-3492 &0\\
0&0&0&2108 \end{array}
\right)\frac{\mu \text{m}^6 \text{GHz}}{L^6}.\nonumber
\label{eq101603}
\end{eqnarray}
when $\theta=0$. For $\theta = \pi/2$ we have 
 \begin{eqnarray}
\hat{H}_{\text{vdW}}^{(100p,100p)} &=&h\times   \left(\begin{array}{cccc}	
-6292  &	0  &	0 & 	8400 \\	
0  &	4908  &	-2800  &	0 \\	
0  &	-2800  &	4908  &	0 \\	
8400  &	0  &	0  &	-6292 
 \end{array}
\right)\frac{\mu \text{m}^6 \text{GHz}}{L^6}.\nonumber \label{eq101702}
\end{eqnarray}
which shows that the states $|r_+;r_+\rangle$ and $|r_-;r_-\rangle$ are strongly coupled; similarly, the states $|r_+;r_-\rangle$ and $|r_-;r_+\rangle$ are strongly coupled. The above examples show that to use the phase shift from a van der Waals interaction in the weak interaction regime, the two atoms shall be placed along the quantization axis, and the total angular momentum of one atom shall be equal to that of the other atom, and we'd better excite both atoms to the same principal quantum level. If the atoms fly freely during the operation of the quantum gate, the relative angular position of the two atoms can change, which means that it is also necessary to suppress the atomic motion so that the off-diagonal~(i.e., state-flip) coupling is not significant.

\subsection{Noise in a Rydberg quantum gate}\label{sec02E}
The robustness of a Rydberg quantum gate denotes how large the gate fidelity can be in the presence of noise in the external control fields, noise in the environment around the qubit, and the real-space motion of atoms.

The first type of noise is the fluctuation of the frequency and magnitude of the external laser or microwave fields that are used to manipulate the atomic states. When the magnitude of the external coherent fields fluctuates, the Rabi frequency for the state rotation fluctuates. This results in errors in the total pulse area which is relevant to all types of Rydberg quantum gates. Moreover, for those protocols that require constant Rabi frequency during the gate sequence~\cite{Levine2019,Shi2018Accuv1} the deviation of the Rabi frequency from the desired one brings extra errors. Like the magnitude fluctuation, the fluctuation of the frequency of the laser or microwave fields effectively broadens the linewidth of the atomic levels involved in the gate. Fortunately, this latter fluctuation can be made negligible~\cite{Legaie2018,Levine2019,Graham2019}.

The second type of noise is the external background field. The Rydberg electron can extend quite far from the nucleus which effectively makes the atomic energy vulnerable to the fluctuation of the background field. To suppress the sensitivity of Rydberg atoms to the stray electric fields, a microwave control protocol was introduced in Ref.~\cite{Booth2017}.  

The third type of noise comes from the motion of the atoms. Even if both the ground and Rydberg states can be trapped in the same optical trap~\cite{Wilson2019} or all relevant states in a quantum gate can be trapped in one trap~\cite{Cortinas2020}, the atom always moves inside the trap. This leads to effective fluctuation of the laser fields upon the atomic qubit even if the laser field strength does not fluctuate. Furthermore, for most experiments on Rydberg gates the trap is turned off during the Rydberg excitation and the free flight of the atom further exaggerates the issue. Beside the issue of fluctuation of the magnitude of the Rabi frequencies, the motion of the atoms introduces random phases in the laser or microwave fields. This phenomenon, usually termed as motional dephasing, Doppler dephasing, or Doppler broadening, can be modeled by a stochastic decay formula $e^{t^2/T_D^2}$, where $T_D\propto 1/(kv_{\text{rms}})$~\cite{Wilk2010,Graham2019,Saffman2011,Jenkins2012} and $v_{\text{rms}}$ is the r.m.s. speed of the qubit in free flight. There have been two proposals to suppress the Doppler dephasing of the two-photon ground-Rydberg transition shown in~\cite{Shi2020}. For the three-photon and four-photon methods of Rydberg excitation, Ref.~\cite{Ryabtsev2011} and Ref.~\cite{Bariani2012} studied protocols for the suppression of the Doppler dephasing. The Doppler dephasing dominates the motion-induced errors in the Rydberg gates for temperatures around or over $10~\mu$K in current experiments. When the atoms are cooled close to the motional ground states in the traps, the Doppler dephasing will be less detrimental while the photon-recoil induced change of motional states can be a more severe issue for the accuracy of the quantum gates especially when relatively long gate durations are involved~\cite{Robicheaux2021}. 

Different protocols of entanglement and logic gates have different sensitivities to the above noise which we review below.

\section{Rydberg gates and entanglement by the blockade mechanism}\label{sec03}
The most frequently employed method of Rydberg-mediated entanglement is by using the blockade effect which was first proposed in Ref.~\cite{PhysRevLett.85.2208} for single atoms, extended to atomic ensembles in Ref.~\cite{Lukin2001}, and further analyzed with more detailed experimental considerations in~\cite{Protsenko2002,Saffman2005,Ryabtsev2005}. In this type of entanglement methods, only one Rydberg excitation is involved ideally; errors arise when more than one Rydberg excitations appear because of the finiteness of the Rydberg interactions. 

In the following we first review protocols based on quasi-rectangular pulses and then protocols based on pulse shaping. For the first category, we look at protocols from two-qubit to multiqubit entanglement. For the two-qubit operations, we first review methods by dynamical phases and then methods by geometrical phases. The list of published fidelity numbers in Table~\ref{table3} quote the results in the corresponding references; in many references more than one type of quantum gates or entangling operations were studied, and relatively larger fidelities are quoted in Table~\ref{table3}. All the data shown in Table~\ref{table3} come from models using $s,~p$, or $d$-orbital Rydberg states that can be excited via a one or two-photon method. There was an analysis of quantum gate by using circular Rydberg states in Ref.~\cite{Xia2013} which shows that gate fidelities over $99.999\%$ are achievable. Table~\ref{table3} shows data for quantum gates and entanglement, where quantum gates can be used in a quantum computing circuit and to create entanglement between individual atoms. On the other hand, the data in Table~\ref{table3} for ``entanglement'' refer to results from protocols that can induce entanglement from a certain N-qubit product state and are not necessarily capable to generate an N-qubit quantum logic gate, where N is an integer and N$\geq2$. For example, Ref.~\cite{PhysRevLett.102.240502} in Table~\ref{table3} studied generation of multi-atom entanglement from a product state. Though Ref.~\cite{Moller2008} studied quantum gates, the data shown in Table~\ref{table3} is for an operation of entanglement generation instead for creating a quantum gate.   %

\begin{figure*}[ht]
\includegraphics[width=7.0in]
{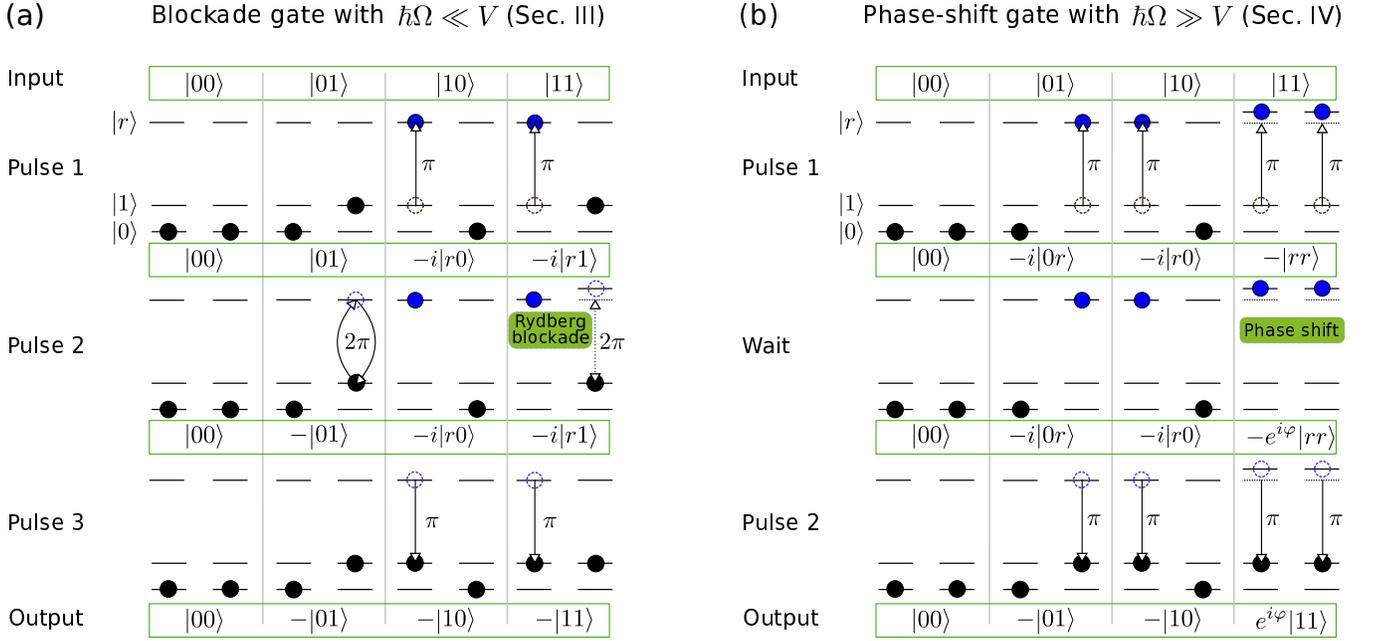}
\caption{(a) The standard Rydberg blockade gate with the condition $\hbar\Omega\ll V$. The first and third pulses are $\pi$ pulses for the transition $|1\rangle\leftrightarrow|r\rangle$. In the second pulse which is a $2\pi$ pulse for inducing a $\pi$ phase change to the input state $|01\rangle$, the input state $|11\rangle$ does not acquire any phase in the strong blockade regime. (b) When $\hbar\Omega\gg V$, the two qubits are simultaneously excited for the transition $|1\rangle\leftrightarrow|r\rangle$ with a $\pi$ pulse, so that all the population in $|1\rangle$ transfers to $|r\rangle$. Then a wait duration $T=\hbar|\varphi/V|$ leads to a phase accumulation $\varphi$ to the state $|rr\rangle$. Afterwards, excitations used in the first pulse are used again so that the qubits are deexcited from Rydberg to ground states. A C$_{\text{Z}}$ gate is realized with $\varphi=\mp\pi$ when $V$ is a repulsive~(attractive) interaction.       \label{figure-originalgate} }
\end{figure*}

\subsection{The original two-qubit phase gates in Ref.~\cite{PhysRevLett.85.2208}}\label{sec03A}
\subsubsection{The Rydberg blockade gate}\label{sec03Aoriginal}
The most known gate protocol was the original Rydberg blockade gate proposed in Ref.~\cite{PhysRevLett.85.2208}. For two atoms each with two ground qubit states $|0\rangle$ and $|1\rangle$, external laser fields are applied to induce the ground-Rydberg transition $|1\rangle\leftrightarrow|r\rangle$. The energy separation between $|0\rangle$ and $|1\rangle$ is $E_{\text{hf}}=h\times 9.2$~GHz for $^{113}$Cs, and $h\times 6.8$~GHz for $^{87}$Rb. So the Rabi frequency for $|1\rangle\leftrightarrow|r\rangle$ shall be much smaller than $E_{\text{hf}}/h$ to avoid exciting $|0\rangle$ to Rydberg state. Of course, one shall also be careful not to excite some Rydberg state $|r'\rangle$ where the frequency separation between $|r'\rangle$ and $|0\rangle$ is near to that between $|r\rangle$ and $|1\rangle$. The error from these unwanted transitions can be suppressed to the order of $10^{-5}$ for typical setups as analyzed in Ref.~\cite{Shi2017}. The method in Ref.~\cite{PhysRevLett.85.2208} is schematically shown in Fig.~\ref{figure-originalgate}(a) which follows as (i) apply a $\pi$ pulse to the control atom to induce the transition $|1\rangle\rightarrow-i|r\rangle$; (ii) apply a $2\pi$ pulse to the target atom to induce the transition $|1\rangle\rightarrow-i|r\rangle\rightarrow-|1\rangle$ if there is no Rydberg blockade; (iii) apply a $\pi$ pulse to the control atom to induce the transition $-i|r\rangle\rightarrow-|1\rangle$. The first and third pulses will change the states only if the input state of the two qubits is $|1\alpha\rangle$ where $\alpha$ is 0 or 1, but the second pulse will change the input state $|01\rangle$, i.e., will induce a $\pi$ phase change. If the blockade interaction of the state $|rr\rangle$ is large enough, the three pulses lead to 
\begin{eqnarray}
  |00\rangle&\rightarrow&|00\rangle\rightarrow|00\rangle\rightarrow|00\rangle,\nonumber\\
  |01\rangle&\rightarrow&|01\rangle\rightarrow-|01\rangle\rightarrow-|01\rangle,\nonumber\\
  |10\rangle&\rightarrow&-i|r0\rangle\rightarrow-i|r0\rangle\rightarrow-|10\rangle,\nonumber\\
  |11\rangle&\rightarrow&-i|r1\rangle\rightarrow-i|r1\rangle\rightarrow-|11\rangle.\label{sec03A01}
\end{eqnarray}
But the actual blockade interaction is finite, so that the second pulse will induce a marginal change as an error to the input state $|11\rangle$ which is $-i|r1\rangle$ at the beginning of the second pulse. Because the transition $-i|r1\rangle\rightarrow-|rr\rangle$ is no longer resonant in the presence of dipole-dipole interaction of the state $|rr\rangle$, $-i|r1\rangle$ will transit to $-|rr\rangle$ with a very small probability. The result is that there will be a small population leakage out of the input state, as well as a phase error to it, which is called the blockade error as analyzed in Ref.~\cite{Shi2018prapp2}. However, most published papers only treat the population error as the blockade error because the phase error is in principle removable by a certain change of the phase in the Rydberg lasers~\cite{Zhang2012}. The blockade error sets a fundamental limit to the achievable gate fidelity of about $0.998$ in the blockade gate~\cite{Zhang2012}.   

The blockade error in the second pulse is sensitive to the interaction $V$ of the state $|rr\rangle$. This can be understood by studying the Hamiltonian 
\begin{eqnarray}
  \hat{H} &=&(\frac{\hbar\Omega}{2} |u\rangle\langle l|+\text{H.c.}) +D |u\rangle\langle u|,\label{sec03A02}
\end{eqnarray}
which can be diagonalized with the following eigenvalues and eigenvectors 
\begin{eqnarray}
  \epsilon_\pm &=& [D\pm \sqrt{(\hbar\Omega)^2+D^2}]/2 , \nonumber\\
   |v_\pm \rangle &=& \left(\frac{\hbar\Omega}{2} |l\rangle + \epsilon_\pm |u\rangle\right)/N_\pm  \label{sec03A03}
\end{eqnarray}
where $N_\pm  = \sqrt{(\hbar\Omega)^2/4 + \epsilon_\pm^2}$, the subscript u and d denote the ``upper'' and ``lower'' states, respectively, and $D$ denotes an energy shift in the state $|u\rangle$. Here, we have $\{|l\rangle,~|u\rangle\}=\{|r1\rangle,~|rr\rangle\}$ and $D$ is the Rydberg interaction $V$. 

For an initial state of $|\psi(t=0)\rangle=-i|r1\rangle$ at the beginning of the second pulse, the state evolves as~\cite{Shi2017}
\begin{eqnarray}
  |\psi(t)\rangle &=&\frac{-2i}{\hbar\Omega}\frac{\epsilon_-N_+e^{-it\epsilon_+/\hbar} |v_+ \rangle - \epsilon_+N_-e^{-it\epsilon_-/\hbar} |v_- \rangle  }{ \epsilon_-  - \epsilon_+ } .\label{detunedTimeevolution}
\end{eqnarray}
At the end of the second pulse, i.e., when $\Omega t=2\pi$, the state becomes 
\begin{eqnarray}
  |\psi(t)\rangle &=&\frac{-2i}{\hbar\Omega}\frac{\epsilon_-N_+e^{-i\frac{2\pi\epsilon_+}{\hbar \Omega}} |v_+ \rangle - \epsilon_+N_-e^{-i\frac{2\pi\epsilon_-}{\hbar \Omega}}|v_- \rangle  }{ \epsilon_-  - \epsilon_+ } ,\nonumber\\
  \label{appstatepulse2}
\end{eqnarray}
then the blockade error, i.e., the population in $|rr\rangle$, is $\epsilon=|\langle rr|\psi(t)\rangle|^2$, which is
\begin{eqnarray}
  \epsilon &=& \frac{\Omega^2}{(V/\hbar)^2+\Omega^2}\sin^2\frac{\pi\sqrt{(V/\hbar)^2+\Omega^2}}{\Omega}.\label{blockadeerror}
\end{eqnarray}
Because deep in the blockade regime we expect $V\gg \hbar\Omega$, we can approximate the above error as $\epsilon\approx \frac{\hbar^2\Omega^2}{V^2}\sin^2\frac{\pi\sqrt{V}}{\hbar\Omega}$. In a practical implementation, the fluctuation in $V$ is severe. Thus one can average $\epsilon$ by taking an integration $\sin^2\frac{\pi\sqrt{V}}{\hbar\Omega}$ in an interval when $\frac{2\pi\sqrt{V}}{\hbar\Omega}$ changes from $2k\pi$ to $2(k+1)\pi$ where $k$ is a very large integer, leading to $\overline{\epsilon}\approx \frac{\hbar^2\Omega^2}{2V^2}$. In order to avoid crosstalk and to measure qubit states with a high fidelity, the qubit spacing $L$ is usually set large enough, so that the dipole-dipole interaction, which drops off as $1/L^3$, is small. When the interaction is small compared to the energy gaps between nearby two-atom Rydberg states coupled via the dipole-dipole interaction, the interaction can be usually described with the second-order perturbation theory. This leads to finiteness of the Rydberg interaction $V$ and, thus, the blockade error $\overline{\epsilon}$ sets a strict limit to the gate fidelity~\cite{Zhang2012}.

The blockade mechanism can be used to realize three-qubit gates. For example, Ref.~\cite{Isenhower2011} analyzed in detail the fidelities for multiqubit C$_k$NOT gates, and showed that a fidelity as high as $99.97\%$ is achievable for $k=3$. Based on the method in Ref.~\cite{PhysRevLett.85.2208}, Ref.~\cite{Shi2018prapp} proposed a protocol to realize the Deutsch gate~\cite{Deutsch1989}, which is a three-qubit gate that by itself constitutes a universal set in the circuit model of quantum computation. Due to the peculiar feature that one type of gate can constitute a universal set, there have been proposals to realize the Deutsch gate in silicon~\cite{Gullans2019} or superconducting systems~\cite{Bakkegaard2019}.

Up until now, most studies on quantum gates and entanglement based on Rydberg interactions focus on alkali-metal atoms. There was one experiment of entanglement between the clock and Rydberg states of alkali-earth strontium atoms in Ref.~\cite{Madjarov2020}. To use the Rydberg blockade for quantum entanglement between nuclear spins in the ground states of divalent AEL atoms, Ref.~\cite{Shi2021fop} proposed a method of Rydberg excitation of ytterbium or strontium atoms with moderate magnetic fields and found that fidelities about 99.8\% are possible for realizing a C$_{\text{Z}}$ gate.

\subsubsection{Definition of fidelity in quantum gates and entanglement by Rydberg atoms}
The characterization of a quantum entangling operation includes fidelity, purity, quantum degree, and entangling capability or entangling power as studied in~\cite{PhysRevLett.77.1413,Poyatos1997,Williams2011}. In quantum information processing the concept of fidelity is frequently used to compare the different degrees of accuracy in different realizations of the same quantum process. In Rydberg atom quantum science, the fidelity for a quantum gate measures how accurately the realized gate operation matches the ideal one, or, more precisely, how faithfully the gate operation can transform an input state to the output state according to the presumed unitary operation. An experimental assessment of the fidelity requires characterization of the density matrix of the qubit system~\cite{Poyatos1997} that involves $O(d^2)$ measurements~\cite{Kim2020} or over. On the other hand, the theoretical study of the fidelity for a quantum operation has been a difficult subject because different formula can satisfy commonly accepted criteria for a reasonable definition. For the example of a two-qubit quantum gate, the input state can be an arbitrary superposition of the four states $\{|00\rangle,~|01\rangle,~|10\rangle,~|11\rangle\}$, where $|0\rangle$ and $|1\rangle$ are the two qubit states and the first~(second) digit in the ket represents the input state for the control~(target) qubit. The gate fidelity is usually defined as an average over the fidelities of operations over all the possible input states. Although there are many definitions of fidelity in, e.g., Refs.~\cite{Jozsa1994,Nielsen2002,Bagan2003,Gilchrist2005,Cabrera2007,Pedersen2007}, we review formulas that were frequently employed for characterizing entanglement by Rydberg interactions.

Many experiments with Rydberg atoms created Bell states~\cite{Wilk2010,Isenhower2010,Zhang2010,Maller2015,Jau2015,Zeng2017,Levine2018,Picken2018,Jo2019,Levine2019,Graham2019,Madjarov2020}. To characterize the quality of the Bell state, the fidelity can be defined as 
\begin{eqnarray}
\mathcal{F} &=& \text{Tr}^2\sqrt{\sqrt{\rho_{\text{id}}}\rho \sqrt{\rho_{\text{id}}}}\label{fidelity01}
 \end{eqnarray}
where $\rho$ and $\rho_{\text{id}}$ are the density matrices for the realized and ideal states, respectively. $\rho$ and $\rho_{\text{id}}$ can be exchanged in the above formula which is a basic property for a valid similarity measurement between two density matrices. Equation~(\ref{fidelity01}) is equivalent to the fidelity defined in~\cite{Poyatos1997}. When $\rho_{\text{id}}=|\Psi\rangle\langle\Psi|$ for a Bell state $|\Psi\rangle$, the above equation simplifies to $\mathcal{F}=\langle\Psi|\rho|\Psi\rangle $~\cite{Jozsa1994,Gilchrist2005,Cabrera2007}, which is a formula frequently used for quantifying how accurately the Bell states were prepared in experiments~\cite{Wilk2010,Isenhower2010,Zhang2010,Maller2015,Jau2015,Zeng2017,Levine2018,Picken2018,Jo2019,Levine2019,Graham2019}. The experimental measurement of the Bell-state fidelity involves measurement of the off-diagonal matrix elements which can be done by parity assessment after applying extra pulses to rotate the states. For theoretical investigation about the average accuracy of an entangling gate by Eq.~(\ref{fidelity01}), one can sample uniformly over the state space spanned by the four input states for the case of two-qubit gates.  

Beside the above formula frequently used in experiments to characterize the quality of Bell state generation, one can compare $\rho$ and $\rho_{\text{id}}$ by defining a ``distance'' between them, $\mathscr{D}(\rho,\rho_{\text{id}})\equiv$Tr$|\rho-\rho_{\text{id}}|/2$~\cite{Gilchrist2005}, where $|A|\equiv\sqrt{A^\dag A}$ for an arbitrary matrix $A$. $\mathscr{D}(\rho,\rho_{\text{id}})$ is zero when $\rho=\rho_{\text{id}}$, which means that their ``distance'' is zero since they coincide with each other. In this case, the fidelity is given by
\begin{eqnarray}
\mathcal{F}  &=& 1- \mathscr{D}(\rho,\rho_{\text{id}}).\label{fidelity03}
\end{eqnarray}
One can find~\cite{Zhang2012} that the fidelity defined in Eq.~(\ref{fidelity03}) is much more sensitive to dephasing errors compared to that in Eq.~(\ref{fidelity01}), and, thus, can be useful to quantify how accurately a state is created~\cite{Beterov2016jpb}.

For quantum information processing, a quantum gate is required, and it is useful to measure how accurately a quantum gate is realized. For the controlled-NOT~({\footnotesize CNOT}), the probability truth table fidelity was frequently employed~\cite{Isenhower2010,Zhang2010,Zeng2017,Levine2019}
\begin{eqnarray}
\mathcal{F}  &=& \text{Tr}(U^\dag \mathscr{U})
\end{eqnarray}
where $U$ is the probability truth table for the ideal gate. For a {\footnotesize CNOT} in the basis $\{|00\rangle,|01\rangle,|10\rangle,|11\rangle\}$, $U$ is
\begin{eqnarray}
U&=& \left(
  \begin{array}{cccc}
    1& 0 & 0&0\\
    0 & 1 &0&0\\
    0 &0 & 0&1\\
    0& 0 & 1&0\\   
    \end{array} 
  \right) ,\label{CNOTtruth}
  \end{eqnarray}
and $\mathscr{U}$ is the probability truth table when the gate is actually realized with some errors, i.e., $\mathscr{U}$ will deviate from the ideal one in Eq.~(\ref{CNOTtruth}) because of various error sources~(see Sec.~\ref{sec02E}). In experiments, the matrix elements in $\mathscr{U}$ can be measured with multiple gate cycles and different input states. Because the {\footnotesize CNOT} truth table fidelity does not fully take into account of dephasing errors, it can be larger than the fidelity defined in other ways as studied in~\cite{Zhang2012}. 

A formula of gate fidelity used in Refs.~\cite{Su2016,Su201701,Su2018cpb,Guo2020,Wu2021pr} adopts the definition given in Ref.~\cite{Nielsen2002},
\begin{eqnarray}
\mathcal{F} &=& \frac{1}{\mathscr{N}+1}+\frac{1}{\mathscr{N}^2(\mathscr{N}+1)} \sum_k\text{Tr}(U X_k^\dag U^\dag \mathcal{E}(X_k)) , \label{fidelity05}
\end{eqnarray}
where $\mathscr{N}=2^d$ for a $d$-qubit gate or entanglement operation, $U$ is the matrix of the quantum gate~[such as the one in Eq.~(\ref{CNOTtruth})], and $\mathcal{E}$ is the quantum gate or entanglement operation we would like to characterize. If we take $d=2$ as an example, the summation over k means to sum over $X_k\in\{{\mathbf{I}},~\sigma_x,~\sigma_y,~\sigma_z\}\otimes\{{\mathbf{I}},~\sigma_x,~\sigma_y,~\sigma_z\}$, where ${\mathbf{I}}$ is the $2\times2$ identity matrix and $\sigma_{x,y,z}$ are the Pauli matrices. To evaluate the fidelity in a quantum gate or entanglement operation, we can calculate the sum over $k$ by evaluating $\mathcal{E}(X_k)$ one by one using the quantum gate or entanglement protocol. 

Another formula of gate fidelity which takes into account both population loss as well as phase errors is~\cite{Pedersen2007}
\begin{eqnarray}
\mathcal{F} &=& \frac{1}{\mathscr{N}(\mathscr{N}+1)}\left[  |\text{Tr}(U^\dag \mathscr{U})|^2 + \text{Tr}(U^\dag \mathscr{U}\mathscr{U}^\dag U ) \right], \label{fidelity04}
\end{eqnarray}
where $\mathscr{N}$ is similarly defined as in Eq.~(\ref{fidelity05}), which is 4 for a two-qubit controlled-$Z$~(C$_{\text{Z}}$) or {\footnotesize CNOT} gate, and $U$ is given in Eq.~(\ref{CNOTtruth}) for the {\footnotesize CNOT} gate. The fidelity defined in Eq.~(\ref{fidelity04}) can capture the phase errors because the term $|\text{Tr}(U^\dag \mathscr{U})|$ is sensitive to the imaginary parts of the output wavefunction of the gate. For this reason, it can be used to characterize the C$_{\text{Z}}$ gate~\cite{Goerz2014,Petrosyan2017,Shi2018Accuv1,Shi2020} where correct phases of the output are important. Only when the phases in the nonzero elements of $\mathscr{U}$ of C$_{\text{Z}}$ are correct can one transform it to an accurate {\footnotesize CNOT}~\cite{Maller2015}, and large phase errors can lead to low fidelities.

Throughout this review, when a fidelity is quoted from the cited reference as shown in Tables~\ref{table3},~\ref{table4}, and~\ref{table5}, the largest, explicitly stated fidelity with Rydberg-state decay at room temperature~(or approximately at room temperature) will be shown unless otherwise specified; when the fidelity quoted is with Rydberg-state lifetimes in cryogenic environments or does not consider the Rydberg-state decay, it will be specified.

\begin{table}  
  \centering
  \begin{tabular}{|c|l|c|}
    \hline  Requirement  & Theoretical fidelity; type of operation; where it's taken   & References \\ \hline
\multirow{3}{*}{\begin{tabular}{c}\text{Basic blockade gates}\\ \text{with the method of}\\\text{Ref.}~\cite{PhysRevLett.85.2208}~(Sec.~\ref{sec03Aoriginal})\end{tabular}} & $99.8\%$; C$_{\text{Z}}$ gates; abstract of Ref.~\cite{Zhang2012}  & Ref.~\cite{Zhang2012}\\  \cline{2-3}	
 & ${\color{blue}99.97}\%$; 4-qubit {\footnotesize C}$_3${\footnotesize NOT} gate; Table~1 on page 763 of Ref.~\cite{Isenhower2011} & Ref.~\cite{Isenhower2011} 	 \\  \cline{2-3}
 & $98.2\%$; Deutsch gate; text below Fig.~4 on page 4 of Ref.~\cite{Shi2018prapp} & Ref.~\cite{Shi2018prapp} 	 \\ \hline
\multirow{3}{*}{\begin{tabular}{c}\text{Optically detuned}\\ \text{Rabi oscillations}\\(Sec.~\ref{sec03detuned01})\end{tabular}}  & 99\%; 2-qubit gate; text below Fig.~5 of page 5 of Ref.~\cite{Han2016} & Ref.~\cite{Han2016}	 \\\cline{2-3}
 & $99.5\%$~(at 4.2~K); 2-qubit gate; top right of page 14 of Ref.~\cite{Shi2018Accuv1}& Ref.~\cite{Shi2018Accuv1} 	 \\  \cline{2-3}
 & $99.994\%$~(at 4.2~K); C$_{\text{Z}}$ gate; caption of Fig.~3 on page 4 of Ref.~\cite{Shi2017}& Ref.~\cite{Shi2017} 	 \\  \hline
Rydberg dressing& $98\%$; C$_{\text{Z}}$ gate; bottom of page S163 of Ref.~\cite{Brion2007jpb} & Ref.~\cite{Brion2007jpb}\\  \hline
Transition slow-down& $99.9\%$; {\footnotesize CNOT} gate; right of page 4 of Ref.~\cite{Shi2020tsd} & Ref.~\cite{Shi2020tsd}\\  \hline
Geometric phases& $94.98\%$; 2-qubit gate~(via cavity); upper left of page 7 of Ref.~\cite{Zhao2018} & Ref.~\cite{Zhao2018}\\  \hline
\multirow{4}{*} {\begin{tabular}{c}\text{Asymmetries in}\\\text{interactions}\\ ~(Sec.~\ref{sec03Bnew2})\end{tabular}}  
 & $99.06\%$; Bell state; lower left of page 3 of Ref.~\cite{Shi2014} & Ref.~\cite{Shi2014}\\  \cline{2-3}
& $99.86\%$; C$_{\text{Z}}$ gate; Table~1 on page 6 of Ref.~\cite{Su2018cpb} & Ref.~\cite{Su2018cpb}\\    \cline{2-3}
& $84\%$; 8-qubit entanglement; text below Eq.~(7) on page 4 of Ref.~\cite{PhysRevLett.102.240502} & Ref.~\cite{PhysRevLett.102.240502}\\  \cline{2-3}
 & $96\%$; 3-qubit gate; text below Eq.~(9) on page 3 of Ref.~\cite{Wu2010} & Ref.~\cite{Wu2010}\\  \hline
\multirow{4}{*} {\begin{tabular}{c}Dissipation\\ ~(Sec.~\ref{sec03dissipation})\end{tabular}} & $99.5\%$;  Bell state; Fig.~2 on page 3 of the SM\tnote{c} of Ref.~\cite{BhaktavatsalaRao2013} & Ref.~\cite{BhaktavatsalaRao2013}\\   \cline{2-3}
 & $99.65\%$;  Bell state~(via cavity); lower right of page 4 of Ref.~\cite{Shao2017cavitydisspation} & Ref.~\cite{Shao2017cavitydisspation}\\  \cline{2-3}
 & $99.17\%$;  2-qubit entanglement; Fig.~5(c) on page 6 of Ref.~\cite{Zhu2020}& Ref.~\cite{Zhu2020}\\  \cline{2-3}
 & $98.8\%$;  20-qubit entanglement; middle right of page 3 of Ref.~\cite{Rao2014} & Ref.~\cite{Rao2014}\\ \hline
\multirow{17}{*}  {\text{\begin{tabular}{c}Pulse shaping\\(from Sec.~\ref{sec03C0m2}\\to Sec.~\ref{sec03C03})\end{tabular}}} 
& $99.7\%$;  2-qubit gate; Table II on page 780 of Ref.~\cite{Muller2011} & Ref.~\cite{Muller2011}\\ \cline{2-3}
& $98\%$;  C$_{\text{Z}}$ gate; Fig. 14 on page 8 of Ref.~\cite{Muller2014} & Ref.~\cite{Muller2014}\\ \cline{2-3}
& ${\color{blue}99.99}\%$;  2-qubit gate; abstract of Ref.~\cite{Goerz2014} & Ref.~\cite{Goerz2014}\\ \cline{2-3}
 & $99.8\%$;  C$_{\text{Z}}$ gate; Fig.~5 on page 7 of Ref.~\cite{Keating2015} & Ref.~\cite{Keating2015}\\ \cline{2-3}
 & ${\color{blue}99.99}\%$;  C$_{\text{Z}}$ gate; left of page 1 of Ref.~\cite{Theis2016} & Ref.~\cite{Theis2016}\\ \cline{2-3}
 & $99.2\%$;  2-qubit gate; text below Fig.~9 on page 8 Ref.~\cite{Wu2017} & Ref.~\cite{Wu2017}\\ \cline{2-3}
 & $99.9\%$;  C$_{\text{Z}}$ gate; text below Fig.~6 on page 6 of Ref.~\cite{Kang2018} & Ref.~\cite{Kang2018}\\\cline{2-3}
 & $99.7\%$;  C$_{\text{Z}}$ gate; left of page 5 of Ref.~\cite{Mitra2020} & Ref.~\cite{Mitra2020}\\\cline{2-3}
& $99.9\%$;  C$_{\text{Z}}$ gate; abstract of Ref.~\cite{Saffman2020} & Ref.~\cite{Saffman2020}\\ \cline{2-3}
 & $99.81\%$;  2-qubit gate; Table~II on page 10 of Ref.~\cite{Kang2020} & Ref.~\cite{Kang2020}\\\cline{2-3}
 & $99.94\%$;  2-qubit gate; abstract of Ref.~\cite{Guo2020} & Ref.~\cite{Guo2020}\\\cline{2-3}
& $99.5\%$;  10-qubit entanglement; text below Eq.~(9), page 4, Ref.~\cite{Moller2008} & Ref.~\cite{Moller2008}\\ \cline{2-3}
& $98\%$;  4-qubit entanglement; Fig.~3(d) on page 3 of Ref.~\cite{Muller2009} & Ref.~\cite{Muller2009}\\   \cline{2-3}
& $98.21\%$; 3-qubit gate; left of page 2039 of Ref.~\cite{Shen:19} & Ref.~\cite{Shen:19}\\\cline{2-3}
& $97.6\%$; Toffoli gate; caption of Fig.~S3 on page 6 of SM of Ref.~\cite{Levine2019} & Ref.~\cite{Levine2019}\\\cline{2-3}
 & $99.7\%$;  Toffoli gate; Fig.~9(b) on page 9 of Ref.~\cite{Guo2020} & Ref.~\cite{Guo2020}\\ \cline{2-3}
 & $99.8\%$;  3-qubit C$_2$NOT gate; Fig.~7(d) of Ref.~\cite{MengLi2021} & Ref.~\cite{MengLi2021}\\  \hline
\multirow{2}{*} {\begin{tabular}{c}\text{Pulse shaping in two}\\\text{ensembles~(Sec.~\ref{sec03Censemble})}\end{tabular}} 
& $99.8\%$;  C$_{\text{Z}}$ gate; text below Fig.~4 on page 3 of Ref.~\cite{Beterov2013} & Ref.~\cite{Beterov2013}\\ \cline{2-3}
 & $99.9\%$; C$_{\text{Z}}$ gate; Fig.~7 on page 9 of Ref.~\cite{Beterov2016jpb} & Ref.~\cite{Beterov2016jpb}\\   \hline
  \end{tabular}
  \caption{  \label{table3} Entanglement and gate fidelities with Rydberg blockade. Labels like ``2-qubit'' and ``3-qubit'' gates denote phase gates that can be transformed to C$_{\text{Z}}$, {\footnotesize CNOT}, or Toffoli gates via single-qubit gates. ``SM'' denotes Supplemental Materials. The order in column 1 is according to their discussion in text, and the order in each block of column 2 is according to the date of publication where 2-qubit gates are first shown. Most data were with Rydberg-state decay rates around $2\pi\times1$~kHz unless otherwise specified; For fidelities at 4~K, Ref.~\cite{Theis2016} showed that C$_{\text{Z}}$ gates with fidelities $99.999\%$ are possible. Among the two types of gates in Ref.~\cite{Isenhower2011}, the one without using asymmetric interactions is quoted. The two methods in Refs.~\cite{Shi2018Accuv1,Shi2017} belong the the category in Sec.~\ref{sec04} but shown here because of their close similarity to the method in Ref.~\cite{Levine2019} which does not state intrinsic fidelities. The blue color highlights the largest fidelities for gates and entanglement with two or more than two qubits.}
  \end{table}


\subsection{Two-qubit phase gates with optically detuned Rabi cycles}\label{sec03detuned01}
It is possible to use detuned Rabi cycles in the regime of excitation blockade to implement Rydberg gates. There are in general two classes of methods. The first class was introduced in Ref.~\cite{Han2016}, which shows that by exciting both the control and target qubits with the same Rabi frequency $\Omega$ and detuning $\delta$ for the transition $|1\rangle\leftrightarrow|r\rangle$, it is possible to find an appropriate ratio between $\Omega$, $\delta$, and the pulse duration so that all the four input states $\{|00\rangle,~|01\rangle,~|10\rangle,~|11\rangle\}$~(only the latter three are excited) return to themselves, with a $\pi$ phase change to $|00\rangle$~(or $|11\rangle$) relative to the other three states. There are intrinsic phase and population errors in the three states where phase change occurs, and a large pulse area is required if a large fidelity is desired. However, the protocol in Ref.~\cite{Han2016} works in the blockade regime and, thus, is relative simple concerning an experimental implementation. 

\begin{figure*}[ht]
\includegraphics[width=6.0in]
{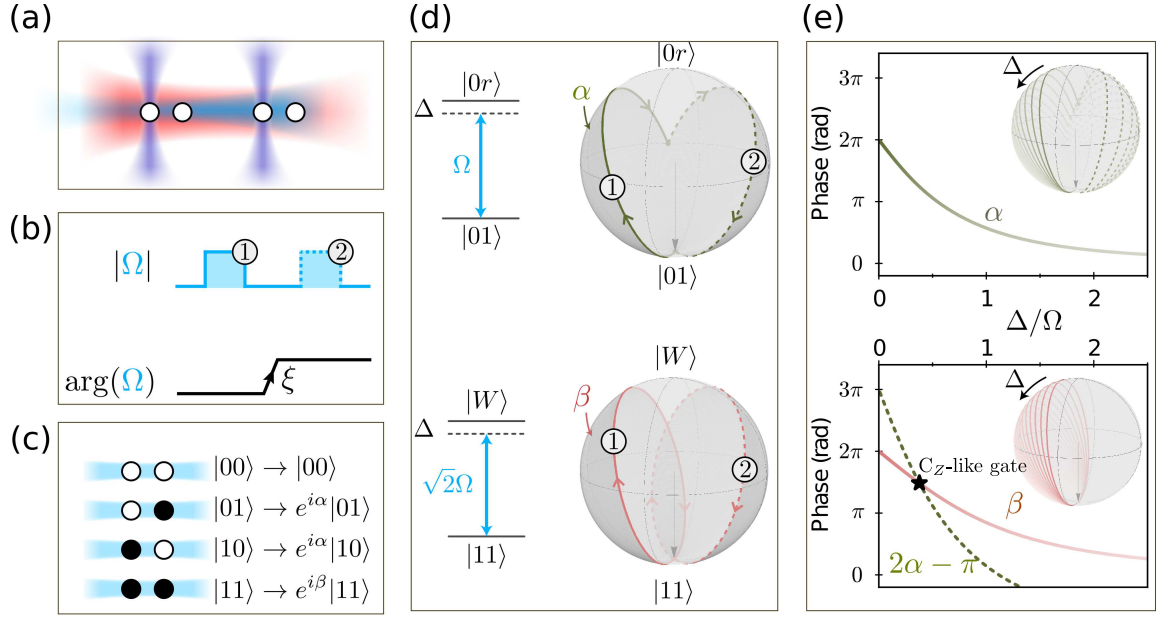}
\caption{Overview of the C$_Z$-like gate via detuned Rabi oscillations which was experimentally studied in Ref.~\cite{Levine2019}. (a) Atoms arranged in pairs are globally driven. (b) In the strong interaction regime $\hbar\Omega\ll V$, Ref.~\cite{Levine2019} studied an extension of the gate method in Ref.~\cite{Shi2018Accuv1} via adding a phase twist $\xi$ in the Rydberg Rabi frequency by modifying phases of the laser fields. (c) The two input states $|01\rangle$ and $|10\rangle$ experience two partial detuned Rabi cycles and are restored upon the completion of the Rydberg excitation, and simultaneously the state $|11\rangle$ is also restored. The time evolutions for $|01\rangle$ and $|10\rangle$ are similar. (d) Both $|01\rangle$ and $|11\rangle$ undergo detuned Rabi oscillations with the same detuning but different effective Rabi frequencies. When $|11\rangle$ finishes one full detuned Rabi cycle, $|01\rangle$ has not yet. After inserting a phase $\xi$ into the Rydberg Rabi frequency and when $|11\rangle$ undergoes another detuned Rabi cycle, $|01\rangle$ undergoes another partial detuned Rabi cycle around another axis on the Bloch sphere. A careful choice of $\xi$ can exactly lead to restoration of the state $|01\rangle$. (e) As shwon in Ref.~\cite{Levine2019}, the dynamical phases $\alpha$ and $\beta$ are determined by the area enclosed by the Bloch sphere trajectory. The figures are adopted from Phys. Rev. Lett. 123, 170503 (2019) with minor modifications.       \label{figure-detunedgate} }
\end{figure*}

Another method by detuned Rabi cycles in the regime of Rydberg blockade is to use the phase accumulated in the detuned Rabi oscillations~\cite{Levine2019}. The phase accumulation in detuned Rabi cycles could be used in the second class of methods~(denoted by \textcircled{2} in Fig.~\ref{figure01}) that involve excitation of two-atom Rydberg states where the blockade error can be completely suppressed~\cite{Shi2017,Shi2018Accuv1}. Specifically, Ref.~\cite{Shi2018Accuv1} showed that by using a single detuned pulse on both qubits, all the four input states $\{|00\rangle,~|01\rangle,~|10\rangle,~|11\rangle\}$ return to themselves perfectly, leading to a phase gate of the form diag$\{|00\rangle,~e^{i\alpha}|01\rangle,~e^{i\alpha}|10\rangle,e^{i\beta}~|11\rangle\}$, which can be repeated several times to form a {\footnotesize CNOT} when assisted by single-qubit gates. Later on, Ref.~\cite{Levine2019} extended the protocol in~\cite{Shi2018Accuv1} to the Rydberg blockade regime where $\Omega, \delta\ll V$, and showed a C$_{\text{Z}}$-like gate implemented by only two Rydberg pulses; the gate method in Ref.~\cite{Levine2019} is shown in Fig.~\ref{figure-detunedgate}. Compared to the method in Ref.~\cite{Han2016}, the method in Refs.~\cite{Shi2018Accuv1,Levine2019} is based on a phase accumulation unequal to $\pi$; compared to the original one-pulse interference method in Ref.~\cite{Shi2018Accuv1}, the method in Ref.~\cite{Levine2019} needs two pulses to implement an entangling gate, but has the benefit of realizing a C$_{\text{Z}}$-like gate with much fewer numbers of Rydberg pulses and smaller error from the fluctuation of Rydberg interaction. Because the gate speed is inverse to the innate Rydberg Rabi frequency~(instead of being inverse to a a higher-order Rabi frequency as frequently required in, e.g., antiblockade), the interference method can be fast and accurate. Because the gap time between the two pulses in the {\footnotesize CNOT} gate of Ref.~\cite{Levine2019} is negligible and in principle can be as small as possible, the method can greatly suppress the Doppler dephasing as well as the Rydberg-state decay, which is why the largest fidelity in ground-state entanglement by Rydberg interaction was realized with quantum interference in detuned Rabi cycles~\cite{Levine2019}.

Because of the C$_{\text{Z}}$-like gate in Ref.~\cite{Levine2019} has a fast speed, we give more details about its implementation. In particular, we would use the picture of detuned Rabi oscillations as adopted in Sec.~\ref{sec03Aoriginal} so as to unveil the details of the state evolution. Also, we will show that a controlled-phase gate with an arbitrary angle $\theta$ is realizable. In the strong blockade regime and when both the control and target qubits experience a detuned Rydberg excitation, the system Hamiltonians are in the form of Eq.~(\ref{sec03A02}) where $D=\hbar\Delta$ and
\begin{eqnarray}
 \{|u\rangle,~|l\rangle\} &=& \{|0r\rangle,~|01\rangle \},~\text{for the input}~|01\rangle,\nonumber\\
 \{|u\rangle,~|l\rangle\} &=& \{|r0\rangle,~|10\rangle \},~\text{for the input}~|10\rangle,\nonumber\\
 \{|u\rangle,~|l\rangle\} &=& \{(|1r\rangle+|r1\rangle)/\sqrt{2},~|11\rangle \},~\text{for the input}~|11\rangle. \label{uldefinition}
\end{eqnarray}
Below, we use the eigenvalues and eigenvectors in Eqs.~(\ref{sec03A03}) for analysis.

The pulse sequence in Ref.~\cite{Levine2019} is designed for realizing $\{|00\rangle,~|01\rangle,~|10\rangle,~|11\rangle\}\rightarrow\{|00\rangle,~e^{i\alpha}|01\rangle,~e^{i\alpha}|10\rangle,~e^{i\beta}|11\rangle\}$ with $2\alpha-\beta=\theta$, where the case $\theta=\pi$ was studied in Ref.~\cite{Levine2019}. If we ignore the transient for the phase change, the schematic of the gate is shown in Fig.~\ref{figure-detunedgate}. As from~\cite{Shi2018Accuv1,Shi2017} or by looking at Eqs.~(\ref{sec03A02}) to~(\ref{appstatepulse2}), the Rydberg pulse for the transition $|1\rangle\rightarrow|r\rangle$ with a Rabi frequency $\Omega$ for a duration $t=2\pi/\overline{\Omega}$ leads to the following state transform
\begin{eqnarray}
 |11\rangle\rightarrow e^{-i\pi(1+\Delta/\overline{\Omega})}|11\rangle, \label{evolutionfor11}
\end{eqnarray}
where $\overline{\Omega} = \sqrt{\Delta^2+2\Omega^2}$; the time evolution for $|01\rangle$ and $|10\rangle$ leads to states in the form of Eq.~(\ref{detunedTimeevolution}). After the first pulse, another pulse of the same duration, but with a Rabi frequency $e^{i\xi}\Omega$, is used. The state $|11\rangle$ experiences another state transform as in Eq.~(\ref{evolutionfor11}), but because the other two input states $|01\rangle$ and $|10\rangle$ are in superpositions of ground and Rydberg states at the beginning of the second pulse, their time evolution results in a state that is lengthy when written out, c.f. Eq.~(\ref{detunedTimeevolution}). In order to have the states $|01\rangle$ and $|10\rangle$ restored upon the completion of the pulse sequence, we need the following condition
\begin{eqnarray}
  |01\rangle&\xrightarrow{\Omega} \text{superposition of }&|01\rangle\text{ and }|0r\rangle \xrightarrow{e^{i\xi}\Omega} e^{i\alpha} |01\rangle,  \label{evolutionfor01}\\
  |10\rangle&\xrightarrow{\Omega} \text{superposition of }&|10\rangle\text{ and }|r0\rangle \xrightarrow{e^{i\xi}\Omega} e^{i\alpha} |10\rangle,   \label{evolutionfor10}
\end{eqnarray}
where the phase $\alpha$ shall satisfy
\begin{eqnarray}
 2\alpha+2\pi(1+\Delta/\overline{\Omega}) = \theta+2k\pi,\label{alpharelation01}
\end{eqnarray}
with an integer $k$, which means that $\alpha$ is a function of $\theta$; the case $\theta=\pi$ corresponds to a C$_{\text{Z}}$-like gate. Equations~(\ref{evolutionfor01}) and~(\ref{evolutionfor10}) have an identical form, so we focus on, e.g., the time evolution involving $|01\rangle$. Because the intermediate states in both Eqs.~(\ref{evolutionfor01}) and~(\ref{evolutionfor10}) are messy, we would like to avoid directly writing them out. However, one can easily show that the establishment of Eq.~(\ref{evolutionfor01}) is equivalent to the condition
\begin{eqnarray}
e^{i\alpha}  \exp\left\{it\left[(e^{-i\xi}\Omega |0r\rangle\langle 01|+\text{H.c.}) /2 +\Delta |0r\rangle\langle 0r| \right]\right\} = \exp\left\{-it\left[(\Omega |0r\rangle\langle 01|+\text{H.c.}) /2 +\Delta |0r\rangle\langle 0r| \right]\right\}
\end{eqnarray}
by understanding that the left hand side of the equation above denotes the unitary operator of the time-reversal evolution from the state $e^{i\alpha} |01\rangle$ back to the intermediate state of Eq.~(\ref{evolutionfor01}). Using the above the equation, we have 
\begin{eqnarray}
e^{i\alpha} ( \epsilon_-e^{it \epsilon_+}  - \epsilon_+ e^{it \epsilon_-}) &=& \epsilon_-e^{-it \epsilon_+}  - \epsilon_+ e^{-it \epsilon_-} ,\nonumber\\
e^{i(\alpha+\xi)} ( e^{it \epsilon_+}  - e^{it \epsilon_-}) &=& e^{-it \epsilon_+}  -  e^{-it \epsilon_-} .\label{alpharelation02}
\end{eqnarray}
For a set of parameters $(k,~\theta,~\Omega)$, we can use Eqs.~(\ref{alpharelation01}), and~(\ref{alpharelation02}) to solve $\Delta,~\alpha$, and~$\xi$ so that
\begin{eqnarray}
e^{-i\xi} &=& \frac{ e^{it \epsilon_+}  - e^{it \epsilon_-}}{ e^{-it \epsilon_+}  -  e^{-it \epsilon_-}}\frac{\epsilon_-e^{-it \epsilon_+}  - \epsilon_+ e^{-it \epsilon_-}}{\epsilon_-e^{it \epsilon_+}  - \epsilon_+ e^{it \epsilon_-}} ,\label{alpharelation03}
\end{eqnarray}
where $ \epsilon_\pm = (\Delta\pm \sqrt{\Omega^2+\Delta^2})/2$ according to Eq.~(\ref{sec03A03}). One can also understand why there is only one solution to $\xi$ via the Bloch-sphere formalism in Fig.~\ref{figure-detunedgate}. Though there will be solutions to any $(\theta,~\Omega)$, of particular interest is $\theta=\pi$ with the solution as $(\Delta/\Omega,~\xi,~t\Omega)\approx(0.3773711,~3.902423,~4.292682)$, where the latter two parameters are in units of radians. Since a quantum circuit may require some other controlled-phase gates, some angle $\theta$ different from $\pi$ may be useful. The results in Eq.~(\ref{alpharelation03}) can allow us to calculate solutions with any $\theta$. For example, the solution is $(\Delta/\Omega,~\xi,~t\Omega)\approx(0.7281492,~4.059675,~3.950048)$ with $\theta=\pi/2$; for $\theta=\pi/3$, the solution is $(\Delta/\Omega,~\xi,~t\Omega)\approx(0.9384181,~4.022575,~3.701998)$.

\subsection{Two-qubit entanglement and phase gates by Rydberg dressing}\label{sec03Bnew1}
A useful method to generate ground-state entanglement is by dressing qubit states with Rydberg states so that there can be an effective blockade between the dressed qubit states~\cite{Santos2000,Bouchoule2002}. This method was experimentally demonstrated in Ref.~\cite{Jau2015} where the qubit state $|0\rangle$ was excited to a $p$-orbital Rydberg state $|r\rangle$ with a one-photon Rabi frequency $\Omega$ and detuning $\Delta$. This leads to splitting of the two states $|0\rangle$ and $|r\rangle$ into two dressed states, where one is mainly populated by the ground state and hence can be viewed approximately as a new qubit state $|0'\rangle$. Because $|0'\rangle$ has some mixing of Rydberg states, the state $|0'0'\rangle$ has an inherent Rydberg interaction. This can be used to create entangled state in one step starting from a two-atom product state $|11\rangle$ when microwave-field transition between $|1\rangle$ and $|0'\rangle$ is used on both qubits ~\cite{Jau2015}. The microwave field induces the transition $|11\rangle\leftrightarrow(|10'\rangle+|0'1\rangle)/\sqrt{2}$ since $|0'0'\rangle$ is no longer resonant because of the Rydberg interaction. Because there is some portion of Rydberg state in $|0'\rangle$, the method can not create perfect ground-state entanglement, but has the advantage of avoiding single-site addressing. The achievable entanglement fidelity by Rydberg dressing has a fundamental limit similar to that in the original Rydberg blockade gate as analyzed in Ref.~\cite{Saffman2016}.

Another way to employ Rydberg dressing is given in Ref.~\cite{Brion2007jpb}, where the two qubit states $|0(1)\rangle$ are dressed with Rydberg states with different Rabi frequencies and detunings $\Delta_{0(1)}$, where $|\Delta_0-\Delta_1|\ll V/\hbar$. Besides the phase accumulation due to the ac Stark shifts in $|00\rangle$ and $|11\rangle$, there will be a Rabi-like flopping between $|01\rangle$ and $|10\rangle$ via a fourth-order Rabi frequency derived from the perturbation theory. By choosing appropriate parameters, it is possible to implement a C$_{\text{Z}}$ gate with a fidelity over $99\%$ if Rydberg-state decay, fluctuation of $V$, and motional dephasing are ignored. Taking into account of these issues, fidelities around $90\%$ could be possible~\cite{Brion2007jpb}. 

The dressing method is not only useful in quantum computing, but also useful in quantum simulation as shown in, e.g., Refs.~\cite{Pupillo2010,Henkel2010,Shi2018pra_m}. In particular, Ref.~\cite{Lee2017} experimentally demonstrated that by mapping the symmetric many-body ground state of an atomic ensemble to the photonic state of a cavity quantum electrodynamics system, one can emulate the Jaynes-Cummings~(JC) model by Rydberg dressing. About 15 years prior to the work in Ref.~\cite{Lee2017}, however, Ref.~\cite{Unanyan2002} had theoretically investigated the JC model by Rydberg interactions. When the resonant dipole-dipole interaction is shared by ground-state atoms via Rydberg dressing, many-body entanglement in the form of a W-state can be created via the coupling between internal and vibrational states~\cite{Stojanovi2021}. The potential of the dressing method in quantum information processing is far from being well exploited in both theory and experiments.

Rydberg dressing can also be used in quantum metrology. For example, Refs.~\cite{Bouchoule2002,Gil2014} showed that Rydberg dressing can enable a quadratic interaction term so as to squeeze the collective atomic state. The capability to achieve extreme squeezing lies in the long lifetimes of the Rydberg-dressed ground state that adheres effective and strong interactions from the dressed Rydberg states. Squeezing by resonant Rydberg excitation was also proposed~\cite{Opatrny2012} if the dissipation in the Rydberg states can be neglected.

\subsection{Two-qubit {\footnotesize CNOT} gates via transition slow-down}\label{TSD}
There was a {\footnotesize CNOT} protocol named transition slow-down~(TSD) proposed in Ref.~\cite{Shi2020tsd}. The mechanism of TSD is that when the state of one atom transits back and forth between ground and Rydberg states, the ground-Rydberg transition in a nearby atom is interrupted but not annihilated since the state of the control atom can still be excited back to the ground state. This resembles the quantum Zeno effect that a frequent interruption by measurement of a quantum system can inhibit the presumed time evolution in it~\cite{MIsra1977}. Although only one type of spin-echo assisted quantum gate was shown in Ref.~\cite{Shi2020tsd} by TSD, there can be various ways to use TSD for quantum logic since TSD is rather simple to implement: just excite both qubits simultaneously to Rydberg states with appropriate parameters compatible with a quantum logic gate. In Appendix~\ref{TSD-appendix}, two examples of efficient {\footnotesize CNOT} via TSD that were not studied in Ref.~\cite{Shi2020tsd} are shown; the gates in Appendix~\ref{TSD-appendix} are not as fast as that in Ref.~\cite{Shi2020tsd}, but could be used to understand why a slow-down effect can help to realize a quantum gate.

\subsection{Two-qubit phase gates by geometric phases with nonadiabatic excitation}\label{nonadiabaticgemo}
The methods in Secs.~\ref{sec03Aoriginal},~\ref{sec03detuned01},~\ref{sec03Bnew1}, and~\ref{TSD} depends on dynamical phases; though the {\footnotesize CNOT} gate in~\ref{TSD} is not a phase gate, it can be transformed to a C$_{\text{Z}}$ gate via basis transform as shown in~\cite{Shi2020tsd}. Here, we review gates by geometric phases.

It turns out that geometric phases can emerge even without using adiabatic following, and there are several Rydberg gate protocols using nonadiabatic geometric phases. For a certain two-atom input state which acquires a phase because of the gate sequence, the phase is geometric if the expectation value of the energy of the state~(at any moment during the gate sequence) stays zero in the whole process of the gate sequence. A method in Ref.~\cite{Zhao2017} showed this type of Rydberg gate depending on nonadiabatic geometric phases. The method in Ref.~\cite{Zhao2017} leads to a gate with an entangling power~\cite{Williams2011} less than the maximal value such as in a C$_{\text{Z}}$ or {\footnotesize CNOT}, and thus needs to be repeated several times to become a C$_{\text{Z}}$ or {\footnotesize CNOT} when assisted by single-qubit gates. However, since any two-qubit entangling gate can form a universal set with several single-qubit gates~\cite{Bremner2002}, the fast gate in Ref.~\cite{Zhao2017} may be useful in neutral-atom computing. Later on, Ref.~\cite{Zhao2018} extended the scheme of Ref.~\cite{Zhao2017} to nonadiabatic holonomic quantum gates with Rydberg superatoms. Although possessing a similar gate matrix, the gate in Ref.~\cite{Zhao2018} can become a C$_{\text{Z}}$ gate by choosing appropriate parameters, i.e., can have the maximal entangling power, but the gate in Ref.~\cite{Zhao2017} can not. However, the method in Ref.~\cite{Zhao2018} requires coupling three types of different Rydberg states with the cavity photons, while the method in Ref.~\cite{Zhao2017} only needs to excite one type of Rydberg state in free space. These gates need neither pulse shaping nor adiabatic following, so they be able to achieve a similar fidelity and similar robustness as the original blockade gate analyzed in Ref.~\cite{Zhang2012}. 

\subsection{Multiqubit entanglement and quantum gates with asymmetric Rydberg interactions}\label{sec03Bnew2}
There is an efficient protocol for creating Greenberger-Horne-Zeilinger~(GHZ) state via asymmetric Rydberg interactions as proposed in Ref.~\cite{PhysRevLett.102.240502}. Consider two types of Rydberg states $|s\rangle$ and $|p\rangle$, if the interactions $V_{\text{ss}},~V_{\text{sp}},~V_{\text{pp}}$ for the states $|ss\rangle,~|sp(ps)\rangle,~|pp\rangle$ satisfy the condition $V_{\text{pp}}\ll V_{\text{ss}},~V_{\text{sp}}$, then a GHZ state in an ensemble of $N$ atoms can be created starting from the ground state $|00\cdots0\rangle$ in three steps. (i) First, excite the atoms with lasers for the transition $|0\rangle\xrightarrow[]{\Omega}-i|s\rangle$ with a pulse duration $\frac{\pi}{\sqrt{N}\Omega}$ and condition $\hbar\Omega\ll V_{\text{ss}}$, so that the state becomes
\begin{eqnarray}
  \frac{1}{\sqrt{2}}\left(|00\cdots0\rangle -  \frac{i}{\sqrt{N}}\sum_{j=1}^N|0s_j\cdots 0\rangle \right).\label{sec03Bnew2equation01}
\end{eqnarray}
(ii) Second, excite the atoms for the transition $|0\rangle\xrightarrow[]{\Omega}-i|p\rangle\xrightarrow[]{\Omega}-|1\rangle$ where a duration $\frac{\sqrt{2}\pi}{\Omega}$ is enough for the transition to accomplish~\cite{Shi2018prapp}. When $V_{\text{pp}}\ll\hbar\Omega\ll V_{\text{sp}}$, $\sum_{j=1}^N|0s_j\cdots 0\rangle$ will not be influenced by the laser fields, but the other component $|00\cdots0\rangle$ in Eq.~(\ref{sec03Bnew2equation01}) will go to the state $(-1)^N|11\cdots1\rangle$ since each atom in the ensemble will experience the transition $|0\rangle\xrightarrow[]{\Omega}-i|p\rangle\xrightarrow[]{\Omega}-|1\rangle$, and there is almost no blockade because of the condition $V_{\text{pp}}\ll\hbar\Omega$. (iii) Third, use the excitation scheme in the first step, so that the ensemble state becomes
\begin{eqnarray}
  \frac{1}{\sqrt{2}}\left(-|00\cdots0\rangle + (-1)^N |11\cdots 1\rangle \right).\label{sec03Bnew2equation02}
\end{eqnarray}
The above approach to generating the GHZ state only requires a short time $\frac{\pi}{\Omega}(\frac{2}{\sqrt{N}} +\sqrt{2})$ and thence can greatly suppress the Rydberg-state decay. Using practical parameters, Ref.~\cite{PhysRevLett.102.240502} showed that an 8-atom GHZ state can be rapidly generated with a fidelity $84\%$. 

Another way to use asymmetric interaction was shown in Ref.~\cite{Cano2014}. Using similar symbols as in the last paragraph~(but here $s$ and $p$ do not denote states of special parities), let us consider the case $V_{\text{pp}}, V_{\text{ss}}\gg V_{\text{sp}}$ and the excitation chain $|0\rangle\xrightarrow[]{\Omega_1(t)}-i|e\rangle\xrightarrow[]{\Omega_2(t)}-(|s\rangle+|p\rangle)/\sqrt{2}$ via a low-lying state $|e\rangle$, where simultaneous excitation of the two Rydberg states $|s(p)\rangle$ are achieved by two different laser fields. When resonant fields are used, the large energy shifts in $|ss\rangle,~|pp\rangle$ make them not accessible, but the Bell state $(|sp\rangle+|ps\rangle)/\sqrt{2}$ can be created when the time dependence of $\{\Omega_1(t),~\Omega_2(t)\}$ is applied in the counterintuitive manner~\cite{Bergmann1998,Kral2007,Vitanov2017}. Ref.~\cite{Cano2014} also studied multiparticle entanglement by using asymmetric Rydberg interactions. 

Apart from the above methods, asymmetric Rydberg interaction can be used in a way similar to the original three-pulse Rydberg gate in Ref.~\cite{PhysRevLett.85.2208}, as studied in Ref.~\cite{Wu2010} and Ref.~\cite{Su2018cpb} with two types of asymmetries for designing a multiqubit gate.

Asymmetric interactions are of special importance for the design of the Toffoli gate, where the interaction between the two edge atoms in a three-atom chain is a severe issue. Ref.~\cite{Brion2007pra} detailed a way to excite the three qubits to three different Rydberg states which are chosen in a way that the interaction between the two edge atoms is negligible. Using the same method, it is possible to implement a C$_k$-NOT gate by simultaneously exciting the transition $|0\rangle\rightarrow|r\rangle$ for $k$ control atoms, so that only the state $|11\cdots1\rangle$ can allow a Rydberg excitation~(of another Rydberg state) in the target qubit, i.e., a controlled gate by using $k$ atoms as a control is realizable~\cite{Isenhower2011}. Alternatively, one can also use microwave fields to superpose different Rydberg states to form a new Rydberg state, where the Rydberg interaction is annulled~\cite{Shi2017pra,Young2020Jeremy} which can be used for designing the Toffoli gate~\cite{Shi2018prapp} and large-scale GHZ states~\cite{Young2020Jeremy}. Another interesting asymmetry in Rydberg interactions was studied in Ref.~\cite{Shi2014}, where a spin-exchange interaction was found when two atoms are in Rydberg states of different principal and angular states. The anisotropic interaction~(viewed in the basis after diagonalization) in Ref.~\cite{Shi2014} is in the form of a spin-exchange interaction~(in the basis of atomic eigenstates) that can be used to create Bell states and SWAP gates. 

\subsection{Multiqubit entanglement with dissipation}\label{sec03dissipation}
There are two types of dissipation in an atomic system involving Rydberg interactions. The first is that the presence of multiple Rydberg excitations causes unequal rates of state evolution in the many-body system~\cite{Bariani2012pra01,Bariani2012,Bariani2012pra}, which is in analogy to the inhomogeneous broadening in the problem of the many-body nuclear spin bath of solid-state system~\cite{Liu2010}. By this type of dissipation, there is an efficient way to generate entanglement in spin waves of atomic gas as studied in~\cite{Bariani2012}. However, this type of dissipation was rarely studied because multiple Rydberg excitations were usually avoided in Rydberg atom quantum science. 

The other dissipation frequently studied is about the dissipation of low-lying excited states or high-lying Rydberg states. Early investigation on this dissipation-induced entanglement with Rydberg atoms was made in Refs.~\cite{BhaktavatsalaRao2013,Carr2013}, where Ref.~\cite{Carr2013} will be reviewed in Sec.~\ref{antibloc-C01} for it belongs to the Rydberg antiblockade regime. The method in Ref.~\cite{BhaktavatsalaRao2013} depends on Rydberg blockade and  electromagnetically induced transparency, as in Ref.~\cite{Muller2009} which will be reviewed in Sec.~\ref{sec03C01EIT}. The difference lies in that Ref.~\cite{Muller2009} depends on adiabatic following, while Ref.~\cite{BhaktavatsalaRao2013} depends on dissipation. 

The method in Ref.~\cite{BhaktavatsalaRao2013} is as follows. For each of two atomic qubits, a two-photon transition $|1\rangle\xleftrightarrow[]{\Omega_{\text{1p}}}|p\rangle\xleftrightarrow[]{\Omega_{\text{pr}}}|r\rangle$ is used. Meanwhile, a Raman transition $|0\rangle\xleftrightarrow[]{\Omega_{\text{01}}}|1\rangle$ is used. The two-photon Rydberg transition is dark to the state $|D\rangle\propto(\Omega_{\text{pr}} |1\rangle-\Omega_{\text{1p}} |r\rangle )$, i.e, $|D\rangle$ does not respond to the Rydberg excitation. Because the state $|0\rangle$ is not excited, the two-atom states $|D_\pm\rangle=(|D0\rangle\pm|0D\rangle)/\sqrt{2}$, $|DD\rangle$, and $|00\rangle$ are all dark if no Rydberg interaction is present. However, $|DD\rangle$ involves a finite Rydberg population and thus is no longer dark, and $|D_+\rangle$ and $|00\rangle$ are excited to $|DD\rangle$ in the Raman transition. As a result, only the state $|D_-\rangle$ is dark, and $|D_+\rangle$, $|00\rangle$, and $|DD\rangle$ actually decay through the rapidly decaying state $|p\rangle$. For this reason, only $|D_-\rangle$ survives in the dissipation. This method depends on that the decay rate of $|r\rangle$ shall be much smaller than that of $|p\rangle$.

Later on, the authors of Ref.~\cite{BhaktavatsalaRao2013} extended their scheme to prepare entanglement in $N$ Rydberg atoms~\cite{Rao2014}. Consider the transition chain $|1\rangle\xleftrightarrow[]{\Omega_{{1r_1}}}|r_1\rangle\xleftrightarrow[]{\Omega_{{r_1r_2}}}|r_2\rangle$ involving two Rydberg states $|r_{1(2)}\rangle$, one can show that when the interactions of the states $|r_ir_j\rangle$ for any two atoms with $i,j\in\{1,2\}$ are large compared to $\Omega_{{1r_1}}$ and $\Omega_{{r_1r_2}}$, the initial state $|11\cdots1\rangle$ will only be excited to the state $\sum_{k=1}^N |1r_1^{(k)}\cdots1\rangle/\sqrt{N}$, which is further excited to $|R\rangle\equiv\sum_{k=1}^N |1r_2^{(k)}\cdots1\rangle/\sqrt{N}$, where the superscript $(k)$ labels the atom in the ensemble. Thus, a three-level system is formed so that there is a many-body dark state $|D\rangle\propto \Omega_{{r_1r_2}}|11\cdots1\rangle-\sqrt{N}\Omega_{{1r_1}}|R\rangle $. The bright states have population in $|r_1\rangle$, which can be coupled to a fast-decaying low-lying state so as to obtain a short lifetime. Then, starting from any initial state, the dissipation through $|r_1\rangle$ will leave the population of the many-body state in the dark state $|D\rangle$. Like that in Ref.~\cite{BhaktavatsalaRao2013}, this method also depends on long lifetimes of a Rydberg state~(which is $|r_2\rangle$ in~\cite{Rao2014}).

By using dissipation and Rydberg blockade, Ref.~\cite{Zhu2020} showed that it is possible to prepare entanglement between two atoms of the form $\sum_{j=1}^3|a_jb_j\rangle/\sqrt{3}$, where $|a(b)_j\rangle$ denotes a ground or Rydberg state of the first~(second) atom. For the case of a highly dissipative cavity, entanglement can also be generated between ground and Rydberg-state atoms via dissipation~\cite{Shao2017cavitydisspation}.

The dissipation-based schemes are in principle robust against the fluctuation of the interactions, but it is an open question whether in the presence of atomic motion the Doppler dephasing of the ground-Rydberg transition will degrade their fidelities since long times are required to establish the desired target states. %

\subsection{Rydberg gates and entanglement with pulse shaping}\label{sec03C}
Pulse shaping has been frequently used in theoretical study on Rydberg gates. There are many Rydberg gate protocols based on adiabatic rapid passage (ARP) and stimulated Raman adiabatic passage (STIRAP) shown in, for example, Refs.~\cite{Beterov2020,Saffman2020} and references therein. Also, there are many gate protocols based on the optimal control theory~\cite{Goerz2011,Muller2011,Goerz2014,Levine2019}. ARP usually denotes population transfer between two states by time-dependent control fields, while STIRAP refers to population transfer between two states via a fast-decaying intermediate state~\cite{Bergmann1998,Kral2007,Vitanov2017}. To achieve better gate performance which involves more parameters than the usual study on ARP or STIRAP, optimal control by numerical optimization can be used to find specific time dependence in the external control fields. As experimentally demonstrated in Refs.~\cite{Omran2019,Levine2019}, optimal control is a useful tool in the study of Rydberg atom quantum science.

\subsubsection{Two-qubit phase gates derived from Ref.~\cite{PhysRevLett.85.2208} }\label{sec03C0m2}
In the very first proposal on quantum computing with neutral Rydberg atoms, Ref.~\cite{PhysRevLett.85.2208} not only presented quantum entangling gates with rectangular pulses, but also with adiabatic pulses. By imposing time dependence in the field amplitude and detuning, Ref.~\cite{PhysRevLett.85.2208} showed that with slowly varying fields the atomic state will adiabatically follow the eigenstate of the system Hamiltonian, from the input state to some other states and then to itself finally, and the dynamical phase $\langle\alpha\beta |e^{-i\mathcal{T}\int \hat{H}(t)dt/\hbar}|\alpha\beta \rangle$ for the input state $|\alpha\beta\rangle$ is tunable by choosing appropriate fields, where $\alpha,\beta=0$ or $1$. When only the state $|1\rangle$ is Rydberg excited for both qubits with the same duration, the dynamical phases $\{\varphi_1, \varphi_1,\varphi_2\}$ for $\{|10\rangle,|01\rangle,|11\rangle\}$ can satisfy the condition $\varphi_2-2\varphi_1=\pi$, which corresponds to an entangling gate that can be transformed to a {\footnotesize CNOT} gate with several single-qubit rotations. The original adiabatic version of the blockade gate in Ref.~\cite{PhysRevLett.85.2208} does not require single-qubit addressing. Being free from single-site addressing can simplify experimental implementation which later on motivated a more detailed analysis of the method in Ref.~\cite{Muller2014}.%

The original adiabatic version of the blockade gate in Ref.~\cite{PhysRevLett.85.2208} is not robust to noise in the laser fields and fluctuation of interactions. This is because the dynamical phase is a function of Rydberg interaction and the amplitude and detuning of the Rydberg lasers. The fluctuation of Rydberg interaction can be a major source of error since it is very sensitive a function of the spacing between the qubits. To tackle this issue, Refs.~\cite{Keating2013,Keating2015,Mitra2020} proposed to use dynamical phases in the blockade regime to implement entangling gates, where quantum annealing was studied in~\cite{Keating2013}. In the blockade regime, the two-qubit Rydberg state is not coupled, and the dynamical phase is dependent on detuning and amplitude of the external control fields. Thus, it is possible to implement adiabatic entangling gates by ramping the frequency and amplitude of Rydberg lasers in the multi-atom system. The protocols in~\cite{Keating2013,Keating2015,Mitra2020} are robust against the fluctuation of the Rabi frequencies which results from their strategy to adiabatically evolve a state which is a superposition of ground and Rydberg states. These methods have a fidelity limited by the Rydberg blockade error and the Doppler dephasing; compared to other methods, their fidelity is more limited by the Rydberg-state decay because the speed of the gate shall be slow enough to establish adiabaticity.

In Ref.~\cite{Shen:19} a Rydberg gate based on shortcut to adiabaticity was proposed which leads to a fast gate operation since adiabaticity is no longer required. In this case, a nontrivial phase can arise with both the geometrical and dynamical characters. This method can also create multiqubit gates. %

\subsubsection{Two-qubit {\footnotesize CNOT} gates with electromagnetically induced transparency}\label{sec03C01EIT}
A {\footnotesize CNOT} gate was proposed in Ref.~\cite{Muller2009} based on electromagnetically induced transparency~(EIT). The {\footnotesize CNOT} gate in Ref.~\cite{Muller2009} only needs three laser fields for the target qubit with the following sequence. (i) Apply a $\pi$ pulse to the control qubit for the transition $|1\rangle\rightarrow|r\rangle$. (ii) Apply an adiabatic pulse to induce a state transfer $|0\rangle\leftrightarrow|1\rangle$ in the target qubit conditioned on that there is a blockade effect from Rydberg interactions. (iii) Apply a $\pi$ pulse to the control qubit for the transition $|r\rangle\rightarrow|1\rangle$. The adiabatic pulse in step (ii) is as follows: the states $|0\rangle$ and $|1\rangle$ are coupled to a low-lying state $|p\rangle$ with a large detuning, and the state $|p\rangle$ is coupled with $|r\rangle$ with a matched detuning so that the transition $|r\rangle\leftrightarrow|0(1)\rangle$ is resonant. The field for the transition $|r\rangle\leftrightarrow|p\rangle$ is constant, but that for $|p\rangle\leftrightarrow|0(1)\rangle$ is time-dependent. The time-dependence is designed so that when the two-qubit input state is $|0\alpha\rangle$, where $\alpha=0$ or $1$, the state $|0\alpha\rangle$ follows adiabatically from $|0\alpha\rangle$ to a time-dependent combination of two dark states, and back to itself at the end of the pulse~(as in EIT); when the two-qubit input state is $|1\alpha\rangle$, the Rydberg interaction shifts away the two-photon resonance $|r\rangle\leftrightarrow|0(1)\rangle$, and as a consequence it allows the transition $|0\rangle\leftrightarrow|1\rangle$ to occur via the intermediate state $|p\rangle$ in the target qubit.

It is useful to compare the above method with another {\footnotesize CNOT} method with three pulses proposed in Ref.~\cite{Shi2018prapp}. The method in Ref.~\cite{Shi2018prapp} needs to couple the two qubit states $|0\rangle$ and $|1\rangle$ with the same Rydberg state $|r\rangle$ for the target qubit. For an $s$ or $d-$orbital Rydberg state $|r\rangle$, this requires four laser fields for Rydberg excitation in the target qubit, two for $|0\rangle\leftrightarrow|r\rangle$, and two for $|1\rangle\leftrightarrow|r\rangle$. In comparison, the {\footnotesize CNOT} gate in Ref.~\cite{Muller2009} only needs three laser fields for the target qubit.

The scheme in Ref.~\cite{Muller2009} is specifically useful if the target qubit is replaced by an ensemble of qubits so as to design a  {\footnotesize C-NOT}$^N$ gate, i.e., a {\footnotesize CNOT} between a mesoscopic ensemble of $N$ atoms and a single control atom. As long as the Rabi frequency for $|r\rangle\leftrightarrow|p\rangle$ is much larger than the peak value of the Rabi frequency of $|p\rangle\leftrightarrow|0(1)\rangle$, the eigenstate of the two-qubit system will adiabatically follow after a superposition of the dark states of the system Hamiltonian. There will be an error due to trying to populate the state where all the atoms in the target ensemble are in the Rydberg state since there will be strong interactions when all the atoms in the target ensemble are in Rydberg states. However, as long as the Rabi frequency for $|r\rangle\leftrightarrow|p\rangle$ is large enough, the error can be largely suppressed. 

In Ref.~\cite{Muller2009}, the adiabatic scheme is employed so as to induce the controlled swap process, and it is subjected to the the usual blockade error since it depends on shifting away the resonance condition by Rydberg interaction. When the blockade error is suppressed, the gate fidelity by this scheme can be large when fine tuning of the laser fields is achieved.

\subsubsection{Two-qubit phase gates with analytical derivative removal by adiabatic gate}\label{sec03CDRAG}
The intrinsic errors in Rydberg gates based on the blockade mechanism is limited by the Rydberg-state decay and the blockade error, where the former can be made smaller by using faster Rydberg excitations. But to shrink the blockade error, larger Rydberg interactions are desirable, which can arise for high-lying Rydberg states. The energy spacing between two nearby Rydberg states becomes smaller when the principal quantum number is larger. For fast gate operations in order to suppress the Rydberg-state decay, the Rydberg Rabi frequency $\Omega$ should be large. This can lead to large population leakage to states that are near the optically-targeted Rydberg state. To tackle this issue, Ref.~\cite{Theis2016} showed that by using the method of analytical derivative removal by adiabatic gate~(DRAG)~\cite{Motzoi2009,PhysRevA.88.062318}, the population leakage can be well suppressed when shaped but analytic Rydberg pulses are used for Rydberg excitation. The analysis in~\cite{Theis2016} showed that gate fidelities over $99.99\%$ are realizable at room temperature; if qubits are hosted in cryogenic chambers at $4.2$~K, their gate can have fidelities over $99.999\%$ thanks to the exceedingly short gate durations of several tens of nanoseconds. By comparing the gate fidelity of Ref.~\cite{Theis2016} with those from other methods in Table~\ref{table3}, one can see that DRAG can lead to the largest fidelities for Rydberg gates via the blockade mechanism. However, the analyses in~\cite{Theis2016} were based on a GHz-scale Rydberg interaction in the form of a pure energy shift which is challenging to be found as discussed in Sec.~\ref{sec02C02}, and it is an open question whether the theory in~\cite{Theis2016} can be extended to protocols where strong enough dipole-dipole interaction~(which is not in the form of a pure energy shift) is considered.

\subsubsection{Two-qubit phase gates by geometrical phases with adiabatic excitation}\label{sec03C0m1}
A Berry phase or geometric phase is accumulated when a state dependent on a varying parameter $\phi_r$ evolves slowly around a closed loop of $\phi_r$ . This type of geometric phase only depends on the path along which $\phi_r$ evolves, and is robust against noise of various forms.

It was proposed in Ref.~\cite{Moller2008} that an arbitrary two-qubit controlled-phase gate can be realized by using a Berry phase in two atoms, each with four atomic states, i.e, three ground states $|0\rangle,~|1\rangle,~|2\rangle$ and a Rydberg state $|r\rangle$, where $|0\rangle$ and $|1\rangle$ are qubit states. For one qubit upon which the transitions $|1\rangle\leftrightarrow|2\rangle$ and $|2\rangle\leftrightarrow|r\rangle$ are driven by external fields with Rabi frequencies $\Omega_1$ and $e^{-i\phi_r}\Omega_r$, respectively, the dark state of the three-level system is $|D_1\rangle=\cos\theta|1\rangle-e^{i\phi_r}\theta|r\rangle$, where $\tan\theta\equiv\Omega_1/\Omega_r$~\cite{Shi2020jpb}. With one adiabatic cycle where the wavefunction evolves starting from $|1\rangle$ and ending at $|1\rangle$, the state acquires a Berry phase $\varphi_1=-\int\sin^2\theta d\phi_r$. What is interesting is that when the {\it same} pulse is applied to two nearby atoms, although the single-atom dark state $|D_1\rangle$ no longer exists, Ref.~\cite{Moller2008} pointed out that for large enough blockade interaction there is another dark state $|D_2\rangle$, and there is a time-dependent component in $|D_2\rangle$ with a phase term
\begin{eqnarray}
\mathscr{B}\equiv \langle D_2|(|1r\rangle+| r1\rangle)=\frac{-\sin\theta\cos\theta e^{-i\phi_r}}{\sqrt{\cos^4\theta+2\sin^4\theta}}.
\end{eqnarray}
Then, by appropriately choosing a closed loop for the varying phase $\phi_r$, one can simultaneously map $\{|01\rangle,~|10\rangle\}$ back to themselves with a phase accumulation $\varphi_1$, and map $|11\rangle$ to itself with another phase accumulation $\varphi_2=-\int|\mathscr{B}|^2 d\phi_r$. Here, one shall bare in mind that $\varphi_1$ arises when only one atom responds to the excitation field, while $\varphi_2$ arises when both qubits are optically excited, i.e., if the input state is $|11\rangle$. When single-qubit rotations are applied, a controlled-phase gate diag$\{1,1,1,e^{\varphi_2-2\varphi_1}\}$ is realized with $\varphi_2-2\varphi_1$ tunable in the interval $(0,~2\pi]$. By this method, Fig.~2(b) of Ref.~\cite{Moller2008} showed that Bell states with fidelities around $99.6\%$ can be prepared.

The above Berry-phase gate can suppress amplitude noise of the external control fields, but is still subjected to the blockade error because the two-qubit dark state $|D_2\rangle$ is only valid in the limit of large blockade shift in $|rr\rangle$. The novelty of the scheme in Ref.~\cite{Moller2008} lies in that when it is applied to a multiqubit system, it can prepare large-scale GHZ states. For more details, see Ref.~\cite{Moller2008}; besides, Ref.~\cite{Wu2017} did a more detailed analysis about the scheme in Ref.~\cite{Moller2008} and studied some of its extensions. 

 Berry phases are particularly useful for designing quantum gates in ensemble qubits~\cite{Beterov2013} which will be reviewed in Sec.~\ref{sec03Censemble}.

\subsubsection{Two and three-qubit gates by geometrical phases with nonadiabatic excitation}\label{nonadiapulse}
In Sec.~\ref{nonadiabaticgemo} we have reviewed Rydberg gates by geometric phases without using pulse shaping. Although the gate in~\cite{Zhao2017,Zhao2018} is implemented with constant Rabi frequencies, cyclic state evolution occurs where geometric phases are accumulated. In Ref.~\cite{Kang2018}, another nonadiabatic holonomic Rydberg quantum gate was studied based on defining qubit states by superposition states of two hyperfine ground states. The gate in Ref.~\cite{Kang2018} is a C$_{\text{Z}}$ gate in their qubit basis, and pulse shapes were found by shortcuts to adiabaticity. However, it is an open question whether one can efficiently manipulate~(load and read) quantum information by qubit states that are not eigenstates of the external fields which should be applied for specifying the quantization axis. Later, the authors of Ref.~\cite{Kang2018} considered using auxiliary atoms to realize heralded nonadiabatic holonomic Rydberg gates~\cite{Kang2020}.%

Quantum gates can be efficiently created between a single atom and an ensemble of atoms, shown in Ref.~\cite{Guo2020}, which proposed a three-pulse nonadiabatic non-Abelian geometric Rydberg controlled-$U(\theta,\phi)$ gate, where $U(\theta,\phi)=\sigma_z\cos\theta+(\sigma_x\cos\phi+\sigma_y\sin\phi)\cos\theta$ and the angles $(\theta,\phi)$ are determined by laser fields and dipole transition matrix elements. The scheme in Ref.~\cite{Guo2020} is particularly robust against the fluctuation of the field intensity in the lasers. The gate in Ref.~\cite{Guo2020} can be easily tuned to various forms including C$_{\text{Z}}$ and {\footnotesize CNOT} by adjustment of parameters of laser fields. In particular, Ref.~\cite{Guo2020} showed that high-fidelity~(see Table~\ref{table3}) three-qubit gates are realizable with pulses found by invariant-based inverse engineering and optimal control. Later on, Ref.~\cite{MengLi2021} extended this scheme and studied an interesting multiple-qubit C$_k$U gate based on inverse engineering, shortcut-to-adiabaticity, and optimal control. In particular, Ref.~\cite{MengLi2021} showed that the operation U in the C$_k$U gate can be an arbitrary single-qubit operation in the target qubit, and fidelities about 99.8\%~(at 0~K; see Sec.~III D therein) are achievable for k up to 4 when U is either the NOT or the Hadamard operation.

\subsubsection{Multiqubit entanglement and phase gates by optimal control}\label{sec03C03}
Numerical optimal control can be used to find time-dependent external fields for implementing high-fidelity Rydberg gates. By optimal control, one uses numerical search to locate a time-dependence pattern for the amplitude and frequency of external fields so as to find an optimal fidelity. For the simplest case, one can use the optimal control by minimizing $-F(a,~b,~c,~\cdots)$, where $F$ is the gate fidelity which depends on system parameters such as the decay rates of atomic states, energy gaps between nearby atomic states, detuning of Rydberg lasers, and so on~(denoted by $a,~b,~c,~\cdots$). In practice, one shall bare in mind that an efficient gate should not require a long pulse to work, and one can add extra terms such as the total pulse area required in the functional as a constraint. For a detailed implementation of Rydberg gates by numerical optimization, see, e.g., Refs.~\cite{Goerz2011,Saffman2020}. 

It was shown in Ref.~\cite{Muller2011} that optimal control can be used to design fast two-qubit entangling gates with neutral atoms trapped on atom chips, where a two-qubit fidelity $99.7\%$ was shown to be possible with a gate duration as short as $30$~ns~(see Table 2 of Ref.~\cite{Muller2011}). Later on, Ref.~\cite{Goerz2014} showed that optimal control can lead to high-fidelity two-qubit entangling gates between two optically trapped neutral atoms. Importantly, the nonanalytical pulses in Ref.~\cite{Goerz2014} can lead to accurate gates that are robust to noise in the laser pulses, and it found that gate fidelities $99.99\%$ are reachable. In both Ref.~\cite{Goerz2014} and Ref.~\cite{Muller2011}, the more practical two-photon excitation of Rydberg states were used for the excitation of even-parity Rydberg states without assuming effective two-photon Rabi frequencies derived by the adiabatic elimination of the intermediate state. As can be seen from Table~\ref{table3}, the optimal control in Ref.~\cite{Goerz2014} can work equally well as the method in Ref.~\cite{Theis2016} since both can  lead to the largest gate fidelity compared to other schemes shown in Table~\ref{table3}.

Optimal control has been experimentally proven to be a powerful tool in Rydberg atom quantum science. For example, Ref.~\cite{Omran2019} reported the creation of “Schr\"{o}dinger cat” states of the GHZ type with qubit numbers up to 20 using pulses found by optimal control. In Ref.~\cite{Omran2019}, the initial multi-atom state adiabatically evolves to a superposition state of two ground states of a many-body Hamiltonian, leading to a 20-qubit GHZ state with a fidelity $54.2\%$. Because the energy gap between the ground state and excited states shrinks quickly when the number of atoms increases with the method of Ref.~\cite{Omran2019}, longer times are required to guarantee adiabaticity for a multi-atom system with more atoms, and thus optimal control was used to minimize the pulse duration and maximize the fidelity for the GHZ state. The results in Ref.~\cite{Omran2019} rigorously showed that neutral atoms are competitive to superconducting systems~\cite{Song2019,Wei2020} regarding the capability to create large-scale many-particle entanglement. Besides, the same groups experimentally demonstrated~\cite{Levine2019}, for the first time, a three-qubit neutral-atom Toffoli gate with Rydberg laser pulses found by numerical optimal control softwares. Although the intrinsic fidelity of the Toffoli gate by optimal control was about $97\%$, and the experimental truth-table fidelity was $87\%$ in Ref.~\cite{Levine2019}, their experiment gave a proof-of-principle demonstration of optimal control in quantum information processing with Rydberg atoms. Optimal control can also be used for manipulating states of one atom when many energy levels are involved. For example, Ref.~\cite{Larrouy2020} experimentally demonstrated that by using pulse sequences found from optimal control theory, one can accurately generate a nonclassical superposition state that cannot be prepared with reasonable fidelities using standard techniques.

Although optimal control was frequently used in the study of Rydberg atom quantum technology, an optimal-control based, fast, high-fidelity, and single-step C$_{\text{Z}}$ gate was only recently proposed in Ref.~\cite{Saffman2020}. It showed that with experimentally feasible parameters, a C$_{\text{Z}}$ gate can be rapidly realized by sending global external field to the two qubits. The optimized pulse was found by dividing the external control field to $M$ equal-length segments where $M$ is a moderate integer so as to search an optimal pattern for the amplitude and frequency of the laser fields in each segment. With the spontaneous emission and blackbody radiation from the intermediate $p$-orbital and Rydberg states included, Ref.~\cite{Saffman2020} showed that a fidelity over $0.997$ is achievable in the frequently adopted two-photon excitation of $s$ or $d$-orbital Rydberg states. Compared to previous protocols based on quantum interference proposed in Ref.~\cite{Shi2018Accuv1} and experimentally realized in Ref.~\cite{Levine2019}, the C$_{\text{Z}}$ gate protocol in Ref.~\cite{Saffman2020} can substantially reduce the motion-induced Doppler dephasing because there is no gap time during which a Rydberg atom drifts freely in free space. Thus, if fine tuning of the control lasers can be achieved, optimal control can be of particular help in the near-term Rydberg atom quantum science where noise from the lasers and atomic motion dominate the gate infidelity based on current technology.

\subsubsection{Two-qubit gates and entanglement with two atomic ensembles}\label{sec03Censemble}
There are two main challenges for designing quantum gates by Rydberg blockade in atomic ensembles. First, it is difficult to deterministically excite Rydberg states when the atom number is random, and second, the phase change of the atomic states in response to Rydberg excitation depends on the number of atoms in the ensemble which makes it difficult to deterministically generate quantum entanglement.

{\it Excitation of Rydberg states.--}When we try to load one atom in each optical dipole trap in a large array, there is a possibility to fail, as well as some chance to load more than one atoms in one trap~\cite{Xia2015,Wang2015,Wang2016,Bernien2017}. If one would like to use this type of atomic arrays for storing quantum information where each site has at least one atom, Sec.~\ref{qubitencoding02} shows that we can treat both single-atom states $|0(1)\rangle$ as qubit states, and ensemble states
\begin{eqnarray}
  |\overline{0}\rangle &=& |000\cdots0\cdots000\rangle,\nonumber\\
  |\overline{1}\rangle &=&\frac{1}{\sqrt{N}}\sum_{j=1}^N |000\cdots1_j\cdots000\rangle,\label{eqsec03C04eq01}
\end{eqnarray}
as qubit states as well, where $N$ is the number of atoms in the ensemble. The integer $N$ in Eq.~(\ref{eqsec03C04eq01}) can fluctuate from site to site. Consider the case when all the atoms in one site are near enough, i.e., the trap is deep enough, the Rydberg interaction between any pair of atoms in the trap can be larger than a certain value $V_{\text{min}}$. If a rectangular Rydberg laser pulse is applied for $|1\rangle\leftrightarrow|r\rangle$ with Rabi frequency $\Omega$ for one atom, the actual Rabi frequency for the transition from $|\overline{1}\rangle $ to the ensemble state
\begin{eqnarray}
  |\overline{r}\rangle &=&\frac{1}{\sqrt{N}}\sum_{j=1}^N |000\cdots r_j\cdots000\rangle,\label{eqsec03C04eq02}
\end{eqnarray}
is enhanced by $\sqrt{N}$, where $j=1,2,~\cdots,N$. For a small $N$, the Rydberg blockade effect can hold with a practical value of $N$ in the experiment, i.e., we can still have the condition $\sqrt{N}\hbar\Omega\ll  V_{\text{min}}$ since $\hbar\Omega\ll V_{\text{min}}$. The issue is, how can one excite the state of the atomic ensemble from $|\overline{1}\rangle $ to $|\overline{r}\rangle $ by a given type (and duration) of pulse when $N$ is random? The answer was given in Ref.~\cite{Beterov2011}, which shows that the excitation from $|\overline{1}\rangle $ to $|\overline{r}\rangle $ with $N$ up to 7 can be accurately accomplished with practical atomic parameters and experimental resources. To show the essence of the theory, we consider the Hamiltonian of a one-photon excitation of Rydberg states
\begin{eqnarray}
 \hat{H} &=& \left[\hbar \frac{\Omega(t)}{2} |\overline{r}\rangle\langle \overline{1}|+\text{H.c.}\right] + \hbar\frac{\delta(t) }{2}\left( |\overline{r}\rangle\langle \overline{r}|- |\overline{1}\rangle\langle \overline{1}|\right] ,\label{sec3D4eq01}
\end{eqnarray}
which has two eigenstates
\begin{eqnarray}
|v_+\rangle &=& \left[ \Omega(t)|\overline{r}\rangle +(\sqrt{\Omega^2(t) + \delta^2(t)}-\delta(t))|\overline{1}\rangle \right]/\mathcal{N}_+ ,\nonumber\\
|v_-\rangle &=& \left[ \Omega(t)|\overline{r}\rangle -(\sqrt{\Omega^2(t) + \delta^2(t)}+\delta(t))|\overline{1}\rangle \right]/\mathcal{N}_-,
\end{eqnarray}
with eigenvalues $\epsilon_\pm =\pm\sqrt{\Omega^2(t) + \delta^2(t)}/2 $ and normalization factors $\mathcal{N}_\pm=\sqrt{\Omega^2(t)+[\sqrt{\Omega^2(t) + \delta^2(t)}\mp\delta(t)]^2}$. If initially $\Omega(0)$ is small compared to $\delta(0)$, $|v_+\rangle$ is approximately $|\overline{r}\rangle$, and $|v_-\rangle$ is approximately $|\overline{1}\rangle$. If both $\Omega(t)$ and $\delta(t)$ change smoothly until at the end of the pulse when $\Omega(t)=\Omega(0)$ and $\delta(t)=-\delta(0)$, $|v_-\rangle$ is approximately $|\overline{r}\rangle$. This means that if we start from the state $|\overline{1}\rangle$, the atom will be excited to the state $|\overline{r}\rangle$. However, it does require smooth enough $\Omega(t)$ and $\delta(t)$ with the condition $|d\Omega(t)/dt|/\sqrt{\Omega^2(t) + \delta^2(t)}\ll1$ and $|d\delta(t)/dt|/\sqrt{\Omega^2(t) + \delta^2(t)}\ll1$. This adiabatic method does not require the pulse area to be equal to a certain value, and it is applicable to excite an atomic ensemble with unknown but small enough $N$.

The above method is useful for exciting a $p$-orbital Rydberg state. For $s$- or $d$-orbital Rydberg state, an intermediate state shall be used. Ref.~\cite{Beterov2011} showed that when the detuning at the intermediate state is zero, the standard three-level STIRAP can not excite the ground state $|\overline{1}\rangle$ to the Rydberg state $|\overline{r}\rangle$ because there is no way to define a dark eigenstate like $(\Omega_{p\rightarrow r}|\overline{1}\rangle-\Omega_{1\rightarrow p}|\overline{r}\rangle)/\sqrt{\Omega_{p\rightarrow r}^2+\Omega_{1\rightarrow p}^2}$, where $\Omega_{p\rightarrow r}$ and $\Omega_{1\rightarrow p}$ are the Rabi frequencies for the upper and lower transitions. However, Ref.~\cite{Beterov2011} found that by adding a large, constant detuning to the intermediate state, there is an eigenstate of the many-body Hamiltonian which adiabatically changes from $|\overline{1}\rangle$ to $|\overline{r}\rangle$ when the counterintuitive STIRAP pulse is applied; compared to the two-state case in Eq.~(\ref{sec3D4eq01}), the STIRAP excitation of an ensemble Rydberg state only needs to tune the Rabi frequency without tuning the detuning. More details about the high-fidelity excitation of $|\overline{r}\rangle$ with $s(d)$ and $p$-orbital Rydberg states can be found in Ref.~\cite{Beterov2014} and the review~\cite{Beterov2020}.

{\it Compensation of random phases.--} The adiabatic excitation of an atomic ensemble to a Rydberg state is much more involved compared to the case of single-atom excitation~\cite{Petrosyan2013}. Ref.~\cite{Beterov2013} found that for different atom numbers, the phases of the Rydberg state $|\overline{r}\rangle$ differ; consequently, when deexcited to the ground state, the phase of the state wavefunction depends on the number of atoms. The phase is accumulated because for $N>1$ in the case of STIRAP, the adiabatic following is no longer about a dark state. Instead, the adiabatic state can sometimes have a nonzero energy although most of the time it is dark during the pulse sequence, leading to a phase with both dynamical and geometrical characters. This issue can be solved by shifting the sign of the Rabi frequencies and detuning used for the deexcitation of the Rydberg state. Numerical simulation in~\cite{Beterov2013} showed that the random phase can be exactly compensated, and thus the standard gate protocols based on the Rydberg blockade are applicable when atomic ensembles are used as qubits. Figure 7 on page 9 of Ref.~\cite{Beterov2016jpb} showed that gate fidelities can be $99.9\%$ found by simulated quantum process tomography with $N$ up to 4. More details about the theories on Rydberg gates with ensembles can be found in Ref.~\cite{Beterov2020}.

Beside the above mentioned protocol of quantum entangling gates in two atomic ensembles, another method was shown in Ref.~\cite{Zhao2018} depending on nonadiabatic geometric phases, and Ref.~\cite{Ji2020} showed entanglement generation between two atom ensembles assisted by pulse shaping and by coupling cavity modes with the atoms. The method of Ref.~\cite{Zhao2018} depends on coupling atoms to cavity modes which is challenging, but it does not depend on pulse shaping since one $\pi$ pulse is sufficient. The gate in Ref.~\cite{Zhao2018} is a two-qubit entangling gate that has an entangling power less than the maximal~\cite{Williams2011} as in a C$_{\text{Z}}$ or {\footnotesize CNOT}. But any two-qubit entangling gate can form a universal set with several single-qubit gates, so the method in Ref.~\cite{Zhao2018} could be useful if the required coupling between the Rydberg atoms and cavity modes can be achieved. Moreover, Ref.~\cite{Guo2020} proposed an optimized geometric quantum computation with a mesoscopic ensemble of Rydberg atoms which was reviewed in Sec.~\ref{nonadiapulse}. However, different from the methods in~\cite{Beterov2011,Beterov2013,Beterov2014,Beterov2016jpb,Beterov2020}, the methods in Refs.~\cite{Zhao2018, Guo2020} assume that $N$ is a fixed number.

\begin{table*}[ht]
     \begin{threeparttable}
  \centering
  \begin{tabular}{|c|l|c|}
    \hline  Requirement  & Theoretical fidelity; type of operation; where it's taken   & References \\ \hline
\multirow{2}{*}{\begin{tabular}{c}\text{Interaction-induced}\\\text{phase shift~(Sec.~\ref{sec04A})}\end{tabular}}  & 99.3\%; C$_{\text{Z}}$ gate; Fig.~12(b) on page 15 of Ref.~\cite{Shi2017pra}~\tnote{a}  & Ref.~\cite{Shi2017pra}	 \\ \cline{2-3}
 & $99.4\%$; Barenco gate; middle left of page 6 of Ref.~\cite{Shi2018a}~\tnote{a} & Ref.~\cite{Shi2018a} 	 \\   \hline
\text{ \begin{tabular}{c}\text{Interaction-induced}\\\text{motional change}\end{tabular}} &\begin{tabular}{l}$99.98\%$; 2-qubit gate; Table~1 on page 383 of Ref.~\cite{Cozzini2006}\\ (Error due to Rydberg-state decay not included here)\end{tabular} & Ref.~\cite{Cozzini2006}\\  \hline
\multirow{6}{*} {\begin{tabular}{c}\text{Rabi-like rotations}\\ via F\"{o}rster resonance\\ (with pulse shaping \\
    except of Ref.~\cite{Beterov2018arX})\\ (Sec.~\ref{sec04C})\end{tabular} }& $99.4\%$; {\footnotesize CNOT} gate; Fig.~6(h) on page 6 of Ref.~\cite{Beterov2016} & Ref.~\cite{Beterov2016}\\  \cline{2-3}
& $98.6\%$; Bell state; left column on page 8 of Ref.~\cite{Beterov2016} & Ref.~\cite{Beterov2016}\\  \cline{2-3}
& $97.9\%$; Bell state; Fig.~7 on page 7 of Ref.~\cite{Beterov2018} & Ref.~\cite{Beterov2018}\\  \cline{2-3}
& $99.8\%$; C$_{\text{Z}}$ gate; upper right of page 4803 of Ref.~\cite{Liao2019} & Ref.~\cite{Liao2019}\\  \cline{2-3}
& $99\%$; C$_{\text{Z}}$ gate; Fig.~8 on page 8 of Ref.~\cite{Sun2020} & Ref.~\cite{Sun2020}\\  \cline{2-3}
& $98.3\%$; Toffoli gate; text below Eq.~(2) on page 6 of Ref.~\cite{Beterov2018arX} & Ref.~\cite{Beterov2018arX} \\\hline
\multirow{4}{*}{\begin{tabular}{c}\text{Dark states formed}\\ via F\"{o}rster resonance \\(with pulse shaping) \\ (Sec.~\ref{sec04D})\end{tabular} }  & {\color{blue}99.995}\%; C$_{\text{Z}}$ gate; text below Fig.~3 on page 4 of Ref.~\cite{Petrosyan2017} & Ref.~\cite{Petrosyan2017}\\  \cline{2-3}
 & $99.96\%$; 2-qubit gate;\tnote{b}~ abstract of Ref.~\cite{Yu2019} & Ref.~\cite{Yu2019}\\    \cline{2-3}
 &  ${\color{blue}99.9}\%$; Toffoli gate; Fig.~6(c) on page 9 of Ref.~\cite{Khazali2020}~\tnote{c}  & Ref.~\cite{Khazali2020}\\  \cline{2-3}
 & $98\%$; 21-qubit C$_{20}$-NOT gate; Fig.~6(c) on page 9 of Ref.~\cite{Khazali2020}~\tnote{c}  & Ref.~\cite{Khazali2020}\\ \hline
  \end{tabular}
  \caption{  \label{table4} Fidelities of quantum gates or entanglement via multiqubit Rydberg excitation. Most numbers of fidelities quoted were calculated with some decay rates around $2\pi\times1$~kHz for the Rydberg state in the corresponding references unless otherwise specified. The blue color highlights the largest fidelities for gates and entanglement with two or more than two qubits.} \begin{tablenotes}
    \item[a] The quoted data assume that the qubits are cooled near to the vibrational ground states of trapped atoms.
    \item[b] The lower part of page 23085 of Ref.~\cite{Yu2019} indicates that the decay rate of the Rydberg-state decay used in the estimate is $2\pi\times0.2$~kHz; the gate in Ref.~\cite{Yu2019} is not an entangling gate.
    \item[c] Fig.~6(c) on page 9 of Ref.~\cite{Khazali2020} showed fidelities with C$_k$-NOT gates~(which was termed Toffoli gate therein) with k from 1 to 20, where the Toffoli gate with $k=3$ has a fidelity about $99.9\%$. Ref.~\cite{Khazali2020} also studied the fan-out gates C-NOT$^k$ which can have high fidelities similar to those of the C$_k$-NOT gates.
      \end{tablenotes}
     \end{threeparttable}
  \end{table*}


\section{Rydberg gates by exciting multiqubits to Rydberg states }\label{sec04}
There are Rydberg gates based on populating multi-atom Rydberg states, which is in contrast to the protocols in Sec.~\ref{sec03} where multi-atom Rydberg states should not be populated. In this section, a full population of multi-atom Rydberg state is required in the methods in Sec.~\ref{sec04A}, while a partial population is involved in the methods of Secs.~\ref{sec04B},~\ref{sec04C}, and~\ref{sec04D}. Besides the gate methods reviewed below, there are Rydberg gates based on partially populating two-atom Rydberg states via the detuned Rabi oscillations~\cite{Shi2017,Shi2018Accuv1} that have been reviewed in Sec.~\ref{sec03detuned01} because of their close relevance to the other methods therein. Table~\ref{table4} lists the published theoretical fidelities with the places denoted about where to find those numbers; in many references more than one type of quantum gates or entangling operations were studied, and a relative larger fidelity is quoted in the table. 

\subsection{Two-qubit phase gates with dynamical phases from van der Waals interactions}\label{sec04A}
When two atoms are both in Rydberg states, the interaction between them can induce phase shifts, which can be used to entangle two atoms shown in Fig.~\ref{figure-originalgate}(b). The original proposal~\cite{PhysRevLett.85.2208} needs to excite both qubits to Rydberg states if they are initialized in $|11\rangle$, then wait for some time $\pi\hbar/V$, and finally deexcite them back to ground states. For a frozen $V$, this leads to a $\pi$ phase shift to the input state $|11\rangle$. Because only the qubit state $|1\rangle$ is excited, the states $|01\rangle$ and $|10\rangle$ also pick up a $\pi$ phase, leading to a controlled-Z gate. A variation of this method was experimentally demonstrated in Ref.~\cite{Jo2019}. Section~\ref{sec02C02} shows that when the two qubits are placed along the quantization axis and excited to the same Rydberg state, a van der Waals interaction appears as an energy shift from a second-order perturbation theory. The energy shift is well suited for this protocol, but it also means that the interaction is limited because the two-atom separation should be sufficiently large. The entanglement fidelity $59\%$ reported in the experiments of Ref.~\cite{Jo2019}, however, pointed to the possibility to employ interaction-induced dynamical phase for quantum information processing in future if the error due to the fluctuation of qubit spacing can be suppressed.

Ref.~\cite{Shi2017pra} showed that it is possible to realize a fast Rydberg gate by the method of~\cite{PhysRevLett.85.2208}. The basic idea is to place two qubits not too far away in the van der Waals regime, and excite both qubits to a certain superposition of Rydberg states where the interaction is annulled by state mixing via microwave fields. Then, a GHz-rate excitation by microwave fields is used to excite the Rydberg state to other states where a strong interaction arises, and a short wait duration induces a $\pi$ phase shift. However, only by sufficiently cooling and using deep enough trap can the error from the distance fluctuation of qubits be suppressed. With a trap depth $1.5~$mK~\cite{Kaufman2012} and qubits cooled to the motional ground states in the traps, Ref.~\cite{Shi2017pra} analyzed that a fidelity about $99\%$ could be achieved for a C$_{\text{Z}}$ gate. Though this fidelity is smaller than most fidelities shown in Table~\ref{table4}, the gate has a fast speed. It is an open question whether some robust protocols like those in~\cite{Beterov2016,Beterov2018} can be combined with the exotic annulled van der Waals interaction in~\cite{Shi2017pra} for fast and robust entanglement.

Ref.~\cite{Shi2018a} showed that by appropriately superposing different Rydberg states, one can design the Barenco gate by using the dynamical phase from the Rydberg interaction. Each Barenco gate is characterized by two specific angles and can form a universal set by itself, i.e., based on the Barenco gate alone, the universal quantum computation can be achieved~\cite{Barenco1995}. The fidelity of the Barenco gate is mainly limited by the Rydberg-state decay and fluctuation of qubit spacings, which depend on the gate durations and further on the two angles possessed in each Barence gate. By assuming cooling qubits to the motional ground states in the optical traps, the fidelity of a Barenco gate can approach $99.7\%$ shown in Fig.~6(a) of Ref.~\cite{Shi2018a} if qubits are hosted in chambers at 4.2~K. The gate fidelities shown in Table~\ref{table4} for the Barenco gate are data with Rydberg-state decay at room temperature. 

Compared to other methods, the Rydberg gates by phase shifts from van der Waals interactions are quite sensitive to the fluctuation of interatomic distances. Implementation of these protocols needs sufficiently cooled qubits if a high fidelity is desired.

\subsection{Two-qubit phase gates via change of motional states by Rydberg interactions  }\label{sec04B}
There is one gate proposal using vibrational states of atoms in dipole traps for Rydberg-mediated entanglement~\cite{Cozzini2006} which is based on a mechanism used for entangling trapped ions~\cite{Garcia-Ripoll2003}. Instead of using Rydberg interactions to induce a phase shift by a blocked Rabi transition or by a phase shift $\sim Vt/\hbar$, Ref.~\cite{Cozzini2006} analyzed a method to use the Rydberg interaction between two atoms trapped in two traps to modify their vibrational states in their traps. By choosing appropriate parameters and assuming ground-state vibration, it is possible to induce a phase shift to the state $|11\rangle$ by letting the atoms visit the two-atom Rydberg state for a little while, while the other three input states will not pick up any extra phase because they do not experience Rydberg interaction and thus their motional states are not altered. The method has a fidelity limited by the Rydberg state decay, blockade error, as well as an extra error from the thermal effect of the qubits unless ground-state cooling is achieved, and Table 1 of Ref.~\cite{Cozzini2006} showed that a fidelity $99.98\%$~(which does not include Rydberg-state decay) is achievable with available experimental techniques. However, it is unclear whether the gate can still obtain a large fidelity in the presence of the fluctuation of $V$ for qubits whose motional states are far from the motional ground states.  

\subsection{Two and three-qubit phase gates by F\"{o}rster-resonance-induced Rabi rotations}\label{sec04C}
For two atoms in a Rydberg state $|r_1r_2\rangle$, the electric dipole-dipole interaction can flip $|r_1r_2\rangle$ to the state $|r_3r_4\rangle$, where the parity of $|r_1\rangle$ shall be different from that of $|r_3\rangle$, and the parity of $|r_2\rangle$ shall be different from that of $|r_4\rangle$. However, there can be some energy difference between the two states $|r_1r_2\rangle$ and $|r_3r_4\rangle$, and $|r_1r_2\rangle$ can be flipped to a lot of states other than $|r_3r_4\rangle$ which is the intrinsic nature of the dipole-dipole interaction. An important special case is the F\"{o}rster resonance when two dipole-coupled states $|r_1r_2\rangle$ and $|r_3r_4\rangle$ have exactly the same energy, and other dipole-coupled states are largely detuned. Then, one can isolate the dipole interaction $|r_1r_2\rangle\leftrightarrow|r_3r_4\rangle$ from the manifold of dipole-coupled states. This can be experimentally achieved by dc Stark shifts~\cite{PhysRevLett.47.405,PhysRevLett.80.249,PhysRevLett.80.253,Westermann2006,PhysRevLett.108.113001,PhysRevA.93.042505,PhysRevA.93.012703} or by radio-frequency control~\cite{Tretyakov2014}. This Stark-tuned dipole-dipole interaction between two individual Rydberg atoms was first observed and reported in~\cite{Ryabtsev2010}. Later on, three-body F\"{o}rster resonances~\cite{Faoro2015,Tretyakov2017,Cheinet2020} and even four-body F\"{o}rster resonances were reported~\cite{Liu2020}. 

The F\"{o}rster resonance can be used to generate entangling gates between two~\cite{Beterov2016,Beterov2018} or three qubits~\cite{Beterov2018arX}. For two-qubit gates, the method consists of four steps which can be understood by looking at the input state $|11\rangle$ since the other three input states do not involve Rydberg interactions. (i) First, excite the state $|11\rangle$ to the Rydberg state $-|r_1r_2\rangle$ whose interaction is small compared to the Rabi frequency of the Rydberg excitation. (ii) Second, tune the energy difference between $|r_1r_2\rangle$ and $|r_3r_4\rangle$ adiabatically, from some large value to zero, and then to some large value again. The pulse is designed so that the transition $-|r_1r_2\rangle\rightarrow i|r_3r_4\rangle$ is realized~(we assume zero phase in the dipole-coupling Rabi frequency) as in a $\pi$-pulse Rabi transition. (iii) Third, repeat the last step so that $i|r_3r_4\rangle$ transitions back to $|r_1r_2\rangle$. (iv) Fourth, repeat the first step so that the input state $|r_1r_2\rangle$ returns to $-|11\rangle$. Because the adiabatic following of the F\"{o}rster resonance does not occur for the other three input states, a C$_{\text{Z}}$ gate map $\{|00\rangle, |01\rangle, |10\rangle, |11\rangle\}\rightarrow-\{-|00\rangle, |01\rangle, |10\rangle, |11\rangle\} $ can be realized. Compared to the method of directly using the $\pi$ phase shift in the Rabi-like dipole-dipole flip $|r_1r_2\rangle\leftrightarrow|r_3r_4\rangle$, the adiabatic tuning of the F\"{o}rster resonance has a much smaller error from the fluctuation of the interatomic distance, as numerically demonstrated in Refs.~\cite{Beterov2016,Beterov2018} which showed that high-fidelity {\footnotesize CNOT} gates can be prepared; note, however, that these numbers for the fidelity assume constant dipole-dipole interactions. As shown in Fig.~8 on page 7 of Ref.~\cite{Beterov2018}, there will be extra errors in the gate fidelity from the fluctuation of the interatomic distance which leads to fluctuation of the the dipole-dipole interactions.

In Ref.~\cite{Beterov2018arX} a fast three-qubit Toffoli gate was proposed by using phase shift from resonant Borromean three-body interactions. The data below Eq.~(2) of Ref.~\cite{Beterov2018arX} shows that an intrinsic gate fidelity $98.3\%$ is realizable with realistic parameters for the dipole-dipole interactions, where the gate fidelity is limited mainly by the Rydberg-state decay. However, it is an open question whether there can be an adiabatic three-qubit gate protocol like the two-qubit gates in~\cite{Beterov2016,Beterov2018} where the error due to the fluctuation of the interatomic distance can be partially suppressed.

It is also possible to combine F\"{o}rster resonance with pulse shaping for Rydberg gates, as shown in Refs.~\cite{Liao2019,Sun2020}. The difference between Ref.~\cite{Liao2019} and~\cite{Sun2020} is that adiabatic following of a dark state was used in~\cite{Sun2020}, but the scheme in Ref.~\cite{Liao2019} can obtain geometric phases for certain input states with either adiabatic or nonadiabatic excitations. 

Although the F\"{o}rster-resonance method has a fidelity limited by the finiteness of Rydberg interactions, Rydberg-state decay, and Doppler dephasing, it has several special attractions. A special benefit of using F\"{o}rster-resonant Rabi-like rotations for Rydberg quantum gates is that the distance $L$ between two qubits can be quite large so as to easily excite two or three qubits to Rydberg states. This leads to three positive consequences. First, the fluctuation of the dipole-dipole coupling strength $C_3/L^3$ will only have a small fluctuation ratio $3\delta L/L$ when the distance fluctuation $\delta L$ is much smaller than $L$; this scaling compares favorably to gates by a van der Waals interaction $C_6/L^6$ where a distance fluctuation leads to a change of interaction $6\delta L/L$. Second, the large distance drastically reduces crosstalk between qubits when single-qubit Rydberg addressing is used. Third, it can be used to couple two qubits far away in a dense qubit array. For example, the lattice constant in the optical lattice in the experiment of~\cite{Graham2019} was $3.1~\mu$m , and the parameters used therein are unlikely able to entangle two qubits separated by more than three lattice periods; however, it was shown in~\cite{Beterov2016,Beterov2018} that {\footnotesize CNOT} gates in two qubits separated by $L=25~\mu$m can have a fidelity $99.4\%$ by the F\"{o}rster-resonance method~(see Fig.~6(h) on page 6 of Ref.~\cite{Beterov2016}). Thus, the F\"{o}rster-resonance method can be useful for large-scale quantum computing where entanglement of two well-separated qubits is required. The main challenge is to accurately isolate the desired F\"{o}rster-resonant processes.

\subsection{Two-qubit and multiqubit gates by dark states formed with F\"{o}rster resonance }\label{sec04D}
There is a special way to use adiabatic following of a dark state formed by F\"{o}rster resonance for Rydberg-mediated entanglement~\cite{Petrosyan2017}. Consider a three pulse sequence of the original method shown in Eq.~(\ref{sec03A01}) and Fig.~\ref{figure-originalgate}(a), where the key step is the blockade that leads to no phase shift to the input $|11\rangle$ during the $2\pi$ pulse. Now, consider the case that the $2\pi$ pulse is replaced by an adiabatic pulse that has the same pulse area, but can allow an adiabatic following after the dark state $|D\rangle\propto 2V|r_11\rangle-\Omega(t)|r_3r_4\rangle$ of the relevant Hamiltonian 
\begin{eqnarray}
  \hat{H} &=&\left[\frac{\hbar\Omega(t) }{2} |r_1r_2\rangle\langle r_11|+ V|r_1r_2\rangle\langle r_3r_4|\right]+ \text{H.c.},\label{sec04C01}
\end{eqnarray}
where $|r_1\rangle$ is the Rydberg state excited during the first pulse on the control qubit, and $|r_2\rangle$ is the Rydberg state that is excited during the adiabatic pulse on the target qubit. Even if the dipole-dipole coupling strength $V$ fluctuates from shot to shot, the dark state persists for each cycle of the gate where $V$ is assumed to be constant. By smoothly varying $\Omega(t)$ from 0 to some large value, and back to 0 again, the state $|r_11\rangle$ follows after the dark state $|D\rangle$ and ends at $|r_11\rangle$ without any phase change in it. This is remarkable in that although the two-atom Rydberg state is populated during the adiabatic pulse, no phase accumulates because the eigenenergy of the dark state is zero in the rotating frame. After the adiabatic pulse, another pulse is applied for restoring the control qubit to the ground state, resulting in a C$_{\text{Z}}$ gate as in Eq.~(\ref{sec03A01}). The method in~\cite{Petrosyan2017} is not only insensitive to the fluctuation of $V$ in different gate cycles, but comes with less blockade error compared to the initial Rydberg blockade gate proposed in~\cite{PhysRevLett.85.2208}. As a consequence, it is possible to have an extremely large intrinsic fidelity $99.995\%$ for a C$_{\text{Z}}$ gate with this method. The fidelity reported there is in principle robust against the change of interactions from shot to shot, but the influence on the adiabatic following by the change of $V$ and the Doppler dephasing due to the free flight of qubits within one gate cycle is unclear.

\begin{figure*}[ht]
\includegraphics[width=6.0in]
{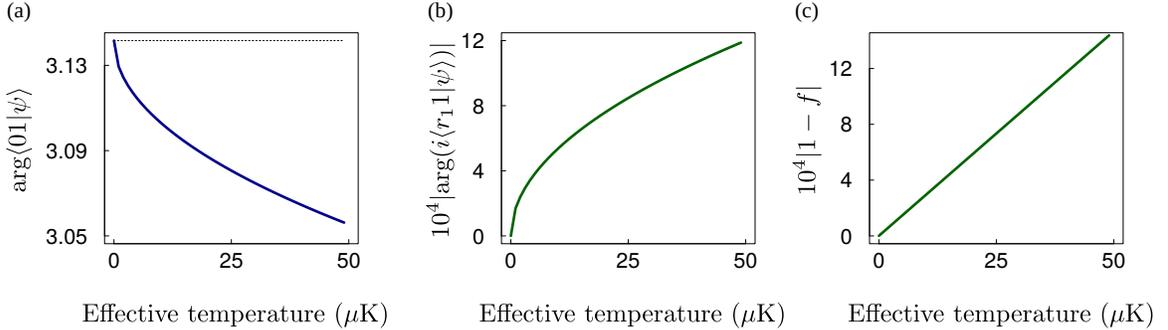}
\caption{Results of numerical simulation for the state transform from $\{|00\rangle,~|01\rangle,~-i|r_10\rangle,~-i|r_11\rangle\}$ to $\{|00\rangle,~-|01\rangle,~-i|r_10\rangle,~-i|r_11\rangle\}$ in the second pulse of the gate proposed in Ref.~\cite{Petrosyan2017} for different atomic temperatures $T_a$; different temperatures lead to different atomic speeds $v_{\text{rms}}=\sqrt{k_BT_am}$, and a larger speed results in larger dephasing of the ground-Rydberg transition. The pulse is a shifted Gaussian one of duration $T_t=0.29~\mu$s. (a) The phase accumulated by the input state $|01\rangle$ at the end of the adiabatic pulse. Ideally the phase is $\pi$ indicated by the horizontal dashed line. (b) The phase acquired~(scaled by $10^4$) by the state $-i|r_10\rangle$ upon the adiabatic pulse. Ideally it should be zero. (c) The error~(scaled by $10^4$) in the fidelity $f$ for the quantum operation of the adiabatic pulse averaged over the four input states.    \label{figure-petrosyn} }
\end{figure*}

To investigate the robustness of the adiabatic following of Ref.~\cite{PhysRevLett.85.2208}, we consider the input states $\{|00\rangle,~|01\rangle,~-i|r_10\rangle,~-i|r_11\rangle\}$ at the beginning of the the adiabatic pulse. The desired output states for the adiabatic pulse are $\{|00\rangle,~-|01\rangle,~-i|r_10\rangle,~-i|r_11\rangle\}$. By using $\Omega=\Omega_m [e^{-(t-T_t/2)^2/(2\sigma^2)}  -  e^{-T_t^2/(8\sigma^2)} $ and a dipole-dipole coupling coefficient $C_3/(2\pi)=-33$GHz$\mu m^3$ of the states $\{|r_1\rangle,~|r_2\rangle\}\equiv \{95p_{3/2},84p_{3/2}\}$~(see Table~I on page 5 of Ref.~\cite{Petrosyan2017}), we numerically study the fidelity $f$ of the operation of the adiabatic pulse by using the formula of Eq.~(\ref{fidelity04}). Because the $^{133}$Cs $p$-orbital Rydberg states are usually excited by one-photon excitation with 319~nm lasers~\cite{Hankin2014}, it is in general not easy to have a large Rabi frequency. So, we consider a peak Rabi frequency $\Omega_m/(2\pi)=7.643$~MHz and $T_t=0.29~\mu$s which is 10 time longer than the adiabatic pulse used in Appendix A2 of Ref.~\cite{Petrosyan2017}. We suppose that the speed of the atom along the direction of the Rydberg laser field is $v_{\text{rms}}=\sqrt{k_BT_am}$, where $k_B$ is the Boltzmann constant, $T_a$ is the atomic temperature, and $m$ is the mass of the atomic qubit. Figure~\ref{figure-petrosyn}(a) shows that the state $|01\rangle$ obtains some phase error upon the adiabatic excitation pulse; the dashed line shows the correct phase $\pi$. For the state $-i|r_11\rangle$, Fig.~\ref{figure-petrosyn}(b) shows that the phase error is below $1.2\times10^{-3}$~radians for $T_a\leq50~\mu$K. The population errors in both $|01\rangle$ and $-i|r_11\rangle$ are negligible. The total error for the operation of the adiabatic pulse averaged over the four states is shown in Fig.~\ref{figure-petrosyn}(c). Note that the errors shown in Fig.~\ref{figure-petrosyn}(c) should be smaller than the final gate error since the operation fidelities in the first and third pulses of the gate of Ref.~\cite{Petrosyan2017} are not included. The data in Fig.~\ref{figure-petrosyn}(b) rigorously show that the adiabatic following for the input state $|11\rangle$ is robust against the fluctuation of the interaction as well as the Doppler dephasing. But the data in Fig.~\ref{figure-petrosyn}(c) indicate that the Doppler dephasing still hampers the operation fidelities since the adiabatic following does not take effect for the other two input states $|01\rangle$ and $|10\rangle$ nor in the first and third pulses.

The method in~\cite{Petrosyan2017} can be extended in various ways. For one example, Ref.~\cite{Yu2019} used a similar method to study a quantum gate with the Rydberg interactions of both the direct dipole-dipole type and a van der Waals type, and found that fidelities 99.96\% are realizable with a Rydberg-state decay rate $2\pi\times0.2$~kHz for a certain gate with a map diag$\{1,~-1,~-1,~1\}$~(see Fig.~2 on page 23083 of Ref.~\cite{Yu2019}). This gate induces no entanglement regarding the entangling power~\cite{Williams2011} and can be easily created by using two single-qubit $\pi$-phase gates on the qubits. However, it should be useful to extend the method in Ref.~\cite{Yu2019} to design an entangling gate. For another example, Ref.~\cite{Khazali2020} extended the scheme in~\cite{Petrosyan2017} to protocols for creating multiqubit C$_k$-NOT and C-NOT$^k$ gates. Detailed analysis in~\cite{Khazali2020} showed that for a multiqubit system, there is still a dark state useful for entanglement generation via the adiabatic following. Figure~6(c) in Ref.~\cite{Khazali2020} shows that a fidelity about $98\%$ with $k$ up to 20 is possible. Ref.~\cite{Khazali2020} also extended the scheme to superconducting systems because they also possess the Ising-type interactions like the F\"{o}rster resonance of Rydberg atoms. It is worth mentioning that the Ising-like coupling shared in different physical systems has been well recognized in the study of multiqubit quantum gates~\cite{Rasmussen2020}. Finally, we note that for all these gates, the relatively long gate durations can lead to large Rydberg-state decay. For example, the gate fidelity has a large sensitivity to the Rydberg decay rate as analyzed in Ref.~\cite{Yu2019}~(see Fig.~5 on page 23088 therein).

\section{Rydberg gates and entanglement by antiblockade}\label{sec05}
It was found in~\cite{Ates2007,Amthor2010} that when the Rydberg interaction $V$ is exactly equal to the detuning of a Rydberg Rabi frequency, it appears as a resonance in the rotating frame so that the two-atom Rydberg state can be fully excited. The antiblockade-induced resonance can also appear in more than two atoms~\cite{Pohl2009,Mazza2020,Gambetta2020} which can lead to experimentally observable many-body effects~\cite{Bai2020}. In Table~\ref{table5}, we list the theoretical fidelities in the literature with the places denoted to point out where the fidelity number appears in the corresponding reference; in many references more than one type of quantum gates or entangling operations were studied, and a relatively large fidelity is quoted in the table. 

\begin{figure*}[ht]
\includegraphics[width=7.0in]
{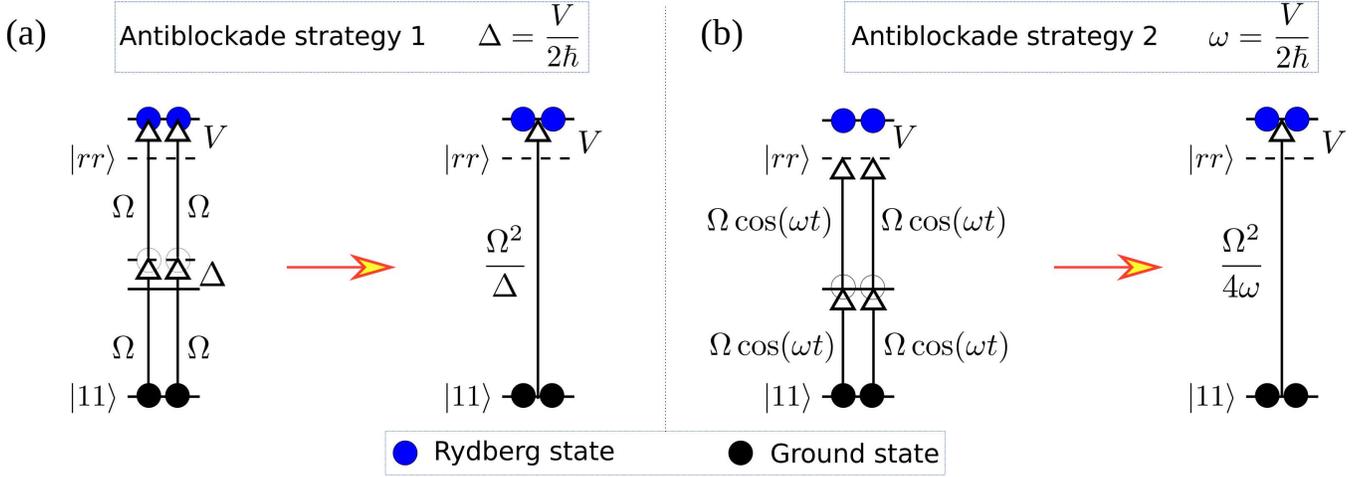}
 \caption{Two typical strategies to use the antiblockade effect. (a) By sending to two qubits the same laser fields for Rydberg excitation with Rabi frequency $\Omega$ and detuning $\Delta$, the state $|11\rangle$ can be excited to $|rr\rangle$ with an effective Rabi frequency $\Omega^2/\Delta$ when $V=2\hbar\Delta$. (b) If the laser fields are resonant but periodically varying so that the Rydberg Rabi frequency obtains a time dependence as $\Omega\cos(\omega t)$, the state $|11\rangle$ can be excited to $|rr\rangle$ with an effective Rabi frequency $\Omega^2/(4\omega)$ when $V=2\hbar\omega$. These two methods are the representative strategies, based on which there are various antiblockade methods reviewed in Sec.~\ref{sec05}.       \label{figure-antiblockade} }
\end{figure*}

Briefly speaking, there are two general strategies to realize the antiblockade, by matching the interaction with frequency detuning of the lasers, and by matching the interaction with a modulation frequency of the amplitude of the laser fields, as schematically shown in Fig.~\ref{figure-antiblockade} with their most basic protocols. The first strategy has been widely studied so that several ways have been found to use it for entanglement. Overall, there are at least five ways to induce entanglement by antiblockade. 

\subsection{Compensating Rydberg interactions by detunings in Rabi frequencies}\label{antibloc-A}
\subsubsection{Simultaneous excitation of multi-atoms with detuned laser fields}\label{antibloc-A01}
Early study on Rydberg-mediated entanglement with antiblockade usually focused on using detuning in the Rydberg Rabi oscillations to compensate the Rydberg interaction, so that a higher-order Rydberg Rabi frequency arises from antiblockade. An early investigation of antiblockade on quantum entanglement was made in Ref.~\cite{Zuo2010}, which presented protocols for entanglement~(such as the GHZ states) between ground and Rydberg states. The model in Ref.~\cite{Zuo2010} assumes only nearest-neighboring van der Waals interaction in a one-dimensional lattice, and thus it is an open question whether there is a similar model for fast multi-atom entanglement with realistic physical parameters in the antiblockade regime. Later on, Refs.~\cite{Shao2014,Tian2015,Su2016,Zhu2019SLSu} did a more detailed analysis, and showed that by setting the detuning equal to the Rydberg blockade, one can populate a two-atom Rydberg state with a Rabi frequency from the second-order perturbation theory; using the same method, by setting the effective detuning equal to the Rydberg blockade in three atoms, one can populate the three-atom Rydberg state with a Rabi frequency derived from the perturbation theory. This strategy can be repeated so more atoms can be populated in Rydberg states with a higher-order Rabi frequency, which can lead to four-atom GHZ states with a high fidelity $96.80\%$ if Rydberg-state decay is ignored~\cite{Shao2014}~(see text below Eq. (9) on page 831 of Ref.~\cite{Shao2014}). The perturbation used in Refs.~\cite{Shao2014,Su2016} led to ac Stark shifts that can be canceled by adding additional control fields. To avoid adding extra control fields, Ref.~\cite{Su2017} introduced a scheme to modify the detuning used in the excitation in specific ways so that two-atom Rydberg excitation can still be realized in one step, and Ref.~\cite{Su2018} introduced a scheme to realize an $N$-atom Rydberg excitation in one step with ac Stark shifts canceled with specially modified detunings. Notice that these approaches to Rydberg antiblockade can lead to many interesting schemes. For example, by using Rydberg antiblockade, Ref.~\cite{Li2018} studied three-qubit gates and quantum error correction by a type of ``Unconventional Rydberg pumping'', Refs.~\cite{Su2020,Zheng2020} studied a novel nondestructive parity meter in ground-state atoms, Ref.~\cite{Xing2020} studied a three-qubit controlled-phase gate with a tunable phase, Refs.~\cite{Zheng2020GHZ,Haase2021} showed methods to inter-convert the GHZ and W states, and Ref.~\cite{LiuBJ2020} studied nonadiabatic noncyclic geometric phases in a Rydberg gate based on which Ref.~\cite{Guo2021} designed a scheme to distinguish different types of GHZ states with fidelities over 99.9\% for GHZ states up to five qubits. Because the different interactions between different pairs of atoms in a multi-atom system can be different, by matching the detunings in the lasers with specific Rydberg interactions, interesting dynamics can be identified. For example, Ref.~\cite{Yin2020} used this strategy to show that high-fidelity three and four-qubit controlled-NOT gates can be created.

\subsubsection{Sequential excitation of atoms with detuned laser fields}\label{antibloc-A02}
The second way to compensate the interaction $V$ is to sequentially and quickly excite one atom after another, with the detuning of each pulse specifically designed to compensate the Rydberg interaction. This means that it is no longer necessary to use adiabatic elimination to obtain an effective Rydberg Rabi frequency, so that the required time for Rydberg excitation is much shorter than those in Refs.~\cite{Shao2014,Su2016,Su2017,Su2018}. The method to entangle two and three atoms by this fast antiblockade strategy was analyzed in detail in Ref.~\cite{Su201701}.

Because the requirement to compensate the interaction, there are relatively few protocols based on pulse shaping to implement quantum gates based on antiblockade. However, there are two examples, shown in Refs.~\cite{Tian2015,Yan2021}, that used pulse shaping and antiblockade for entanglement generation, where Ref.~\cite{Yan2021} is not quoted in Table~\ref{table5} because it did not show achievable gate fidelities by their scheme.

\begin{table*}[ht]
     \begin{threeparttable}
  \centering
  \begin{tabular}{|c|l|c|}
    \hline  Requirement  & Fidelity;\tnote{a}~ type of operation; where it's taken   & References \\ \hline
\multirow{9}{*}{\begin{tabular}{c}\text{Higher-order}~$\Omega$~\text{via}\\\text{matching the detuning}\\ \text{in laser fields with}\\ \text{Rydberg interactions}\\ (Sec.~\ref{antibloc-A01})\end{tabular}}  & 99.68\%; Bell state~(via pulse shaping); lower left of page 4 of Ref.~\cite{Tian2015} & Ref.~\cite{Tian2015}	 \\ \cline{2-3}
 & $99.95\%$; Bell state; Fig.~10(a) of page 7 of Ref.~\cite{Su2017} & Ref.~\cite{Su2017} 	 \\ \cline{2-3}
 & $99.73\%$; 2-qubit entanglement; Fig.~9 of page 1943 of Ref.~\cite{Zhu2019SLSu} & Ref.~\cite{Zhu2019SLSu} 	 \\ \cline{2-3}
& $98\%$; 2-qubit entanglement; Fig.~5(b) of page 5 of Ref.~\cite{LiuBJ2020} & Ref.~\cite{LiuBJ2020}	 \\ \cline{2-3}
 & $96.54\%$; 3-qubit gate; lower right of page 829 of Ref.~\cite{Shao2014} &Ref.~\cite{Shao2014}\\  \cline{2-3}
 & $96.79\%$; 3-qubit gate; text below Fig.~6 of page 6 of Ref.~\cite{Su2018} &  Ref.~\cite{Su2018}	 \\ \cline{2-3}
 & $97.34\%$; 3-qubit entanglement; upper right of page 8 of Ref.~\cite{Li2018} & Ref.~\cite{Li2018}\\   \cline{2-3}
 & $98.87\%$; Toffoli gate; middle right of page 5 of Ref.~\cite{Xing2020} & Ref.~\cite{Xing2020}\\  \cline{2-3}
 & ${\color{blue}99.65}\%$; Toffoli gate; Table~1 on page 35583 of Ref.~\cite{Yin2020} & Ref.~\cite{Yin2020}\\  \hline
\text{Sequential excitation} & $99.9\%$; 2-qubit state; lower right of page 8 of Ref.~\cite{Su201701}~({\bf fast})\tnote{b} & Ref.~\cite{Su201701}\\  \hline
\multirow{3}{*} {\begin{tabular}{c}\text{Ground-state blockade}\\ (Sec.~\ref{antibloc-B})\end{tabular}}& $99.32\%$; Bell state~(via pulse shaping); text below Fig.~4 on page 4 of Ref.~\cite{Shao2017ground} & Ref.~\cite{Shao2017ground}\\  \cline{2-3}
 & $99.89\%$; Bell state~(via cavity); text below Fig.~5 on page 4 of Ref.~\cite{Shao2017} & Ref.~\cite{Shao2017}\\  \cline{2-3}
 & $99.66\%$~(at $0$~K); C$_{\text{Z}}$ gate; text below Fig.~6 on page 6 of Ref.~\cite{Shao2020}~({\bf fast})\tnote{c} & Ref.~\cite{Shao2020}\\  \hline
\multirow{4}{*} {\begin{tabular}{c}\text{Dissipation with}\\ \text{asymmetric interactions}\\(Sec.~\ref{antibloc-C01})\end{tabular}} & $99.88\%$; Bell state; upper right of page 3 of Ref.~\cite{Carr2013} & Ref.~\cite{Carr2013}\\  \cline{2-3}
 & $99.91\%$; 2-qubit entanglement; text below Fig.~2 on page 4 of Ref.~\cite{Shao2014pra} & Ref.~\cite{Shao2014pra}\\  \cline{2-3}
 & $99.47\%$; 3-qubit entanglement; text below Fig.~4 on page 4 of Ref.~\cite{Chen2017} & Ref.~\cite{Chen2017}\\    \cline{2-3}
 & $99.09\%$; 3-qubit entanglement~(via cavity); lower left of page 1642 of Ref.~\cite{Li:18} & Ref.~\cite{Li:18}\\    \hline
\multirow{8}{*} {\begin{tabular}{c}Dissipation\\(Sec.~\ref{antibloc-C02})\end{tabular}} & $99.9\%$;  Bell state; Fig.~2 on page 3 of Ref.~\cite{Su2015} & Ref.~\cite{Su2015}\\  \cline{2-3}
 & $99\%$;  Bell state~(via cavity); text above Fig.~7 on page 5 of Ref.~\cite{Chen2018} & Ref.~\cite{Chen2018}\\ \cline{2-3}
 & $99.7\%$;  2-qubit entanglement; end of Sec.~4 on page 2300 of Ref.~\cite{Li:18oe} & Ref.~\cite{Li:18oe}\\ \cline{2-3}
 & ${\color{blue}99.98}\%$;  2-qubit entanglement; text above Fig.~3 on page 10124 of Ref.~\cite{Jin:21} & Ref.~\cite{Jin:21}\\  \cline{2-3}
 & $98.24\%$;  3-qubit entanglement~(via cavity); lower left of page 5 of Ref.~\cite{Shao201702} & Ref.~\cite{Shao201702}\\  \cline{2-3}
 & $99\%$;  6-qubit entanglement; upper right of page 4 of Ref.~\cite{Wintermantel2020} & Ref.~\cite{Wintermantel2020}\\\cline{2-3}
 & $99.24\%$; 3-qubit entanglement; abstract of Ref.~\cite{Yang2021} & Ref.~\cite{Yang2021}\\ \cline{2-3}
 & $98\%$;  3-qubit entanglement; abstract of Ref.~\cite{Li2021} & Ref.~\cite{Li2021}\\ \hline
\multirow{5}{*}  {\begin{tabular}{c}\text{Compensating Rydberg}\\ \text{interactions by using}\\ \text {oscillating}~$\Omega$~(Sec.~\ref{antibloc-D}) \end{tabular}}& $99.35\%$; {\footnotesize CNOT} gate; middle left on page 4 of Ref.~\cite{WU2020126039} & Ref.~\cite{WU2020126039}\\  \cline{2-3}
 & $99.6\%$;  C$_{\text{Z}}$ gate; text below Eq.~(10) on page 1203 of Ref.~\cite{Wu:20} & Ref.~\cite{Wu:20}\\ \cline{2-3}
 & $98\%$;  Bell state; Fig.~9(c) on page 8 of Ref.~\cite{Li2020} & Ref.~\cite{Li2020}\\ \cline{2-3}
 & $99\%$; C$_{\text{Z}}$ gate; lower left of page 7 of Ref.~\cite{Wu2021}~({\bf fast})~\tnote{b} & Ref.~\cite{Wu2021}\\  \cline{2-3}
 & $99.1\%$; SWAP gate; texts below Fig.~3 and Fig.~4(a) on pp. 816 and 817 of Ref.~\cite{Wu2021pr} & Ref.~\cite{Wu2021pr}\\  \hline
  \end{tabular}
  \caption{  \label{table5} Fidelities of quantum gates and entanglement with antiblockade. The order of the list follows mainly after the order of the discussions in the text. Two-qubit operations are first shown, then multiqubit ones according to their publication dates. Here, ``2-qubit'' and ``3-qubit'' are labeled for phase gates that can be transformed to C$_{\text{Z}}$, CNOT, or Toffoli gates via single-qubit gates. Most numbers of fidelities quoted were calculated with some decay rates around $2\pi\times1$~kHz for the Rydberg state in the corresponding references unless otherwise specified. The blue color highlights the largest fidelities for gates and entanglement with two or more than two qubits.  }
  \begin{tablenotes}
    \item[a] Most references show more than one type of gates or entanglement, where an operation with larger fidelity is quoted. 
    \item[b] Here, ``fast'' only means that they do not depend on Rydberg excitation with a Rabi frequency derived with a perturbation theory via the antiblockade, while a practical experimental implementation has a speed limited by experimentally feasible parameters.
        \item[c] In Ref.~\cite{Shao2020} a relative slow gate was shown when taking realistic data though the theory does not involve high-order Rydberg Rabi frequencies derived by the antiblockade conditions. The text below Fig.~4 on page 5 of Ref.~\cite{Shao2020} shows that state decay at 0~K was considered. 
    \end{tablenotes}
     \end{threeparttable}
  \end{table*}


\subsection{Blocking transitions from certain two-atom ground states by antiblockade}\label{antibloc-B}
The second method of quantum gates with antiblockade is to use Rydberg antiblockade to design an effective ground-state blockade, as proposed in Ref.~\cite{Shao2017ground} with van der Waals interaction and Ref~\cite{Shao2020} with isolated dipole-dipole interactions. 
\subsubsection{Ground-state blockade via Rydberg antiblockade with van der Waals interactions }\label{antibloc-B02}
Consider the Raman transition with an effective Rabi frequency $\Omega_{\text{hyper}}$ between $|0\rangle$ and $|1\rangle$ via a largely detuned low-lying $p$-orbital state, there will be the following transition chain
\begin{eqnarray}
|00\rangle\leftrightarrow(|01\rangle+|10\rangle)/\sqrt{2}\leftrightarrow|11\rangle.
\end{eqnarray}
However, if one excites $|11\rangle$ to a two-atom Rydberg state $|rr\rangle$ via the Rydberg antiblockade mechanism~\cite{Shao2014,Su2016,Su2017,Su2018} so that $|11\rangle\leftrightarrow|rr\rangle$ is realized with a Rabi frequency $\Omega_{\text{Ryd}}$ and detuning $\Delta$, there will be Stark shifts $(\Delta\pm\sqrt{\Delta^2+\Omega_{\text{Ryd}}^2})/2$ for $|11\rangle$, which can be large compared to $\Omega_{\text{hyper}}$ so as to block the transition from $(|01\rangle+|10\rangle)/\sqrt{2}$ to $|11\rangle$. This means that there will be an isolated transition from $|00\rangle$ to the Bell state $(|01\rangle+|10\rangle)/\sqrt{2}$, so that entanglement can be generated rapidly. This idea is similar to the dressing method reviewed in Sec.~\ref{sec03Bnew1}. The ground-state blockade mechanism can also entangle three atoms~\cite{Shao2017ground} and can be used to generate entanglement by dissipation in the context of cavity quantum electrodynamics~\cite{Shao2017}.

\subsubsection{Ground-state blockade via Rydberg antiblockade with dipole-dipole interactions}\label{antibloc-B02}
Another ground-state blockade mechanism with isolated dipole-dipole interactions similar to the one in~\cite{Shao2017ground} was proposed in Ref.~\cite{Shao2020}. The dipole-dipole interactions employed in Ref.~\cite{Shao2020} couple three two-atom Rydberg states, and by coupling each of two nearby atoms for the transition $|0\rangle\leftrightarrow|r\rangle\leftrightarrow|1\rangle$ with appropriate Rabi frequencies and detunings~(see Fig.~1 of Ref.~\cite{Shao2020}), it is possible to create a transition between $|11\rangle$ and $(|1r\rangle+|r1\rangle)/\sqrt{2}$ while the other states $|00\rangle,~|01\rangle$, and $|10\rangle$ are decoupled because of being largely detuned effectively by the dipole-dipole interaction and detuning in the laser fields. This means that by applying the control for a full Rabi-like transition between $|11\rangle$ and $(|1r\rangle+|r1\rangle)/\sqrt{2}$, a C$_{\text{Z}}$ gate can be created. However, the antiblockade can fail due to the fluctuation of interatomic distance, and it is unclear about to what extent will the breakdown of the Rydberg antiblockade due to this fluctuation hamper the ground-state blockade.

\subsection{Quantum gates and entanglement by dissipation }\label{antibloc-C}
\subsubsection{Entanglement with dissipation and asymmetries in Rydberg interactions}\label{antibloc-C01}
The third method of quantum gates by antiblockade is to use asymmetric Rydberg interactions and dissipation, as proposed in Ref.~\cite{Carr2013}. The asymmetric Rydberg interaction can be used in the excitation blockade regime as reviewed in Sec.~\ref{sec03Bnew2}. To use it in the antiblockade regime, Ref.~\cite{Carr2013} assumed that the interactions $V_{\text{ss}},~V_{\text{sp}}$ and $V_{\text{pp}}$ for the two-atom Rydberg states $|ss\rangle,~|sp(ps)\rangle$ and $|pp\rangle$ satisfy the condition $V_{\text{pp}}= V_{\text{ss}}\neq V_{\text{sp}}$, which can be realized by choosing different Zeeman substates $|s\rangle$ and $|p\rangle$ within one Rydberg level. Then, by using Rydberg excitation fields with a detuning that exactly compensates the Rydberg interaction $V_{\text{pp}}= V_{\text{ss}}$, one can populate the states $|ss\rangle$ and $|pp\rangle$ respectively from $|00\rangle$ and $|11\rangle$ via the excitation $|0\rangle\xrightarrow[]{\Omega_{\text{eff}}}-i|s\rangle$ and $|1\rangle\xrightarrow[]{\Omega_{\text{eff}}}-i|p\rangle$, where $\Omega_{\text{eff}}=2\hbar\Omega^2/V_{\text{ss}}$ is a Rabi frequency from the second-order perturbation theory. The decay of the states $|ss\rangle$ and $|pp\rangle$ will populate the states $|00\rangle,~|01\rangle,~|10\rangle$ and $|11\rangle$ with an equal probability, but among which only $|00\rangle$ and $|11\rangle$ will be reexcited to the two-atom Rydberg states which enables recycling. Then, it is possible to create an entangled state $(|01\rangle+e^{i\alpha}|10\rangle)/\sqrt{2}$ with an undetermined phase. By adding a Raman transition between $|0\rangle$ and $|1\rangle$, Ref.~\cite{Carr2013} showed that the maximally entangled state with $\alpha=\pi$ can be created with a high fidelity of similar value in the original Rydberg blockade method if the fluctuation of Rydberg interactions is ignored. The method in~\cite{Carr2013} can also generate anti-ferromagnetic state where a purity of $96\%$ was found in a 4-atom system. Later on, Ref.~\cite{Shao2014pra} extended the scheme to prepare an entangled state in the form of, in the two-atom case, $\sum_{k=1}^3|g_kg_k\rangle/\sqrt{3}$, i.e., a maximally entangled state between three ground states $|g_k\rangle$ with $k=1,2,3$. Besides depending on dissipation and antiblockade, Ref.~\cite{Shao2014pra} requires the interactions $V_{ij}$ of the Rydberg states $|r_ir_j\rangle$ to satisfy the condition $V_{ii}=V_{jj}\neq V_{ij}$, where $i,j\in\{1,2,3\}$ and $i\neq j$. When the Rydberg interactions between three atoms satisfy specific relations, more complicated entanglement can arise. For example, Ref.~\cite{Chen2017} showed that a three-qubit singlet state can be created, and Ref.~\cite{Li:18} showed that a three-qubit $W$ state can be created by coupling atoms with cavity modes.

\subsubsection{Entanglement with dissipation}\label{antibloc-C02}
The methods reviewed in Sec.~\ref{antibloc-C01} depend on both dissipation and asymmetries in the Rydberg interactions. Soon after the introduction of the methods of Ref.~\cite{Carr2013}, Ref.~\cite{Su2015} put forward a concise scheme using dissipation and antiblockade to achieve the goal of Refs.~\cite{Carr2013,Shao2014pra}. Ref.~\cite{Su2015} showed that two transitions are sufficient, namely, a Rydberg excitation $|0\rangle\xrightarrow[]{}-i|r\rangle$ (detuned by $\Delta$) and a resonant Raman transition $|0\rangle\xrightarrow[]{}-i|1\rangle$. An appropriate choice of $\Delta$ can block the transition from $|D\rangle\equiv (|01\rangle-|10\rangle)/\sqrt{2}$ to Rydberg states, but any two-atom states not identical to $|D\rangle$ can be excited to Rydberg states that will then decay to $|D\rangle$. Then, $|D\rangle$ becomes the only stable state in the process of dissipation. Later on, it was found that by choosing appropriate detunings, interactions between the states, and dissipation, entanglement with various forms can be created, shown in Refs.~\cite{Li:18oe,Wintermantel2020,Yang2021,Li2021}. Dissipation can also be used to generate entanglement with dipole-dipole flip-flop processes as shown in Ref.~\cite{Jin:21}. Unlike that in Sec.~\ref{antibloc-C01}, this way does not require asymmetric interactions and may be easier to be experimentally tested if the fluctuation of interactions can be sufficiently suppressed. As in Sec.~\ref{antibloc-C01}, the scheme based on dissipation and antiblockade could also be used in the context of cavity quantum electrodynamics~\cite{Shao201702,Chen2018}.  

\subsection{Compensating Rydberg interactions by periodically changing Rydberg Rabi frequencies }\label{antibloc-D}
The fourth method of quantum entanglement via antiblockade is to use periodically changing Rydberg Rabi frequencies to compensate the Rydberg interactions. In Ref.~\cite{WU2020126039}, it was shown that instead of using the mismatch between the laser frequency and atomic transition frequency to compensate the Rydberg interactions as in the above methods, laser fields with amplitudes that are periodically alternating can effectively compensate the Rydberg interactions. Because the laser fields used in the Rydberg excitation have wavelengths of hundreds of nanometers, one can still use rotating-wave approximation to derive a Hamiltonian for the atom-laser interaction,
\begin{eqnarray}
  \hat{H} &=&\frac{\hbar\Omega}{2} |r\rangle\langle 1|+\text{H.c.},\label{anti-eq01}
\end{eqnarray}
where $|1\rangle$ and $|r\rangle$ are the ground and Rydberg states, respectively, and $\Omega=\Omega_m\cos(\omega t)$, where the engineered modulation has an angular frequency $\omega$ that is orders of magnitude smaller than the central frequency of the lasers. Then, using the same strategy for deriving Eq.~(\ref{anti-eq01}) and the rotating-wave approximation conditioned on $\omega\ll \Omega_m$, one can again use a rotating-frame transform to derive
\begin{eqnarray}
  \hat{H} &=&\left(\frac{\hbar\Omega_m}{4} |r\rangle\langle 1|+\text{H.c.}\right) +\hbar\omega|r\rangle\langle r|,\label{anti-eq02}
\end{eqnarray}
where the last term above can also be $-\hbar\omega|r\rangle\langle r|$ if another rotating frame is employed. The last term in Eq.~(\ref{anti-eq02}), then, becomes an effective detuning that can be used to compensate the Rydberg interaction involved in the two-atom state $|rr\rangle$. Based on this formalism, Ref.~\cite{WU2020126039} showed C$_{\text{Z}}$ and {\footnotesize CNOT} gate protocols by using the time-dependent Rabi oscillations with two atoms. The scheme in Ref.~\cite{WU2020126039} leads to an effective Rabi frequency via the second-order perturbation theory, where a detailed analysis showed that high-fidelity C$_{\text{Z}}$ gates can be realized~\cite{Wu:20}. To realize a faster gate, Ref.~\cite{Wu2021} proposed to excite one qubit via constant Rydberg Rabi frequency, but to excite the other qubit via a periodically alternating laser fields with a Hamiltonian of Eq.~(\ref{anti-eq02}). Then, resonant Rydberg excitations can be realized based on this method with fidelities over $99\%$ for the two-qubit C$_{\text{Z}}$ gate~\cite{Wu2021}. This exotic antiblockade strategy can also be used together with dissipation for entanglement generation~\cite{Li2020}. 

The antiblockade by periodical driving can be used to construct certain quantum gates that were rarely studied by other means of Rydberg-mediated gate methods. For example, Ref.~\cite{Wu2021pr} used the periodical driving and a modified antiblockade scheme to study a controlled-SWAP gate in three qubits, where the state of one control qubit control the SWAP operation of the states of two target qubits. Ref.~\cite{Wu2021pr} also studied a SWAP gate by the same mechanism. Importantly, both the two gate protocols in Ref.~\cite{Wu2021pr} require only one step. These gates can obtain relatively large fidelities; for example, the texts below Fig.~5 and Fig.~7(a) on page 818 indicate that their C-SWAP gate can have a fidelity 98.7\% at room temperature. To consider the achievable fidelity of their schemes, Ref.~\cite{Wu2021pr} showed that with experimentally affordable cooling of atomic qubits, motion-induced Doppler dephasing only lead to an extra error smaller than 1\%.

Periodically modulating the amplitudes of laser fields can be used for antiblockade in ways other than compensating the interaction. For example, Ref.~\cite{Jin:21} considered using the strategy for speeding up entanglement creation via dissipation; the number quoted from Ref.~\cite{Jin:21} in Table~\ref{table5}, however, does not involve dissipation. It is also of interest to study many-body dynamics by periodical driving of Rydberg atoms~\cite{Mallavarapu2021}.

Though the antiblockade mechanism can lead to high-fidelity Rydberg gates in theory, it is challenging to realize it in experiments because the distance and relative angular positions between qubits in real space can fluctuate, which leads to off-diagonal Rydberg interactions shown in Sec.~\ref{sec02C02} as well as change of the diagonal interactions. Except rare cases such as Ref.~\cite{Zheng2020}, most studies of Rydberg antiblockade did not explore errors from the fluctuation of Rydberg interactions. It demands efforts to realize the theoretical fidelities of quantum gates in the antiblockade regime by, for example, adopting ground-state cooling of qubits or using sufficiently deep traps.

\begin{table*}[ht]
     \begin{threeparttable}
  \centering
  \begin{tabular}{|c|c|c|c|c|}
    \hline
       Category&   \multicolumn{1}{|c|}{Rydberg blockade } &   \multicolumn{3}{|c|}{Frozen interaction}\\
    \cline{1-5}
     Feature&   \multicolumn{1}{|c|}{Robust to the fluctuation of $V$} &   \multicolumn{3}{|c|}{Sensitive to the fluctuation of $V$~\tnote{a}}\\
      \cline{1-5}
      \multirow{2}{*}{  Type of $V$}&  Dipole-dipole, or van der Waals, or in a&   \multicolumn{2}{|c|}{Van der Waals}& \multirow{2}{*}{} \\\cline{4-4}
   &  regime with both types of interactions &  &  \multicolumn{2}{|c|}{Dipole-dipole}\\
    \cline{1-5}
    Application &  \begin{tabular}{c}Blocking Rydberg excitation of one atom by\\the interaction from a nearby Rydberg atom\end{tabular} &   Phase shift & Antiblockade& \begin{tabular}{c}Two-atom state flip\end{tabular}  \\
    \cline{1-5}
    Proposed in&  \multicolumn{2}{|c|}{ \cite{PhysRevLett.85.2208}~(individual atoms); \cite{Lukin2001}~(ensembles of atoms)} & \cite{Ates2007}&  (Intrinsic) \\
      \cline{1-5}
     \begin{tabular}{c}Experiments\\ ({\footnotesize CNOT},~C$_{\text{Z}}$,\\ Toffoli)\end{tabular} & \begin{tabular}{c}Two atoms: \cite{Isenhower2010,Zhang2010,Maller2015,Zeng2017,Levine2019,Graham2019}\\  Three atoms~(Toffoli): \cite{Levine2019}  \\            Two photons: \cite{Tiarks2019}\end{tabular}   &   &  &\\
      \cline{1-1}\cline{2-5}
     \begin{tabular}{c}Experiments\\ (Individual-atom\\Entanglement)\end{tabular} & \begin{tabular}{c}Two atoms: \cite{Wilk2010,Isenhower2010,Zhang2010,Maller2015,Zeng2017,Levine2018,Picken2018,Levine2019,Graham2019,Madjarov2020}\\        Twenty atoms: \cite{Omran2019}\\
         Two photons: \cite{Tiarks2019}\end{tabular}   &   \cite{Jo2019}& &\\\hline
     \begin{tabular}{c}   Achievable gate\\or entanglement \\fidelity in theory \end{tabular}  &\begin{tabular}{c}2-qubit: $99.99\%$\\ 4-qubit: $99.97\%$~\tnote{b}\\(see Table~\ref{table3})\end{tabular}&  &\begin{tabular}{c} 2-qubit:$99.98\%$\\ 3-qubit: $99.65\%$ \\~(see Table~\ref{table5})\end{tabular} &\begin{tabular}{c} 2-qubit:$99.995\%$\\ 3-qubit: $99.9\%$ \\~(see Table~\ref{table4})\end{tabular} \\
   \hline    
  \end{tabular}
     \caption{ Summary of quantum entanglement and logic gates based on two-body Rydberg interactions of neutral atoms. There are in general two categories, the strong blockade regime and the frozen interaction regime, where the former was widely studied because of its robustness to the fluctuation of Rydberg interactions.   \label{table6}  }
     \begin{tablenotes}
  \item[a] For the gates in the frozen interaction regime, the methods in Secs.~\ref{sec04C} and~\ref{sec04D} can also be robust against the fluctuation of interatomic interactions.
  \item[b] A four-qubit gate is quoted here because Table~1 on page 763 of Ref.~\cite{Isenhower2011} listed data for gates C$_k$NOT with $k\geq3$. The fidelity of Toffoli gate C$_2$NOT should be larger than 99.97\% according to Eq.~(2) of Ref.~\cite{Isenhower2011}. 
    \end{tablenotes}
     \end{threeparttable}
\end{table*}

\section{Challenges}\label{discussions}
\subsection{Fidelity of Rydberg gates}
The most visited experimental method used for Rydberg-mediated entanglement is via the blockade mechanism, as summarized in Table~\ref{table6}. This is because it is robust against the fluctuation of the Rydberg interaction $V$ from shot to shot; within each experimental cycle, $V$ will also change, and the blockade method is also robust to it. To suppress the blockade error, only very large $V$ can be useful because the blockade error scales as $\sim\hbar^2\Omega^2/(2V^2)$ shown around Eq.~(\ref{blockadeerror}). The blockade error can be made negligible by using very large direct dipole-dipole interaction. Although the direct dipole-dipole interaction is large for high-lying Rydberg states, the inter-state energy spacing between two nearby Rydberg levels is small. This makes it challenging to suppress leakage error unless prestigious pulse shaping is employed~\cite{Theis2016} if very high Rydberg states are used. Another challenge to realize gates with a large $V$ lies in that when placing two qubits too close, it is difficult to avoid crosstalk concerning the Rydberg excitation of a selected atom. For this reason, the value of $V$ is well below $h\times100$MHz in typical experiments with alkali-metal atoms~\cite{Wilk2010,Isenhower2010,Zhang2010,Maller2015,Jau2015,Zeng2017,Levine2018,Picken2018,Levine2019,Graham2019} although the entanglement experiment with strontium atoms used a much larger interaction $V/h\sim 130$~MHz~\cite{Madjarov2020}.

Even if the blockade error is suppressed, the motional dephasing is another stumbling block. Motion-induced dephasing is prevailing because the atom is not static even if it is cooled to the motional ground state in the dipole trap. If the atoms are released during the Rydberg pulses, the motional effect within each gate cycle is actually not random, but is dependent on a linear effect as studied in detail in Ref.~\cite{Shi2020}. The average effect of multiple gate cycles, however, appears as a dephasing, which is in close analogy to the nuclear spin dephasing in solid-state systems~\cite{Liu2010}. By using the linear effect of the phase accumulation due to the atomic drift, it was shown that either by using extra laser fields for phase compensation during the gap times, or by using a ``V''-type dual-rail Rydberg excitation method, one can almost completely suppress the motional dephasing~\cite{Shi2020}.

The current stage in experiments is still about removing technical imperfections~\cite{Levine2018,Levine2019,Graham2019}. Only after these issues are tackled will the suppression of the blockade error and the dephasing error be interesting. However, given the high fidelity reported in~\cite{Madjarov2020} with rapid control thanks to the large Rabi frequencies realized, there is possibility to realize ground-state entanglement by Rydberg interactions with a fidelity limited only by the two factors mentioned above. With all the strengths and weaknesses for each entangling method discussed, however, there are hope as well as extreme challenges toward a realistic large-scale neutral-atom quantum computer.

\subsubsection{Suppressing blockade error by spin echo}\label{sec06A1}
One method to suppress the blockade error is by spin echo as shown in Ref.~\cite{Shi2018prapp2}. To understand the spin-echo method, one should first understand that the fluctuation of $V$ in Sec.~\ref{sec03A} denotes that in different runs of the gate, the fluctuation of the two-qubit spacing leads to randomness of $V$. In a first approximation, although the initial value of $V$ at the beginning of each gate cycle can differ from the presumed blockade interaction $V_0$, it does not change much during the gate sequence. Thus, one can start to understand the spin echo by first assuming a frozen interaction $V=V_0$, and then analyze fluctuation of the interaction around $V_0$ in different gate cycles.

The essence of the spin-echo Rydberg gate is to break the second $2\pi$ pulse in Fig.~\ref{figure-originalgate}(a) into three pulses. First, a $\pi$ pulse is used to induce the transition $|1\rangle\leftrightarrow|r\rangle$ for the target qubit with a Rabi frequency $\Omega$. For the input state $|11\rangle$, the relevant Hamiltonian is given in Eq.~(\ref{sec03A02}) with $\{|u\rangle,~|l\rangle\}=\{|rr\rangle,~|r1\rangle\}$ and $D=V_0$. Second, a microwave $\pi$ pulse is used to transfer the Rydberg state $|r\rangle$ to another Rydberg state $|r'\rangle$ with a Rabi frequency $i\Omega_\mu$, where $\hbar\Omega_\mu\gg|V|$. The microwave fields induce a Rydberg state flip $|r\rangle\rightarrow|r'\rangle$ for both the control and target qubits. The interactions of the states $|rr\rangle$ and $|r'r'\rangle$ are $V_0$ and $V_0'$, respectively, where $V_0$ and $V_0'$ have opposite signs, i.e., one is attractive and the other is repulsive. Third, a $\pi$ pulse is used to induce the transition $|r'\rangle\rightarrow|1\rangle$ with a Rabi frequency $\Omega'$, where $\Omega'=\Omega V_0'/V_0$. The Hamiltonian for the third pulse here is
\begin{eqnarray}
  \hat{H}' &=&(\frac{\hbar\Omega'}{2} |r'r'\rangle\langle r'1|+\text{H.c.}) +V_0' |r'r'\rangle\langle r'r'|.\label{sec03B01}
\end{eqnarray}
If we ignore the duration of the microwave pulse, the state evolution in response to the three pulses described above is as $|\psi'\rangle=e^{-i\hat{H}'\frac{\pi}{\hbar\Omega'}}\hat{\mathscr{M}}e^{-i\hat{H}\frac{\pi}{\hbar\Omega}}|\psi\rangle$, where $|\psi\rangle$ is the initial two-qubit state and $\hat{\mathscr{M}}$ represents the time evolution operator during the microwave pulse. One can easily verify that $|\psi'\rangle=|\psi\rangle$ because $\hat{H}'\frac{\pi}{\hbar\Omega'}+\hat{H}\frac{\pi}{\hbar\Omega}=0$ when we replace $|r'\rangle$ and $\langle r'| $ by $|r\rangle$ and $\langle r|$ in $\hat{H}'$.

The above discussion about the spin echo has ignored the duration of the microwave pulses, and no phase appears in the qubit states. To induce a C$_{\text{Z}}$ gate and to include the finite duration of the microwave pulses, one should correctly set the phase for the fields used in the pulses. A detailed analysis can be found in Ref.~\cite{Shi2020jpb}, where the fluctuation of $V$ around $V_0$ was studied. One can see that for the spin echo to work, the key step is to know the value of $V_0'/V_0$. If the spacing and relative angular position of the qubits do not significantly change between the first and third pulses in the echo sequence, the value $V_0'/V_0$ is simply the ratio of the two $C_6$ coefficients of the two states $|r'r'\rangle$ and $|rr\rangle$. Even if the traps are shallow so that the shot-to-shot fluctuation of the initial locations of the qubits in their traps is large, the ratio between the qubit interactions $V'$ and $V$ in the third and the first pulses are still given by the ratio of the $C_6$ coefficients as long as the qubit flight is slow during the pulse sequence. This means that the spin echo sequence is robust to the fluctuation of the interaction. The numerical analysis in Ref.~\cite{Shi2018prapp2} showed that a large fidelity over $99.99\%$ is achievable with Rydberg-state decay at $4.2$~K~(see the lower left on page 6 of Ref.~\cite{Shi2018prapp2}).

Compared to the methods in Sec.~\ref{sec03}, the spin-echo method can suppress the blockade error and the suppression is robust against the fluctuation of the Rydberg interaction. To our knowledge, the spin-echo Rydberg gate in Ref.~\cite{Shi2018prapp2} is the only method in the blockade regime that can suppress the blockade error without using pulse shaping. Even so, neither the spin-echo method nor the methods in Sec.~\ref{sec03} can suppress the Doppler dephasing and it is an open question whether one can find an experimentally friendly way to suppress the blockade error as well as the motional dephasing. 
\subsubsection{Suppressing blockade error by pulse shaping}\label{sec06A2}
One possible solution to the issue of blockade error may come from pulse shaping. There are in general two strategies to use pulse shaping for suppressing the blockade error. First, the method of analytical derivative removal by adiabatic gate reviewed in Sec.~\ref{sec03CDRAG} can remove the blockade error in the blockade regime. Notice that the blockade error in the standard Rydberg gate is intrinsic since the Rydberg interaction $V$ can not be too large. If very high Rydberg states are used, the energy levels are too dense which makes it challenging to selectively address the desired Rydberg state. Section~\ref{sec03CDRAG} has reviewed a pulse-shaping method to use high enough Rydberg states for Rydberg gates, where a high fidelity 99.99\% is achievable. The comparison in Table~\ref{table3} shows that the pulse-shaping strategy is among the most accurate gate protocols. Though promising, the theory in Sec.~\ref{sec03CDRAG} still depends on large enough interatomic interactions. If the atoms are placed too closely, selectively addressing~(as required in information loading and reading) may be challenging.

Second, adiabatic pulses with F\"{o}rster-resonant dipole-dipole interactions in Refs.~\cite{Beterov2016,Petrosyan2017} as reviewed in Secs.~\ref{sec04C} and~\ref{sec04D} can remove the blockade error. These two strategies are robust against the fluctuation of the interatomic interactions as shown in Fig.~\ref{figure-petrosyn} for a numerical simulation with the example of Ref.~\cite{Petrosyan2017}. Experiments have not exploited Rydberg-mediated entanglement via the F\"{o}rster-resonant dipole-dipole interactions so far.  

Even if the blockade error is suppressed either by spin echo or by pulse shaping, they require multi-pulses to work which inevitably leads to excess Doppler dephasing as in the original Rydberg blockade gate. It is an open question whether there is any method to simultaneous suppress the blockade error and the motional dephasing for designing an efficient two-qubit quantum entangling gate with the maximal entangling power~(such as the C$_{\text{Z}}$ and {\footnotesize CNOT})~\cite{Williams2011}. With the many pulse-shaping methods as reviewed in Sec.~\ref{sec03C}, we hope that certain pulse-shaping methods may lead to suppression of both the blockade error and the dephasing error.

\subsubsection{Suppressing blockade error and motional dephasing by gates with collectively-encoded qubit}\label{sec06A3}
Another possible solution to the above issues is to explore quantum entangling gates with qubits defined by, for example, the single-ensemble collective states as reviewed in Sec.~\ref{qubitencoding} and shown in Fig.~\ref{figure-encoding}(c). Almost all Rydberg quantum gates reviewed in Sec.~\ref{sec03},~\ref{sec04}, and~\ref{sec05} use the first method for defining qubits as reviewed in Sec.~\ref{qubitencoding} and shown in Fig.~\ref{figure-encoding}(a), i.e., using hyperfine-Zeeman substates from single atoms as qubit states~(or its superatom version as in Sec.~\ref{sec03Censemble}). For a large-scale quantum computer, the many qubits defined by many atoms may lead to a rapid scaling of the Rydberg-state decay and motion-induced dephasing errors in the hyperfine-qubit method. But if one ensemble can carry the information of many qubits, then there can be novel ways to shrink the Rydberg-state decay and motional dephasing.

\subsubsection{Suppressing blockade error by the method of Ref.~\cite{Levine2019}  }\label{sec06ALevine}

As reviewed in Sec.~\ref{sec03detuned01}, the interference gate explored in~\cite{Levine2019} also has a blockade error because it depends on the Rydberg blockade mechanism. The supplemental material of Ref.~\cite{Levine2019} discussed a mechanism to suppress the blockade error based on a static $V$ throughout the gate sequence, which shows that a finite $V$ leads to a modification of the detuning by an amount of $\Omega^2\hbar/(2V)$ by the renormalization of $(|1r\rangle+|r1\rangle)/\sqrt{2}$ concerning its excitation to the two-atom Rydberg state $|rr\rangle$. In other words, the value of $\Delta$ in Eqs.~(\ref{evolutionfor11})-(\ref{alpharelation03}) should be updated according to the renormalization $(|1r\rangle+|r1\rangle)/\sqrt{2}$. This means that in theory the blockade error can be suppressed in the gate of~\cite{Levine2019}. 

The method for suppressing the blockade error discussed in~\cite{Levine2019} is based on that $V$ is a van der Waals interaction which does not change from shot to shot, nor does $V$ change during the gate sequence within one gate cycle. In practical experiments, however, $V$ does fluctuate, which makes the blockade error irremovable. Even if $V$ is frozen, the renormalization of the state $(|1r\rangle+|r1\rangle)/\sqrt{2}$ in Eq.~(\ref{uldefinition}) results in admixing of the state $|rr\rangle$ in it, which leads to a state dynamics slightly different from that shown in Sec.~\ref{sec03detuned01}. But for sufficiently large $V$, the admixing of the state $|rr\rangle$ is negligible. This means that the possibility to suppress the blockade error for the gate in~\cite{Levine2019} is intimately related with the capability to realize large van der Waals interaction $V$ and to suppress the motion of the qubits. 

\subsubsection{Suppressing blockade error and motional dephasing by adequate cooling of alkali-earth-like atoms }\label{sec06A4}
The discussions in Secs.~\ref{sec06A1},~\ref{sec06A2},~\ref{sec06A3}, and~\ref{sec06ALevine} focus on a theoretical attack of the gate errors when alkali-metal atoms are used for qubits. If qubits show negligible position fluctuation, the blockade error or dephasing error can be absent by the methods reviewed in Secs.~\ref{sec04} and~\ref{sec05} which are not based on the blockade mechanism. Nevertheless, it is difficult to cool alkali-metal atoms to an extent where the position fluctuation is negligible for Rydberg quantum gates.

Another route toward neutral-atom quantum computation is by using atoms with two valence electrons and a nonzero nuclear spin, i.e., some alkaline-earth-metal or lanthanide atoms, which we call alkaline-earth-like~(AEL) atoms. Compared to alkali-metal atoms, divalent AEL atoms can be more easily cooled~\cite{Yamamoto2016,Saskin2018,Cooper2018,Covey2019,Madjarov2020,Norcia2018}, their nuclear spin states can be preserved during cooling~\cite{Reichenbach2007}, and both their ground and Rydberg states can be trapped in one trap~\cite{Wilson2019}. These in principle mean that AEL qubits can show less position fluctuation if they are used in Rydberg-mediated entanglement generation. However, previous studies on quantum computing with AEL atoms~\cite{Jaksch1999,Calarco2000,Hayes2007,Daley2008,Gorshkov2009,Daley2011,Daley2011qip,Pagano2019,Jensen2019} rarely used Rydberg interactions; though a recent Rydberg blockade experiment~\cite{Madjarov2020} with AEL atoms reported a two-atom entanglement fidelity $99.5\%$, the entanglement in~\cite{Madjarov2020} was between a Rydberg state and a metastable state $(5s5p)^{3}P_0$ of the nuclear-spin-free strontium-88. It is an open question whether the versatile and accurate control over AEL atoms can significantly suppress the blockade error and motional dephasing for high-fidelity quantum entanglement between stable states.
 
\subsection{Three-qubit gates}
To execute a computation task with a quantum circuit, not only high-fidelity two-qubit entangling gates are needed, but also high-fidelity error correction schemes should be at hand, in which the three-qubit Toffoli gate~\cite{Williams2011} plays an important role~\cite{Nielsen2000}. Rydberg Toffoli gates have been extensively studied in theory~\cite{Brion2007pra,Shi2018prapp,Levine2019,Isenhower2011,Khazali2020}, but a high-fidelity realization is difficult in experiment. Though there are Toffoli or other three-qubit gates listed in Tables~\ref{table4} and~\ref{table5}, these methods depend on frozen interactions or well isolated dipole-dipole processes which demand extreme efforts to realize in real experiments. Until now, there was one experimental study of Toffoli gate reported in Ref.~\cite{Levine2019} where the blockade mechanism was used so as to be robust against the fluctuation of Rydberg interactions. Although both two and three-qubit gates were realized by global Rydberg pulses, Ref.~\cite{Levine2019} reported an experimental Toffoli-gate fidelity 87.0\% which is about 10\% smaller than their theoretical estimate; in comparison, the realized two-qubit gate fidelity was 97.4\% in Ref.~\cite{Levine2019}, which is less than 3\% smaller than their theoretical estimate. This contrast shows the difficulty to carry out high-fidelity three-qubit Rydberg gates.  

The challenge to realize the Toffoli gate lies in that for a three-atom system lying on one line, there will be some residual dipole-dipole interaction between the two outermost atoms when they are both in the Rydberg states, which not only leads to blockade error, but also induces unwanted phase shifts. Typical solutions to this problem include (i) to excite the three qubits to three different Rydberg states, so that the interaction between the two edge atoms is negligible~\cite{Brion2007pra}; (ii) to choose an appropriate gate duration so as to effectively annul the induced phase shift~\cite{Shi2018prapp}; (iii) to use microwave fields to induce zero van der Waals interaction~\cite{Shi2018prapp,Shi2017pra}; (iv) to use optimal control to find appropriate time-dependent Rabi frequency and detuning in the pulse~\cite{Levine2019}. Theoretically, the last method has a fidelity bounded below $98\%$ as analyzed in~\cite{Levine2019}, and although the methods (i-iii) in principle can lead to large fidelities, the residual blockade between the edge atoms in method (i) and the fluctuation of $V$ in methods (ii,iii) set an intrinsic challenge in their implementation. There is also an adiabatic method using dark states for implementing a high-fidelity Toffoli gate~\cite{Khazali2020}, as well as a fast Toffoli gate protocol based on three-atom F\"{o}rster resonance~\cite{Beterov2018arX}, where the fidelity depends on how well one can isolate the chosen dipole-dipole processes. 

\section{Conclusions}\label{concl}
There are in general three mechanisms to generate quantum entanglement between individual atoms by Rydberg interactions, namely, the blockade mechanism, the excitation of multi-atom Rydberg states, and the antiblockade mechanism. The theoretical entanglement and gate fidelities using them are listed in Tables~\ref{table3},~\ref{table4}, and~\ref{table5}, from which we can see that the largest theoretical fidelity for two-qubit gates is 99.995\% with the method of excitation of multi-atom Rydberg state, and the largest fidelity for multiqubit gates is 99.97\% via the blockade mechanism. The total numbers of entanglement fidelities over or equal to 99.9\% (with qubits hosted at room temperature) are eight, two, and five in Tables~\ref{table3},~\ref{table4}, and~\ref{table5}, respectively. Though there are five fidelities over or equal to 99.9\% in Table~\ref{table5}, the analyses leading to them ignored the fluctuation of interactions. Even if the fluctuation of atomic spacing is absent, the comparison between Tables~\ref{table3},~\ref{table4}, and~\ref{table5} shows that in theory, it seems more possible to reach high-fidelity quantum gates and entanglement via the blockade mechanism, i.e., via the methods reviewed in Sec.~\ref{sec03}.

The fundamental errors for Rydberg-mediated entanglement via the blockade mechanism stem from Rydberg-state decay, rotation errors due to imperfect blockade, and motional dephasing in the ground-Rydberg rotations. The Rydberg-state decay can be made smaller and smaller, but not removable. Whatever the method is, there is always motional dephasing during the control of the ground-Rydberg transition by using electromagnetic radiations. Although fortunately there are several theories coping with the motional dephasing, the blockade error is a real issue once large fidelity is desired. This is because the fidelity by using Rydberg blockade effect is always hampered by the blockade error, which is difficult to be completely suppressed because it is challenging to accurately address the two qubits without crosstalk if they are too close to each other; so, large enough qubit spacing is necessary which limits the magnitude of the blockade interaction. The fidelity by using the frozen interactions in Secs.~\ref{sec04} and~\ref{sec05} can avoid the blockade error; but most of them~(except the ones as in Refs.~\cite{Beterov2016,Petrosyan2017}) depend on negligible fluctuation of the qubit spacing. These methods are physically interesting though it is an outstanding challenge to suppress the position fluctuation of qubits to a level where the motion-induced dephasing and change of interaction disappear. Given the fact that many Rydberg-mediated gates have been proposed as reviewed in Secs.~\ref{sec03},~\ref{sec04}, and~\ref{sec05}, there are enough reasons to expect new physics and new protocols on Rydberg gates in near future that can indeed push the Rydberg-mediated quantum computing to a practical level.

\section*{ACKNOWLEDGMENTS}
The author thanks the anonymous referees for their valuable comments and suggestions, and thanks T. A. B. Kennedy and Yan Lu for discussions. This work is supported by the National Natural Science Foundation of China under Grants No. 12074300 and No. 11805146, the Natural Science Basic Research plan in Shaanxi Province of China under Grant No. 2020JM-189, and the Fundamental Research Funds for the Central Universities.

\appendix{}
\section{Entanglement of photons via Rydberg interactions }\label{sec03D}
Rydberg interaction is not only useful in quantum information processing, but also in quantum optics~\cite{Dudin2012,Peyronel2012,Firstenberg2013,Li2013,Gorniaczyk2014,Baur2014,Tiarks2014,Li2016pan,Thompson2017,Busche2017,Paris-Mandoki2017,Ripka2018,Liang2018,Li2019}; for a review, see Ref.~\cite{Firstenberg2016}. In particular, it was shown that by using Rydberg blockade one can achieve photonic quantum entanglement by mapping single photons to Rydberg polaritons~\cite{Paredes-Barato2014}. In a Rydberg polariton, the Rydberg excitation is shared in many atoms in a collective way, and the blockade interaction induces phase shifts to the polariton, which can be transfered to the output photon when released in the formalism of EIT~\cite{Fleischhauer2002}. The key step for realizing this method has been experimentally tested~\cite{Busche2017}. Later on, Ref.~\cite{Khazali2015} proposed an improved version of the photon-photon gate. Although both methods depend on the dual-rail coding~\cite{Chuang1995}, the difference between~\cite{Paredes-Barato2014} and~\cite{Khazali2015} lies in that the qubit states $|0(1)\rangle$ are all stored in Rydberg dark-state polaritons in~\cite{Paredes-Barato2014}, but only one of the two qubit states is stored in the Rydberg polaritons in~\cite{Khazali2015} and the other is stored in ground-state polaritons. In principle, these methods can induce high-fidelity photon entanglement if the efficiency for loading and retrieving of single photons is high. However, the dephasing of spin wave is a critical issue in a practical implementation~\cite{Dudin2012,Jenkins2012,Schmidt-Eberle2019}. 

Another method for photon entanglement is to pass a single-photon pulse through a medium where there is already a Rydberg collective excitation in the medium, so that the Rydberg interaction between the flying Rydberg polariton and the static Rydberg excitation induces a phase shift~\cite{Friedler2005,Gorshkov2011}. The first photon-photon quantum entangling gate based on Rydberg interaction was experimentally realized with this method, as reported in Ref.~\cite{Tiarks2019}. Because there is inevitable overlap between the static and flying Rydberg excitation, heavy dissipation arises which puts fundamental limit to the achievable fidelity~\cite{Gorshkov2011}. The benefit of this method, however, lies in that it is relatively simple to manipulate the system and there is less motional dephasing because only one of the two qubits is stored in the method of~\cite{Tiarks2019}. 

There are also entanglement methods by coupling Rydberg atoms with cavity photons~\cite{Hao2015,Das2016} via the scheme proposed in Ref.~\cite{Duan2004}, where entanglement can emerge between a stored photon (in the form of collective spin wave) and a flying photon.

\section{Transition slow-down}\label{TSD-appendix}
The mechanism of transition slow-down~(TSD) proposed in Ref.~\cite{Shi2020tsd} belongs to the Rydberg blockade regime. It involves Rydberg excitation of both qubits, and the qubits should be placed close enough so that strong dipole-dipole interactions can appear. Because of the particular applicability of TSD in realizing the {\footnotesize CNOT} gate, we show two examples of {\footnotesize CNOT} below via TSD. Though these examples have gate speeds not as fast as that in Ref.~\cite{Shi2020tsd}, they can illustrate why a ``slow-down'' effect can help in designing a quantum gate.

\begin{figure*}
\includegraphics[width=6.1in]
{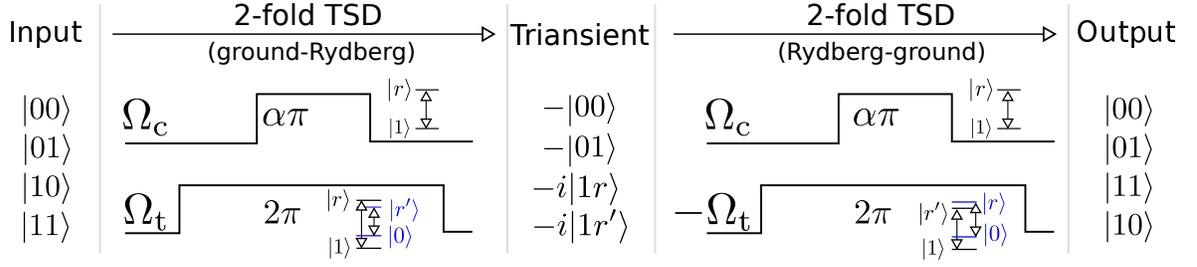}
 \caption{A two-step {\footnotesize {\footnotesize CNOT}} realized by a 2-fold TSD in a ground-Rydberg transition, where either the first or the second step consists of a pulse sequence described around Eqs.~(\ref{appBeq01}) and (\ref{appBeq02}) with $\alpha=\sqrt{15}$. In each step, the pulse areas for the control and target qubits are $\alpha\pi$ and $2\pi$, respectively. Here $|r\rangle$ and $|r'\rangle$ are different Rydberg eigenstates with different principal quantum numbers. For the target qubit, the Rabi frequencies for $|0(1)\rangle\leftrightarrow|r'(r)\rangle$ are both $\Omega_{\text{t}}$ in the first step, while those for $|1(0)\rangle\leftrightarrow|r'(r)\rangle$ are both $-\Omega_{\text{t}}$ in the second step.  \label{figure01-app} }
\end{figure*}

The first method uses a pulse sequence shown in Fig.~\ref{figure01-app}, which consists of two pulses. We study the effect of the first pulse. The pulse for the control qubit is applied during $t\in [t_0,~t_0+t_\pi)$, and the pulse for the target qubit is applied during $t\in[0,~2t_\pi)$, where $t_0\in[0,~t_\pi)$ and $t_\pi=\pi/\Omega_{\text{t}}$. Starting from an initial two-atom state $|11\rangle$, the wavefunction at $t=t_0$ becomes 
\begin{eqnarray}
 |\psi(t_0)\rangle &=&\cos(\Omega_{\text{t}}t_0/2) |11\rangle -i\sin(\Omega_{\text{t}}t_0/2) |1r\rangle. \label{appBeq01}
\end{eqnarray}
During $t\in[t_0,~t_0+t_\pi)$, the system Hamiltonian is
  \begin{eqnarray}
    \hat{H} &=&\hbar[ (\Omega_{\text{c}}|r1\rangle + \Omega_{\text{t}}|1r\rangle)\langle11|/2+\text{H.c.}]+\hat{H'},\label{appBeq02}
\end{eqnarray}
  where $\hat{H'}$ includes excitation from $|1r\rangle$ and $|r1\rangle$ to $|rr\rangle$ and dipole-dipole interaction involving $|rr\rangle$ and nearby dipole-coupled states. In the strong interaction regime, $\hat{H'}$ can be discarded, and $ \hat{H}=\hbar\bar\Omega |\mathscr{R}_+\rangle\langle11|/2 +\text{H.c.}$, where $\bar\Omega\equiv\sqrt{\Omega_{\text{c}}^2+\Omega_{\text{t}}^2}$ and $|\mathscr{R}_+\rangle\equiv (\Omega_{\text{c}}|r1\rangle + \Omega_{\text{t}}|1r\rangle)/\bar\Omega$. When $\bar\Omega t_\pi = 4k\pi$ is satisfied with an integer $k$, the wavefunction at $t=t_0+t_\pi$, given by exp$[-i  \hat{H}t_\pi/\hbar]|\psi(t_0)\rangle $, is equal to Eq.~(\ref{appBeq01}). The state for the target qubit becomes $|1r\rangle$ at $t=2t_\pi$. So, a $2\pi$ pulse leads to the excitation from $|11\rangle$ to $|1r\rangle$. The smallest $\alpha\equiv\Omega_{\text{c}}/\Omega_{\text{t}}$ satisfying the above condition is $\alpha=\sqrt{15}$. By this, a fast and high-fidelity {\footnotesize CNOT} is shown in Fig.~\ref{figure01-app}, where two Rydberg states $|r\rangle$ and $|r'\rangle$ are involved. As can be easily verified, the pulse sequence simply leads to the state mapping $\{|00\rangle, |01\rangle, |10\rangle, |11\rangle \}\rightarrow -\{|00\rangle, |01\rangle, i|1r'\rangle, i|1r\rangle \}$ when the transitions $\{|1\rangle, |0\rangle\}\leftrightarrow\{|r\rangle, |r'\rangle\} $ are excited by a time $2\pi/\Omega_{\text{t}}$ for the target qubit with Rabi frequencies $\Omega_{\text{t}}$, within which the transition $|1\rangle\leftrightarrow|r\rangle$ is excited by a time $\pi/\Omega_{\text{t}}=\alpha\pi/ \Omega_{\text{c}}$ for the control qubit. Subsequently, the transitions $\{|0\rangle, |1\rangle\}\leftrightarrow\{|r\rangle, |r'\rangle\} $ are excited by a time $2\pi/\Omega_{\text{t}}$ for the target qubit with Rabi frequencies $-\Omega_{\text{t}}$ while the excitation for the control qubit stays the same as in the first step. Then one can find that the input-output map corresponds to a {\footnotesize CNOT} gate. The gate has an intrinsic fidelity bounded by the Rydberg-state decay and the residual blockade error $\sim(\Omega_{\text{c(t)}}/V)^2$, where the latter can be made negligible when very large native dipole-dipole interaction is used. The Rydberg-state decay is $\epsilon=T_R/\tau$, where $T_R$ is the expected time for the atoms to stay in the Rydberg state averaged over the four input states, and $\tau$ is the lifetime of the Rydberg state. By using $t_0=0$, we find that the times for the input states $|00\rangle$ and $|10\rangle$~(or, equivalently, $|01\rangle$ and $|11\rangle$) to be in the Rydberg state are respectively $\pi/(2\Omega_{\text{t}})$ and $\pi/(\Omega_{\text{t}})$ if $\alpha=\sqrt{15}$ for either of the two pulses in Fig.~\ref{figure01-app}, so we have $T_R=3\pi/(4\Omega_{\text{t}})$ when there is no gap time between the two pulses for the target. For $\tau\sim330~\mu$s corresponding to s-orbital rubidium states with principal quantum numbers around $100$ at room temperature~\cite{Beterov2009}, the Rydberg-state decay leads to a gate infidelity $\epsilon=T_R/\tau\approx1.1\times10^{-3}$ if $\Omega_{\text{t}}/(2\pi)=1$~MHz, i.e., a fidelity $99.89\%$ is realizable. A higher fidelity is possible with larger Rabi frequencies.

\begin{figure*}
\includegraphics[width=5.5in]
{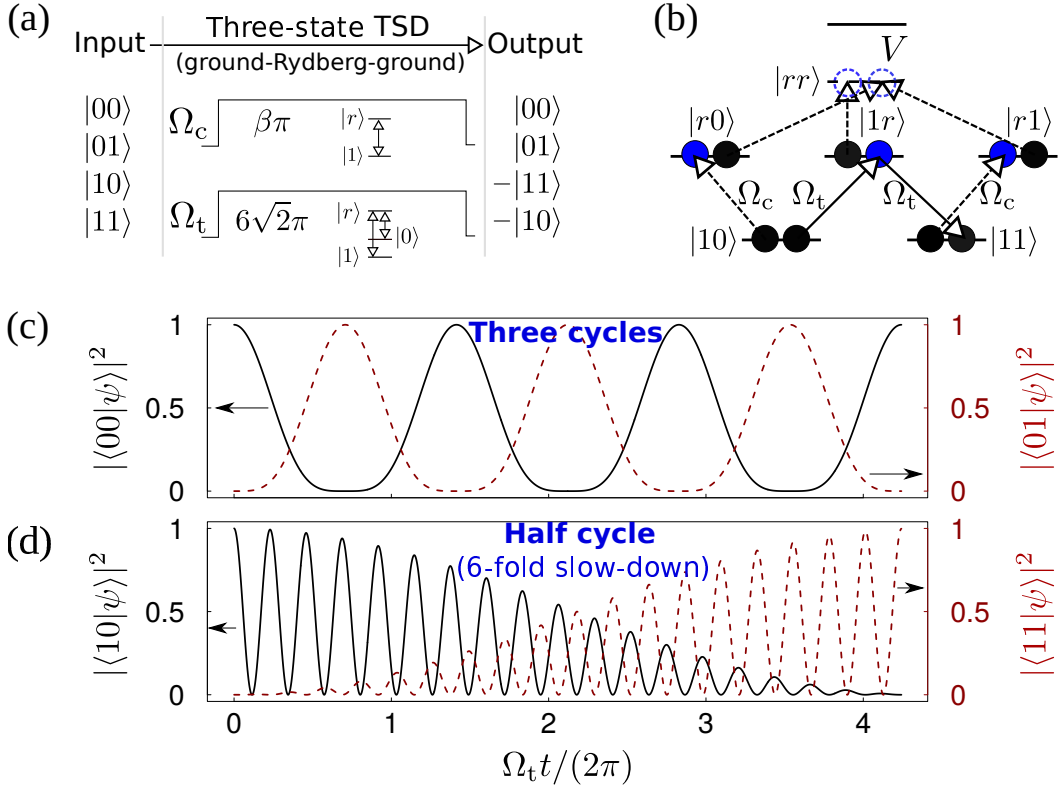}
 \caption{A three-state transition slow-down and a one-shot {\footnotesize CNOT}. (a) A three-state TSD is realized by sending to each qubit one pulse with equal duration. The pulse areas are $\beta\pi$ and $6\sqrt{2}\pi$ for the control and target qubits, respectively. (b) Relevant transitions for the input states $|10\rangle$ and $|11\rangle$. Transitions involving $|00\rangle$ and $|01\rangle$ are not shown because they do not involve dipole-dipole interactions. That the arrows point to the left or right does not mean transitions with change of angular momentum $\pm\hbar$, but only for clarifying the protocol.  (c) If the input state is $|00\rangle$, it transitions to $|01\rangle$ and back within a time of $2\sqrt{2}\pi/\Omega_{\text{t}}$. Three cycles are shown with a pulse duration $6\sqrt{2}\pi/\Omega_{\text{t}}$. (d) With the same excitation, the input state $|10\rangle$ just goes to $|11\rangle$ at the end of the pulse. The population error at the end of the pulse is zero in (c) and $1.8\times10^{-3}$ in (d). The contrast between (c) and (d) shows that the transition between $|0\rangle$ and $|1\rangle$ in the target qubit is slowed down by 6 folds. Here $\Omega_{\text{c}}=4.245739\Omega_{\text{t}}$ is used corresponding to $\beta\approx36.0$. Figure (b) is adopted from Phys. Rev. Appl. 14, 054058~(2020). \label{figure02-app} }
\end{figure*}

Another TSD method in three states can lead to a one-shot {\footnotesize CNOT}. We study a ground-Rydberg-ground transition chain $|0\rangle\leftrightarrow|r\rangle\leftrightarrow|1\rangle$, i.e., a state swap between the two qubit states $|0\rangle$ and $|1\rangle$ via a Rydberg state $|r\rangle$. The three-state TSD is implemented by exciting the control and target qubits with respective Rabi frequencies $\Omega_{\text{c}}$ and $\Omega_{\text{t}}$ for the same duration, with transition $|1\rangle\leftrightarrow|r\rangle$ for the control qubit and $|0\rangle\leftrightarrow|r\rangle\leftrightarrow|1\rangle$ for the target qubit. Because $|0\rangle$ is not excited for the control qubit,  the Hamiltonian is simply $\hbar\Omega_{\text{t}}(|00\rangle+|01\rangle)\langle 0r|/2+$H.c. for the input states $|00\rangle$ and $|01\rangle$. One can show that a superposition of these two states is preserved when the excitation only lasts for a duration $t=2\sqrt{2}k_1\pi/\Omega_{\text{t}}$ with an integer $k_1$~\cite{Shi2018prapp}. For the remaining input states, we consider the ordered basis $\{|1r\rangle,|r1\rangle,|r0\rangle,|11\rangle,|10\rangle \}$ and the following Hamiltonian
\begin{eqnarray}
  \hat{H}&=&\frac{\hbar}{2} \left(\begin{array}{ccccc}
0&0&0& \Omega_{\text{t}}&\Omega_{\text{t}}\\
0&0&0& \Omega_{\text{c}}&0\\
0&0&0&0& \Omega_{\text{c}}\\
\Omega_{\text{t}}& \Omega_{\text{c}}&0&0&0\\
\Omega_{\text{t}}&0& \Omega_{\text{c}}&0&0
  \end{array}\right),
  \label{eq03}
\end{eqnarray}
where  we ignore the two-Rydberg state $|rr\rangle$ with reasons shown below Eq.~(\ref{appBeq02}). By diagonalizing Eq.~(\ref{eq03}), one can show that
\begin{eqnarray}
 |10\rangle &=& (-|\mathscr{R}_1\rangle+|\mathscr{R}_2\rangle-|\mathscr{R}_3\rangle+|\mathscr{R}_4\rangle )/2,\nonumber\\
 |11\rangle &=& (|\mathscr{R}_1\rangle-|\mathscr{R}_2\rangle-|\mathscr{R}_3\rangle+|\mathscr{R}_4\rangle )/2,\label{eq04}
\end{eqnarray}
where $|\mathscr{R}_k\rangle$ with $k=1-4$ are four eigenvectors of Eq.~(\ref{eq03}) with respective eigenvalues $(\Omega_{\text{c}},~-\Omega_{\text{c}},~\overline\Omega,-\overline\Omega)/2$~[the fifth eigenvector does not enter Eq.~(\ref{eq04})], where $\overline{\Omega}\equiv\sqrt{\Omega_{\text{c}}^2+2\Omega_{\text{t}}^2}$. Starting from the input state $|\psi(0)\rangle=|10\rangle$, the wavefunction $|\psi(t)\rangle$ at $t$ becomes
\begin{eqnarray}
\frac{1}{2} (-e^{-\frac{it\Omega_{\text{c}}}{2}}|\mathscr{R}_1\rangle+e^{\frac{it\Omega_{\text{c}}}{2}}|\mathscr{R}_2\rangle-e^{\frac{-it\bar\Omega}{2}}|\mathscr{R}_3\rangle+e^{\frac{it\bar\Omega}{2}}|\mathscr{R}_4\rangle ),\nonumber
\end{eqnarray}
which means that $|\psi(t)\rangle=-|11\rangle$ if $\Omega_{\text{c}}t=4k_2\pi$ and $\overline{\Omega} t=2(2k_3+1)\pi$ can be satisfied with two integers $k_2$ and $k_3$ for a pulse duration $t$. Then, one can show that the same condition leads to that the input state $|11\rangle$ becomes $-|10\rangle$. To summarize, if a set of integers $(k_1,k_2,k_3)$ can be found for the following conditions
\begin{eqnarray}
\Omega_{\text{t}}t=2\sqrt{2}k_1\pi,~\Omega_{\text{c}}t=4k_2\pi,~\overline{\Omega} t=2(2k_3+1)\pi,\label{eq06}
\end{eqnarray}
the following gate similar to the {\footnotesize CNOT} is realized
\begin{eqnarray}
\{|00\rangle, |01\rangle, |10\rangle, |11\rangle\}\rightarrow\{|00\rangle, |01\rangle, -|11\rangle, -|10\rangle\}\nonumber
\end{eqnarray}
with a single pulse for each qubit; application of the phase gate $|1\rangle\rightarrow -|1\rangle$ in the target qubit can recover a {\footnotesize CNOT}. Another way to realize the standard {\footnotesize CNOT} is to add a $\pi$ phase difference for the Rabi frequencies in the transitions $|0\rangle\leftrightarrow|r\rangle$ and $|1\rangle\leftrightarrow|r\rangle$ so that the standard {\footnotesize CNOT} can be realized.

Unlike the two-state TSD, it is difficult to realize a high-fidelity three-state TSD because in order to avoid large Rabi frequencies we shall consider small integers $k_j$, but unfortunately we can not find small enough integers satisfying Eq.~(\ref{eq06}). As a compromise with $k_j<10$, $j=1-3$, we find a large fidelity by using $\Omega_{\text{c}}=4.245739\Omega_{\text{t}}$ associated with $(k_1,k_2,k_3)=(3,9.007,8.993)$. The fact that $k_1$ is integer while $k_2$ and $k_3$ are approximate integers means that $\{|00\rangle, |01\rangle\}\rightarrow\{|00\rangle, |01\rangle\}$ is exact in this one-shot {\footnotesize CNOT}, while $\{|10\rangle, |11\rangle\}\rightarrow-\{|11\rangle, |10\rangle\}$ comes with some error, which is $1.8\times10^{-3}$ by numerics based on Eq.~(\ref{eq03}). This means that the error of the {\footnotesize CNOT} gate fidelity is $\epsilon_1=9\times10^{-4}$ if Rydberg-state decay and the residual blockade error are ignored. The blockade error can be made negligible when large native dipole-dipole interaction is employed, while the Rydberg-state decay remains. The numerical simulation of the 6-fold TSD in three states based on Eq.~(\ref{eq03}) is shown in Fig.~\ref{figure02-app}. We find that the time for the atoms to be in Rydberg states is $T_R=1.06\times2\pi/\Omega_{\text{t}}$ during the three cycles of the transition $|00\rangle\rightarrow|01\rangle\rightarrow|00\rangle$ shown in Fig.~\ref{figure02-app}(c). For the transition in Fig.~\ref{figure02-app}(d), the time for the atoms to experience Rydberg population is $2T_R$. So the decay error of the {\footnotesize CNOT} is $\epsilon_2=3T_R/(2\tau)$ when averaging over the four input states, where $\tau$ is the lifetime of the Rydberg state. This means that if $\Omega_{\text{t}}/(2\pi)=1$~MHz, then $\epsilon_2=4.8\times10^{-3}$ if we consider $\tau\sim330~\mu$s as in the two-pulse {\footnotesize CNOT} by the two-state TSD, leading to an infidelity $5.7\times10^{-3}$ for the one-pulse {\footnotesize CNOT}. Although this error is about five times larger than that by the two-state TSD with similar Rabi frequencies, the one-shot {\footnotesize CNOT} does not need extra switch on and off of the external control field and thus can avoid errors due to the switching of lasers to the largest extent. To decrease the Rydberg-state decay for larger gate fidelity, one can use a cryogenic chamber to host the qubits or use stronger laser fields for faster Rabi rotations.

Compared to other methods, TSD can lead to a {\footnotesize CNOT} gate with a minimal number of pulse switchings which are related with Rydberg population loss in the standard Rydberg gates~\cite{Maller2015}. Moreover, the Doppler dephasing can be suppressed in the one-shot TSD-based {\footnotesize CNOT} because there is no time required to leave a Rydberg atom in free flight that can lead to strong motional dephasing~\cite{Shi2020}. The protocol shown in Fig.~\ref{figure02-app} is the only one-pulse {\footnotesize CNOT} gate based on the blockade mechanism; there are, of course, some one-shot {\footnotesize CNOT} proposals based on the antiblockade effect reviewed in Sec.~\ref{sec05}.

%


\end{document}